\documentclass{aa}  
\usepackage{graphicx}
\usepackage{placeins}
\usepackage{txfonts}
\usepackage[colorlinks,citecolor=blue,urlcolor=blue,filecolor=blue,linkcolor=blue]{hyperref}
\usepackage{amsmath}
\usepackage{amssymb}
\usepackage{upgreek}
\usepackage{natbib}
\usepackage{xcolor}

\begin{document} 

    \title{New supernova remnant candidates in the LOFAR Two Metre Sky Survey}


   \author{K. Tsalapatas
          \inst{1}
          \and
          M. Arias
          \inst{1}
          \and T. Shimwell
          \inst{2}
          \and K. Rajwade 
          \inst{2,4}
          \and M. J. Hardcastle
          \inst{3}
          \and A. Drabent
          \inst{5}
          }

   \institute{Leiden Observatory, Leiden University, Einsteinweg 55, 2333 CC Leiden, The Netherlands\\
              \email{tsalapatas@strw.leidenuniv.nl}
         \and
             ASTRON Netherlands Institute for Radio Astronomy, Oude Hoogeveensedijk 4,7991 PD Dwingeloo, The Netherlands 
             \and
            Department of Physics, Astronomy and Mathematics, University of Hertfordshire, College Lane, Hatfield AL10 9AB, UK
        \and
            Astrophysics, University of Oxford, Denys Wilkinson Building, Keble Road, Oxford OX1 3RH, UK
            \and
            Thüringer Landessternwarte, Sternwarte 5, D-07778 Tautenburg, Germany
        }

   \date{}

 \abstract{\textit{Context:} In spite of their key role in galaxy evolution and several decades of observational efforts, the census of supernova remnants (SNRs)  in our Galaxy remains incomplete. Theoretical predictions based on the local supernova rate estimate the expected number of SNRs in the Galaxy to be $\gtrsim$ 1000. By contrast, the number of detected SNRs amounts to about 300. High-resolution, wide-area radio surveys at low frequencies are ideal tools with which to find missing SNRs, given the prominence of these sources at low radio frequencies.\\
 \textit{Aims:} We aim to find missing SNRs using proprietary data from the LOFAR Two-Metre Sky Survey (LoTSS) at 144~MHz.\\
 \textit{Methods:} We used LoTSS total intensity maps of two Galactic regions, one with $39^\mathrm{o} < l < 66^\mathrm{o}$ and $|b|< 2.5^\mathrm{o}$, and the other with $145^\mathrm{o} < l < 150^\mathrm{o}$ and $|b| < 3^\mathrm{o}$, in addition to mid-infrared (MIR) data from the Wide-field Infrared Survey Explorer (\textrm{WISE}) all-sky survey to search for SNR candidates. \\
 \textit{Results:} We report the discovery of 14 new SNR candidates selected on the basis of their morphology at 144~MHz and a lack of MIR emission. We also follow up on 24 previously reported SNR candidates, inferring their spectral index between the LoTSS frequency (144~MHz) and the frequency at which they were reported. We confirm that 6 of these 24 sources have a non-thermal spectral index, whereas another 4 have a thermal index and are thus rejected as SNR candidates; our study is inconclusive regarding the remaining 14 sources. The majority of the new SNR candidates are small in angular size ($<20'$) and have low surface brightness at 1~GHz ($\lesssim 10^{-21}$~W~m$^{-2}$~Hz$^{-1}$~sr$^{-1}$). Additionally, most of them are located in a Galactic region with $39^\mathrm{o} < l < 60^\mathrm{o}$ and $|b|< 2.5^\mathrm{o}$.\\
\textit{Conclusions:} The high resolution and sensitivity of LoTSS observations has resulted in the detection of 14 new SNR candidates. In order to unambiguously confirm the SNR nature of these candidates, follow-up X-ray observations are required with facilities such as \textit{eROSITA}.
 }

   \keywords{Supernova remnants  --
                LOFAR --
                LoTSS -- Galactic plane
               }

   \maketitle


\section{Introduction}\label{Ch:1}
\noindent Supernova remnants (SNRs) occur when an exploded star's ejecta interacts with its surrounding medium. Supernova explosions are the result of either the core collapse that marks the end of stellar evolution for massive stars (M$\gtrsim$ 8 M$_{\odot}$, \citealp{Woosley_2005}) or the thermonuclear explosion of white dwarfs that reach the Chandrasekhar limit via mass accretion or merger events (also known as type Ia supernovae; \citealp[see][ for a review on their progenitors]{Maoz_2014}). 
In the case of core-collapse supernovae, the explosion produces a compact object ---either a neutron star or a black hole \citep[see \citealp{Fryer_2013} and ][]{Heger_2003}---, whereas type Ia explosions do not result in a compact remnant \citep{Maoz_2014}. If the compact object is a pulsar, the SNR can also contain a pulsar wind nebula (PWN).

Despite their different physical mechanisms, supernovae of both explosion types are surprisingly similar in terms of the energy they deposit into the interstellar medium (ISM), 
which is typically of the order of $\sim$ 10$^{51}$~erg \citep{vink12}. This enormous amount of energy greatly affects the ISM through complex feedback mechanisms \citep[e.g. \citealp{Melioli2006}, \citealp{Martizzi_2015}][]{Lucas_2020}. Furthermore, the energy released is sufficient to accelerate cosmic rays up to $\sim 10^{15}$ eV if the acceleration efficiency is 5-10 \% \citep{Strong_2010}. For these reasons, SNRs play a crucial role in galactic evolution.

When it comes to the number of SNRs in our Galaxy, there is significant disagreement between the number of observed SNRs ($\sim$300, \citealp{Green_2022}, hereafter G22) and the number of SNRs expected on the basis of stellar population modelling \cite[$\gtrsim$ 1000, e.g.][]{Ranasinghe_2022}. The catalogue maintained and updated by D. Green\footnote{\url{https://www.mrao.cam.ac.uk/surveys/snrs/}} is the definitive compilation of Galactic SNRs reported in the literature, and its most recently updated version (G22) contains 303 confirmed SNRs. The number of expected SNRs is related to the supernova rate in the Milky Way, which is thought to be between two and three supernovae per century  \citep[see \citealp{1994ApJS...92..487T},][]{Li_2011}, and to the typical lifetime of an SNR, which is $\leq100,000$~years \citep{Gordon_1999}. This difference has typically been attributed to observational selection effects that hinder the discovery of small SNRs and/or large and faint SNRs \citep{Green_1991}.

The dominant emission mechanism from SNRs at radio wavelengths ---at which the vast majority (>90\%) of confirmed SNRs have been detected--- is synchrotron radiation, which has a negative spectral index of $\alpha \approx -0.5$ \cite[for $S_\nu \propto \nu^\alpha$,][]{Dubner_2015}. However, discovering SNRs in the radio sky can be tricky, because of a variety of observational factors: these include confusion with H\,{\sc ii} regions with similar shell-like structure, the diffuse synchrotron emission of the Milky Way, and the complexity of emission of many Galactic fields, which is both difficult to image and to interpret.

H\,{\sc ii} regions are the main source of confusion with SNRs at radio frequencies. They 
are regions of ionised hydrogen around young O and B stars. H\,{\sc ii} regions 
radiate thermally, with a spectral index of $\alpha\gtrsim-0.1$ \citep{Condon_Ransom_2016}, and, unlike SNRs, they exhibit strong emission at mid-infrared (MIR) wavelengths. Their radio morphology is often shell-like, which is why they can be misidentified as SNRs.
\citet{Anderson2011} showed that in the MIR, H\,{\sc ii} regions characteristically show $24~\mu$m emission due to heated dust that is spatially coincident with the radio emission, surrounded by $8~\mu$m emission from polycyclic aromatic hydrocarbons (PAHs).

The literature on SNR candidates from radio searches is extensive, and the following is an incomplete overview. 
\cite{helfand06} conducted the MAGPIS survey (centred at 1400~MHz with resolution of $\sim$6$''$ and a source-detection threshold of $\sim$$1-2$~mJy) and found 49 SNR candidates; all of them have been followed up and either confirmed
as SNRs or reclassified as H II regions \citep{johanson09, anderson17, hurley-walker19a}. \cite{green14} identified 23 new candidates
in the region covered by the Molonglo Galactic plane Survey (centred at 843~MHz with resolution of $\sim$45$''$ and sensitivity of $\sim$1-2~mJy~beam$^{-1}$), which were also followed up by \cite{hurley-walker19a}. \cite{anderson17} proposed
76 SNR candidates found as part of the THOR+VGPS survey \citep[centred at 1400~MHz with resolution of $\sim$25$''$ and sensitivity of $\sim$1~mJy~beam$^{-1}$;][]{beuther16}, and a further 2 were proposed by \cite{ranasinghe21}. 
Confirming candidate SNRs as fully fledged SNRs is arduous: \cite{Driessen_2018} followed up on six of the \cite{anderson17} candidates and could only confirm one
to be an SNR. 
The Northern sky has been less thoroughly observed for SNR candidates than the Southern sky. The Canadian Galactic plane Survey \cite[CGPS; with data products in two spectral windows: one at 1420~MHz with a resolution of $\sim$1$'$ and a sensitivity of $\sim$0.3~mJy~beam$^{-1}$, and the other at 408~MHz with a resolution of $\sim$3.4$'$ and a sensitivity of $\sim$3~mJy~beam$^{-1}$;][]{taylor03}
remains the best resource to date for studying SNRs in this region in survey form.

The GLEAM survey \cite[centred at 200~MHz with a resolution of $\sim$2$'$ and a sensitivity of $\sim$20~mJy~beam$^{-1}$,][]{hurley-walker17}
showcased the importance of low-frequency surveys in finding the missing SNRs. \cite{hurley-walker19a} followed up on 119 candidates
proposed in the literature, definitively and  tentatively confirming 13 and 2 as SNR candidates, respectively, and \cite{hurley-walker19b}
proposed a further 27 candidates. More recently, \cite{dokara21} searched the GLOSTAR Galactic Plane Survey \cite[centred at 5800~MHz with a resolution of $\sim$18$''$ and a sensitivity of $\sim$20~mJy~beam$^{-1}$;][]{brunthaler21} for new SNRs, and detected 157 candidates, of which 80 are new. \cite{dokara21} reclassified four of the \cite{green19} SNRs as H\,{\sc ii} regions,
and provide evidence that nine of the candidates are non-thermal emitters through polarisation measurements. 
There have been other surveys covering smaller interesting 
Galactic regions, such as that carried out by \citealp{umana21} (centred at 912~MHz with a resolution of $\sim$21$''$ and a sensitivity of $\sim$541~$\mu$Jy~beam$^{-1}$) in the SCORPIO field with the ASKAP telescope \citep{hotan21}. 
\cite{ingallinera19} found three SNRs in an Australia Telescope Compact Array (ATCA) survey of the same region, centred at 2100~MHz with a resolution of $\sim$10$''$ and a sensitivity of $\sim$100~$\mu$Jy~beam$^{-1}$, but noted that more are expected. 

The LOFAR Two Metre Sky Survey \cite[LoTSS,][]{Shimwell_2017,Shimwell_2019,Shimwell_2022} is an ongoing Northern sky survey at 144~MHz with unprecedented sensitivity ($\sim$100~$\mu$Jy~beam$^{-1}$) and angular resolution ($6''$). It is an ideal tool for finding 
new Galactic SNRs, as its high angular resolution and sensitivity tackle the selection effects that hinder the detection of small and young, and old and faint SNRs \citep{Green_1991}. Furthermore, the low-frequency coverage of LoTSS implies that H\,{\sc ii} regions ---which are thermal sources and are therefore less bright at 144~MHz than SNRs---
will be a less persistent source of confusion than they are for surveys at higher frequencies (e.g. the CGPS). 

Even though LoTSS is an ongoing project, and the existing data releases DR1 \citep{Shimwell_2019} and DR2 \citep{Shimwell_2022} comprise pointings away from the Galactic plane, some parts of the Milky Way have already been observed and reduced, and form the basis of this work. We analysed LoTSS total intensity maps of two Galactic regions, with Galactic longitude $39^\mathrm{o} < l < 66^\mathrm{o}$ and latitude $|b|< 2.5^\mathrm{o}$, and $145^\mathrm{o} < l < 150^\mathrm{o}$ and $|b| < 3^\mathrm{o}$, respectively, in order to (i) find new SNR candidates and (ii) follow-up on SNR candidates previously identified in other surveys. In order to consider a source to be an SNR candidate, we require that it has a clear, extended radio morphology in the 144~MHz maps, and that it displays a lack of emission in the MIR. (We note that, for most of these sources, this is the only existing radio detection, and so it is not possible to measure a spectral index value.) A low MIR-to-radio flux density ratio is a common criterion for distinguishing between SNRs and H\,{\sc ii} regions, and has been used in a number of similar studies \citep[e.g. \citealp{Anderson_2017}, \citealp{Hurley_Walker_2019},][]{Ball2023}.

In Sect. \ref{Ch:2} we present the LOFAR and \textrm{WISE} observations that form the basis of this work. In Sect. \ref{Ch:3} we assess the image quality of these fields and in Sect. \ref{Ch:4} we discuss the methods used to analyse them. In Sect. \ref{Ch:5} we present the results of our study: the discovery of 14 new SNR candidates, and the follow-up of 24 candidates previously reported in the literature. In Sect.  \ref{Ch:6} we discuss some properties of the newly discovered candidates, and in Sect. \ref{sec:conclusions} we present our conclusions. 


\section{Data}\label{Ch:2}

\subsection{LoTSS}\label{Ch:2.1}

The LOFAR Two Metre Sky Survey \citep[LoTSS,][]{Shimwell_2017} is an ongoing high-band antenna (HBA) survey of the Northern sky at 120-168 MHz, with a resolution of $6''$ and a point-source sensitivity of approximately 100~$\mu$Jy~beam$^{-1}$. The calibration procedure of the LoTSS fields is discussed thoroughly in \cite{Shimwell_2022}, and treats both direction-independent (DI) and direction-dependent (DD) effects. The resulting data products are full- ($6''$) and low-resolution ($20''$) full-bandwidth images (Stokes I) at a central frequency of 144~MHz, undeconvolved Stokes Q and U image cubes, and undeconvolved Stokes V images.

In this study, we are interested in searching for diffuse emission, and so we use the low-resolution (20$''$) LoTSS images, where the deconvolution of extended sources has higher fidelity \citep[Sect. 3.4 of][]{Shimwell_2022}. The data used here are proprietary LoTSS pointings that will be published as part of data release 3 (DR3).
Galactic pointings are subject to a series of effects that make calibrating them difficult, and so the Galactic fields are usually noisier than typical extragalactic LoTSS fields. The median rms noise of the LoTSS fields under consideration for this work is $\sim$500~$\mu$Jy~beam$^{-1}$.

\subsection{LOFAR project LC18\_027}\label{Ch:2.1.1}
The Galactic field named GalField3 and centred at $(l, b) = (39.0 ^\mathrm{o}$ 00.0$^\mathrm{o}$) was observed in Cycle 18 as part of a pilot survey to search for long-period neutron stars (Rajwade et al., in preparation). The data from this observation run were first reduced with the LoTSS-DR2 calibration pipelines \citep[described in ][]{Shimwell_2022}. However, the deconvolution of the LoTSS-DR2 reduced image was poor, as discussed in Sect. \ref{Ch:4}.

Instead, we used the direction-independent calibrated image reduced as part of the LC18\_027 project for further analysis. The LC18\_027 calibration pipeline uses the standard LOFAR INitial Calibration Software (LINC)~\citep{linc2019}, which computes the gain, phase, and bandpass solutions based on observations of well-known calibrators (3C295 and 3C48 in the case of these observations). The solutions are then applied to the visibilities to create a direction-independent, calibrated measurement set. This image does not account for direction-dependent effects in calibration, and has a resolution of $57''\times43''$. This results in an rms noise of $\sim4$~mJy~beam$^{-1}$, which is an order of magnitude larger than the noise in the LoTSS images. The Galactic diffuse emission can cause various patterns in the radio image. In order to correct for these effects, a reasonable sky model at 150~MHz is essential, which is still a work in progress.  For the remainder of this work, we refer to the image taken from the LC18\_027 project as GalField3~[LC18\_027].

\subsection{\textrm{WISE}}\label{Ch:2.2}

The Wide-field Infrared Survey Explorer \citep[\textrm{WISE},][]{WISE} is a space-based infrared observatory that surveyed the entire sky at 3.4, 4.6, 12, and 22~$\mu$m with an angular resolution of $6''.1$, $6''.4$, $6''.5,$ and $12''.0,$ respectively. For this study, we used the images at 12 and 22~$\mu$m (corresponding to the W3 and W4 survey bands) to check whether the sources identified as candidate SNRs from their 144~MHz emission show mid-infrared emission, which is typical in H\,{\sc ii} and absent in SNRs.

\section{Image quality}\label{Ch:3}
The image quality of the LoTSS pointings is influenced primarily by their elevation \citep{Shimwell_2017}. The elevation of the pointings used in this work are listed in the third column of Table \ref{tbl:scaling_factors}, with the majority of them being low. Additionally, the pointings used in this study are located on the Galactic plane, where the presence of bright, extended sources as well as the diffuse Galactic emission makes it particularly difficult to accurately calibrate and image due to the difficulty of representing these structures in the self-calibration sky model.

\subsection{Pointing quality}\label{Ch:3.2} 
To assess the quality of the individual pointings, we used the integrated-to-peak flux density ratio of point sources as a proxy for the deconvolution performance.
We extracted and catalogued the compact radio sources in the low-resolution fields of interest. We used the Python Blob Detector and Source Finder \citep[pyBDSF,][]{pyBDSF} with the same parameters used in the LoTSS Data Releases. For the source extraction, we only used the central region within $\sim$2.5$^\mathrm{o}$ of the pointing phase centre. This radius is in accordance with the separation between two neighbouring pointing centres, $\Delta \simeq$ 2.58$^\mathrm{o}$ \citep{Shimwell_2017}. Within this region, we expect the recoverable flux density to be the least altered during mosaicking and thus closer to the accuracy of the final full data release. The ratio of the integrated flux density $S_\mathrm{I}$ to the peak flux density $S_\mathrm{p}$ of the compact sources, as a function of signal-to-noise ratio (S/N), is a measure of source extension and is commonly used to estimate whether
or not a source is unresolved. \cite{Franzen_2015} showed that in the case of point sources with independent errors for integrated ($\sigma_{S_\mathrm{I}}$) and peak flux densities ($\sigma_{S_\mathrm{p}}$), the natural logarithm of their ratio, $R=\ln{\frac{S_\mathrm{I}}{S_\mathrm{p}}}$, is Gaussianly distributed and centred at zero. We therefore expect the ratio $\frac{S_\mathrm{I}}{S_\mathrm{p}}$ of point sources to follow a symmetric distribution around 1. The radio population, however, does not consist of point sources and we can expect this symmetric distribution to get skewed for larger values due to fully resolved sources. 

\begin{figure}[h]
   \centering
   \includegraphics[width=9.1cm]{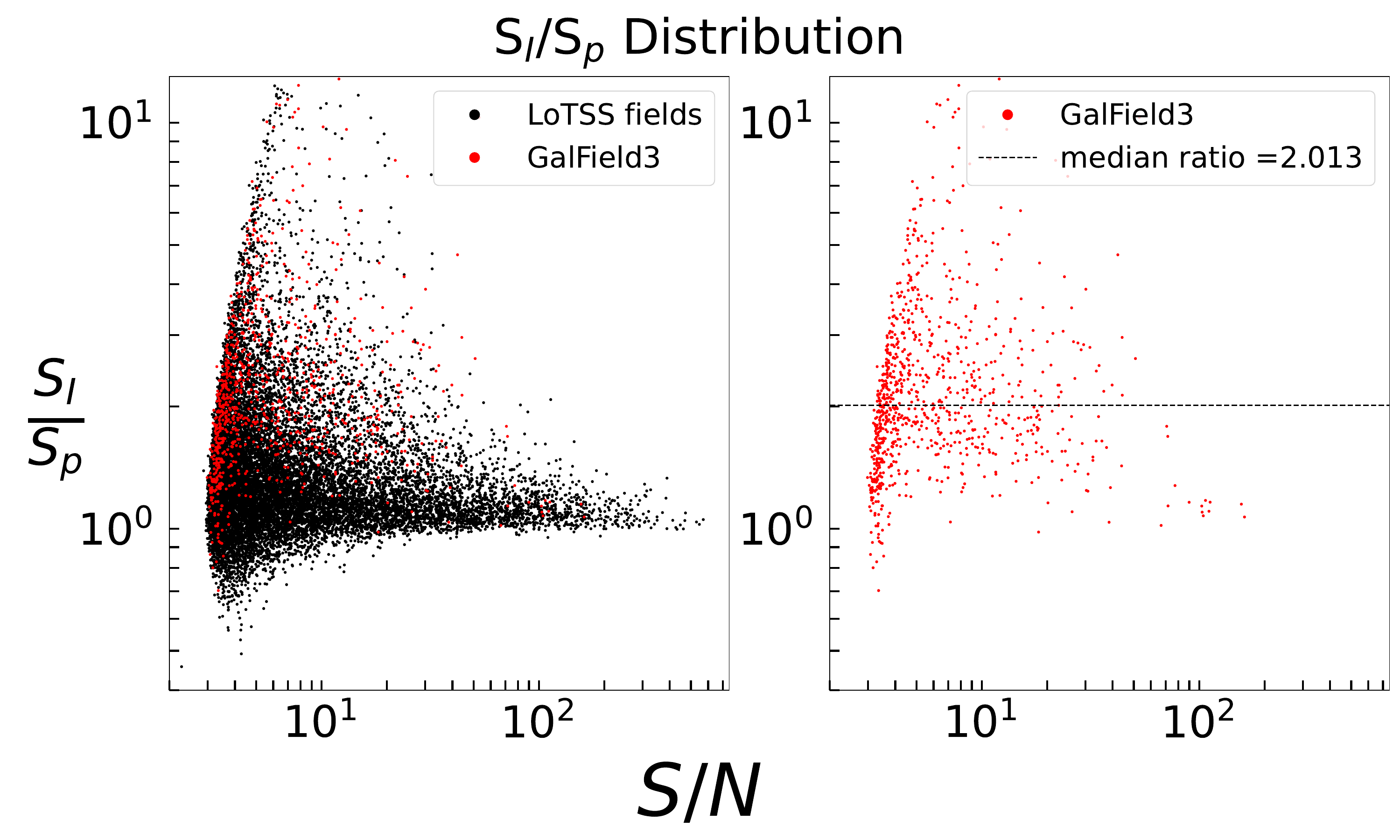}
      \caption{Integrated flux density to peak flux density ratio, as a function of the S/N of all images reduced with the LoTSS-DR2 pipelines. The left panel shows the catalogued sources of every LoTSS field concerned in this study, namely P297+30, P293+20, P289+10, P286+15, and P062+54 in black and the LoTSS reduced image of GalField3 in red. The right panel shows only the catalogued sources from the LoTSS image of GalField3. The black, dashed line marks the median ratio of integrated flux density to peak flux density for the source population, which is 2.013.}
         \label{fig:sources}
   \end{figure}

\begin{figure}[h]
   \centering
   \includegraphics[width= 6 cm]{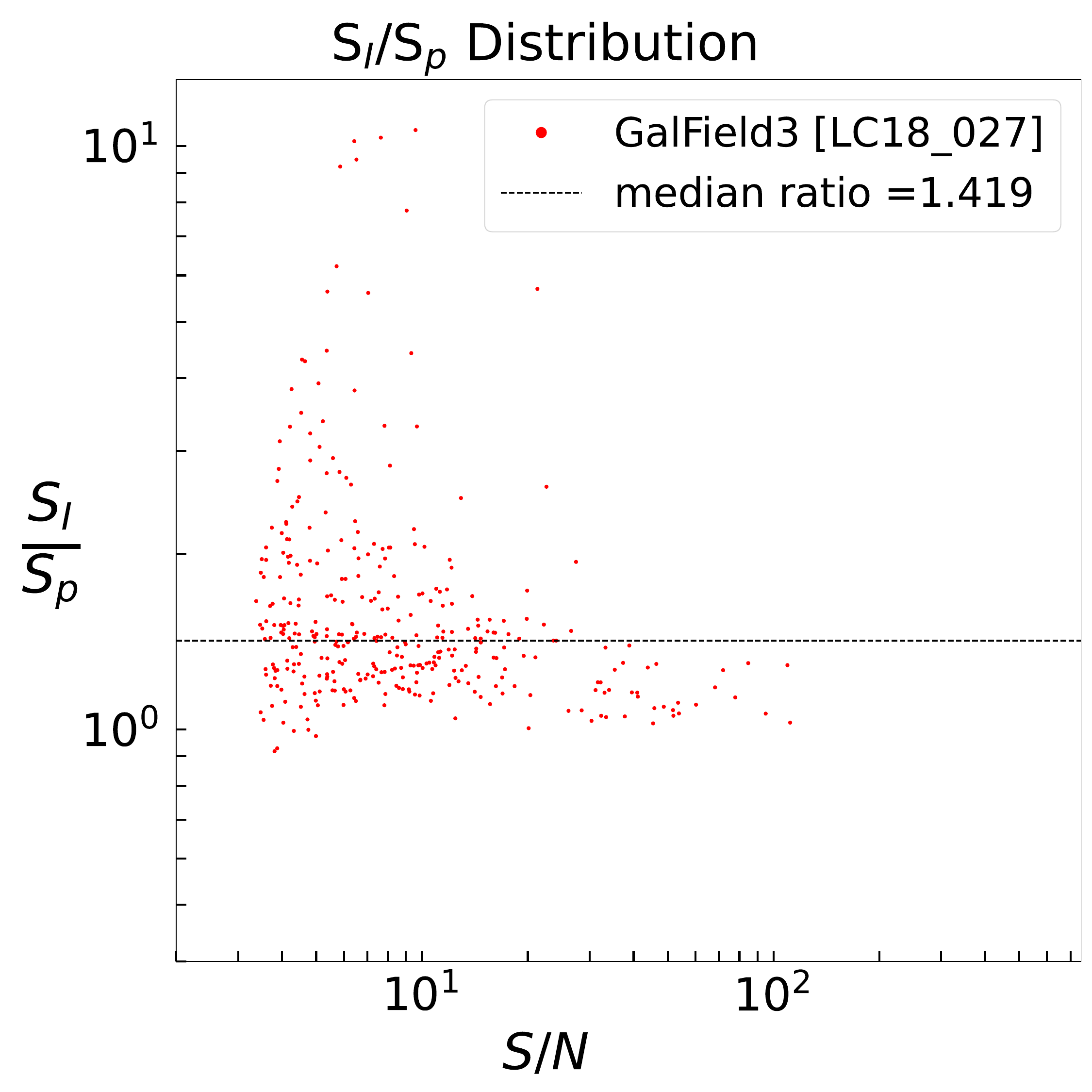}
      \caption{Integrated flux density to peak flux density ratio as a function of the S/N for the GalField3~[LC18\_027] image. The black dashed line marks the median ratio of the source population, which is 1.419.}
    \label{fig:GF3_alt_sources}
   \end{figure}

Every LoTSS pointing behaves as expected (see Fig. \ref{fig:sources}). On the other hand, the LoTSS processed image of GalField3 has an unnaturally high median ratio of $2.013$ and is not symmetrically distributed. We consider this field's current deconvolution poor and untrustworthy and decided to exclude it from the following analysis. We use the LoTSS processed image of this field only to search by eye for new SNRs, as it has a high dynamic range due to its calibration for both DI and DD effects. For the SNRs found in this field, we report their flux densities using the GalField3~[LC18\_027] total intensity map (see Sect. \ref{Ch:2.1.1}). The distribution of the $\frac{S_\mathrm{I}}{S_\mathrm{p}}$
ratio of the LC18\_027 reduced image shows a more normal behaviour (Figure \ref{fig:GF3_alt_sources}).

\subsection{Flux-density scale}
The current, incomplete knowledge of the exact LOFAR beam results in an inaccurate flux density scale \citep[\citealp{Shimwell_2019} and][]{Shimwell_2022}. These errors were quantified by \cite{10.1093/mnras/stw1763}, who also devised a bootstrapping method to align the obtained flux scale with flux scales of other surveys with well-studied accuracy. This method is present in the LoTSS processing pipeline and aligns the LoTSS flux density scale with those of the Very Large Array Low-frequency Sky Survey Redux \citep[VLSSr,][]{VLSSr} and the WEsterbork Northern Sky Survey \citep[WENSS,][]{WENSS} \citep[see Sects. 2 of][ and \citealp{Shimwell_2022}]{Shimwell_2019}. 

In the LoTSS-DR1 paper, \cite{Shimwell_2019} assessed the performance of this method by comparing the integrated flux density of the LoTSS-DR1 catalogued, compact sources with their respective cross-matched sources from the TIFR GMRT Sky Survey-Alternative data release \citep[TGSS-ADR1,][]{TGSS}. \cite{Shimwell_2019} indicated the need to place a 20\% uncertainty on the integrated flux density measurements. We consider this fractional uncertainty an inherited one that applies to every pointing processed with the current LoTSS pipelines.

\subsection{Alignment of the flux density scale of the Galactic pointings}\label{Ch:3.3}

We applied a post-processing step to further improve the accuracy of the flux density in the pointings concerning this study. We used the TGSS-ADR1 (central frequency at 147.5~MHz, \citealp{TGSS}) to calculate scaling factors for each pointing. Namely, we compared the integrated flux densities of the catalogued sources in our pointings to the TGSS-ADR1 catalogue \citep[similarly to the works in \citealp{Botteon2022},][]{Hoang2022}.

We cross-matched our catalogued sources with TGSS-ADR1 using a 10" nearest-neighbour distance. Any LoTSS sources in the cross-matched catalogue that were fitted with multiple Gaussians (coded M in \texttt{pyBDSF}) were removed from further analysis. We fitted a line to the integrated flux densities in the LoTSS and TGSS-ADR1 to derive the scaling factors in each pointing. We employed the Deming regression \citep{deming1943statistical} to linearly fit the data. The Deming regression is used in cases where both the independent and dependent variables come with an error. This is the case here, because both the TGSS-ADR1 catalogue and our \texttt{pyBDSF} catalogued sources report uncertainties on the integrated flux-density measurements.

Before applying the Deming regression, we accounted for the difference in frequencies between LoTSS and TGSS-ADR1. We multiplied the integrated flux densities in TGSS-ADR1 with (144/147.5)$^\alpha$. We chose $\alpha = -0.783,$ which was used in the flux scale alignment in the LoTSS-DR2 paper \citep[see Sect. 3.3 in ][]{Shimwell_2019}, in order to maintain consistency between our analysis and this latter work. After this correction, the scaling factor of each pointing is the inverse of the best-fitted slope. The derived scaling factors are listed in the fourth column of Table \ref{tbl:scaling_factors}.

\subsection{Sensitivity of the fields under consideration}
We estimated the point-source sensitivity of a field from its intensity image by fitting a Gaussian to a histogram of the background pixel values (with values $<10$~mJy~beam$^{-1}$). The best-fitted rms serves as an estimate of the field's sensitivity \citep{Taylor_Carilli_Perley_2008}. Table \ref{tbl:scaling_factors} shows the sensitivity of each field used in this study. The median sensitivity of our fields is 532~$\mu$Jy~beam$^{-1}$. 

\begin{table*}
\centering   
\begin{tabular}{c c c c c}
\hline\hline             
Pointing & centre (l, b) & elevation & Scaling factor &  sensitivity [$\mu$Jy beam$^{-1}$]  \\    
\hline 
P062+54 & (149.5$^\mathrm{o}$, +1.7$^\mathrm{o}$) & 18.7$^\mathrm{o}$ & $1.018 $ & 315 \\
P286+10 & (43.9$^\mathrm{o}$, +1.1$^\mathrm{o}$) & 45.2$^\mathrm{o}$  & $0.996 $ & 1010  \\
P286+15 & (48.2$^\mathrm{o}$, +3.6$^\mathrm{o}$) & 23.1$^\mathrm{o}$  & $1.390 $ &  532 \\

P293+20 & (55.6$^\mathrm{o}$, +0.6$^\mathrm{o}$) & 22.4$^\mathrm{o}$  & $1.271 $ &  479  \\
P297+30 & (66.1$^\mathrm{o}$, +2.5$^\mathrm{o}$) & 37.3$^\mathrm{o}$  & $1.073 $ & 501  \\
GalField3 [LC18\_027] & (39.0$^\mathrm{o}$, +0.0$^\mathrm{o}$)& 36.7$^\mathrm{o}$  & 1.887 & 4694  \\ 
\hline                          \end{tabular}
\caption{Details of the pointings used in this work.}
\label{tbl:scaling_factors}
\end{table*}


\section{Methods}\label{Ch:4}

To find new SNR candidates, we visually examined the LoTSS total intensity maps for diffuse radio emission. For all the radio sources considered, we examined whether they were an already known SNR from the G22 catalogue or had been  reported in the literature as an SNR candidate or a Galactic H\,\textsc{ii} region \citep[e.g.][]{Anderson_2014}. For those sources that had not been previously identified in the literature, we searched the \textrm{WISE} W3 and W4 images for coincident MIR emission. Finally, the sources that did not show strong MIR emission were considered to be new SNR candidates and are reported in Sect. \ref{Ch:5}.

\subsection{Integrated flux density measurements of candidate SNRs}\label{Ch:4.4}
We measure the integrated flux density of our SNR candidates by drawing a DS9 \citep{DS9} polygon region encompassing the outline of the diffuse radio source as closely as possible. If the source has a partially filled centre, we draw a second, inner polygon to exclude the empty region from the measurement. The integrated flux density $S_\mathrm{I}$ in  Jy of the SNR is then given by:
\begin{equation}
     S_\mathrm{I} = \frac{\sum_{i} (I_{i}-I_{\mathrm{background}})}{A}
     \label{eq: flux}
,\end{equation}

\noindent where $I_{i}$ is the value of the i-th pixel in Jy~beam$^{-1}$ within the region that contains the SNR, $I_{\mathrm{background}}$ is the average pixel value in the background region (after subtracting the compact sources contaminating the background region), and $A$ is the number of pixels in a beam. In the cases where there is a radio source overlapping with our SNR, we remove its contribution by replacing the values $I_{i}$ of this polluting source with an average value $I$. To determine this average value $I$, we draw a representative DS9 region within the SNR and take the average of its pixel values.  

\begin{figure}
   \centering
   \includegraphics[width=7 cm]{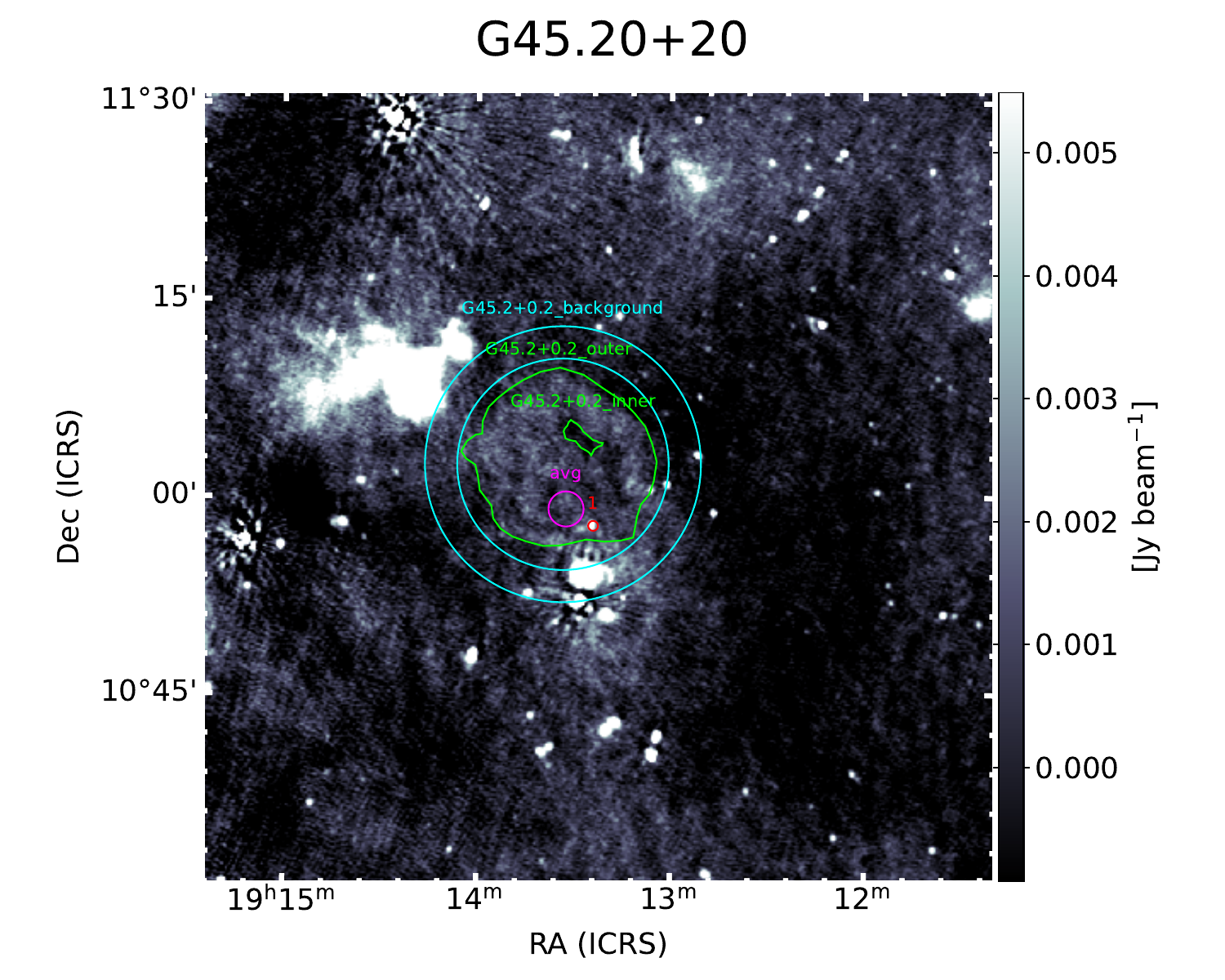}
      \caption{LoTSS image of the candidate SNR G45.20+0.20. Green polygons define the regions that encompass the supernova. The magenta region is the one used to calculate a representative average pixel value of the SNR. The red region encircles a contaminating external radio source. The cyan region defines a representative part of the radio background.}
      \label{fig:SNRexample}
   \end{figure}

The uncertainty of the integrated flux density of the SNR is calculated with the following formula: 

\begin{equation}
     \sigma_{S_{I}} = \sqrt{\sigma_\mathrm{background}^{2}\left( \frac{N}{A} \right) + (\sigma_\mathrm{pointing}S_{I})^{2} } 
     \label{eq:sigma}
,\end{equation}

\noindent where S$_{I}$ is the integrated flux density of the SNR (Eq. \ref{eq: flux}), $\sigma_\mathrm{background}^{2}$ is the variance of the pixel values in the background region, $N$ is the number of pixels in the SNR region, $A$ is the number of pixels in the beam, and $\sigma_\mathrm{pointing}$ is the flux density scale uncertainty of the respective pointing. We adopted the conservative value of their inherited uncertainty, $\sigma_\mathrm{pointing}$ = 0.2. 
Figure \ref{fig:SNRexample} shows the candidate G45.20+0.20 along with the regions defined to measure its flux density and its uncertainty. This example provides a good, visual overview of the method described here (see Table \ref{tbl:new}).

\section{Results} \label{Ch:5}
We find 14 new SNR candidates in the LoTSS survey. The majority of them (12 out of the 14), fall in a region with $39^\mathrm{o} < l < 66^\mathrm{o}$ and $|b|< 2.5^\mathrm{o}$. The remaining two candidates are located at $l\simeq 149^\mathrm{o}$ and $|b| < 2^\mathrm{o}$. None of these sources appear in the G22 catalogue. We further conducted a follow-up study of the 22 SNR candidates out of the 76 reported by \citet{Anderson_2017} that fall within the regions of our study. 

\subsection{New SNR candidates discovered in the LoTSS survey} \label{Ch:5.1}

\begin{table*}
\caption{New SNR candidates in LoTSS.} 
\centering          
\begin{tabular}{l l l l l c c }     
\hline\hline       
Name        &J2000 RA (hms)        & J2000 DEC (dms)       & $\theta_\mathrm{min} \times \theta_\mathrm{max}$ [arcmin] & S$_{144~\mathrm{MHz}}$ [Jy]  & Morphology & Confidence\\ 
\hline
G39.05+0.79 & 18h59m48.43s    & +05d48m15.40s    & 8.1 $\times$ 11.2 &  4.13 $\pm$ 0.84     & F & I       \\
G40.50+0.50   & 19h03m19.00s & +06d57m22.17s & 7.1 $\times$ 8.0 & 1.96 $\pm$ 0.42 & S & III       \\
G41.49$-$0.62 & 19h09m13.16s    & +07d16m53.70s    & 6.9 $\times$ 11.5 &     -      & P & II       \\
G42.95$-$0.30  & 19h11m00.35s    & +08h45m08.50s    & 2.6  $\times$ 2.80   &  3.84 $\pm$ 0.88 & S & III       \\
G44.30+0.50   & 19h10m43.08s & +10d22m37.40s & 2.2  $\times$ 6.6    & 1.35 $\pm$ 0.28      & P & I       \\
G45.20+0.20   & 19h13m33.49s & +11d02m00.04s & 5.3  $\times$ 5.6    & 0.70 $\pm$ 0.29       & F & II       \\
G46.60+0.20   & 19h15m57.95s & +12d12m23.94s & 7.4 $\times$ 8.1    & 2.80 $\pm$ 0.57       & S & I       \\
G47.78+2.02 & 19h11m38.64s & +14d07m55.02s & 2.8 $\times$ 3.7  &  0.59 $\pm$ 0.13       & S & I       \\
G52.90+0.10   & 19h28m42.58s & +17d46m29.60s & 3.3 $\times$ 4.7  & 0.30 $\pm$ 0.06     & P & II        \\
G53.80+0.80   & 19h27m56.36s & +18d52m43.12s & 2.7 $\times$ 7.7  & 0.55  $\pm$ 0.12      & P & II       \\
G55.54+1.96 & 19h27m06.56s & +20d57m32.86s & 10.3 $\times$ 13.2  & 2.77  $\pm$ 1.07       & F & II       \\
G65.38+2.06 & 19h49m11.31s    & +29d48m02.30s    & 7.6 $\times$ 7.7   &  0.59 $\pm$ 0.12       & F & I       \\
G148.20+0.80  & 03h59m08.28s & +54d10m06.31s & 39.0  $\times$ 42.4   &  3.38 $\pm$ 1.70       & S & II       \\
G149.10+1.90  & 04h07m37.89s & +54d12m33.05s & 31.3  $\times$ 43.4  & 3.62  $\pm$ 0.81  & S & II\\
\hline                  
\end{tabular}  
\tablefoot{Reading from left to right, the columns correspond to: the name of the candidate, the right ascension and declination of the candidate centre, its angular size, its integrated flux density  at 144~MHz, its morphology, with (S) denoting shell, (P) denoting partial shell, and (F) denoting filled centre, and our confidence encoded from I to III, with I being the most confident and III being the least confident.}
\label{tbl:new}   
\end{table*}

Table \ref{tbl:new} summarises the properties of the SNR candidates discovered in this study. We assign each candidate a morphological type of either a shell (S) for ring-like structures, filled shell (F) for candidates with filled centres, or a partial shell (P) for sources with irregular structure. We also include a metric of our confidence in each candidate's SNR nature. The code of that metric is I, II, and III, with I meaning that we are very confident in this candidate, II meaning that we are mildly confident in the candidate, and III meaning that our confidence is low. This metric is subjective; it is based on the complexity of the background radio emission in the field, the faintness and morphology of the candidate source, and the MIR emission. We now give some background information on each of the candidate SNRs we have identified. We refer the reader to the figures in Appendix \ref{appendix} for the LoTSS and false-colour MIR maps of the sources.

G39.05+0.79 is located in the close proximity of numerous bright radio sources and in an overall complex region of the Galactic plane. It is a filled-shell-type candidate, without sharply defined boundaries. Its central and northern parts seem brighter in radio. There is some coinciding MIR emission, mostly at the southern part of the candidate but it does not exhibit a similar morphology in radio, indicating that the sources could simply be overlapping. We report a flux density of $S_{144~\mathrm{MHz}}$= 4.13$\pm$0.84~Jy and assign it a confidence code I.

G40.50+0.50 is an almost circular candidate with a shell-like morphology. Some small H\,{\sc ii} regions are located to the west of the candidate. We excluded their contribution during the measurement of the candidate's integrated flux density. We report a flux density of $S_{144~\mathrm{MHz}}$= 1.96$\pm$0.42~Jy and assign it a confidence code III.

G41.49$-$0.62 is a faint, partial shell in the vicinity  of the known (and extremely bright) SNR G41.1$-$0.3. In the LoTSS low-resolution image, the source shows a diffuse, almost straight region to the west that meets with a bow-like structure in the east. This candidate falls within the GalField3 field and unfortunately is not visible in the GalField3~[LC18\_027] image because of artefacts caused by G41.1$-$0.3; the image's dynamic range in the region is quite low because it is not calibrated for direction-dependent effects. For this reason, we do not report an integrated flux density value. We assign G41.49$-$0.62 a confidence code II.

G42.95$-$0.30 is one of the smallest candidates of this study, with a radius of $\sim2.5'$. It is in a noisy region ---with many artefacts--- towards the edge of the field, where the beam effects are most uncertain. We report a flux density of $S_{144~\mathrm{MHz}}$= 3.84$\pm$0.88~Jy and assign it a confidence code III.

G44.30+0.50 is a bow-like structure surrounded by numerous H\,{\sc ii} regions, as is evident in the false colour map in figure \ref{fig:G44.3+0.5}. This source, however, is distinct from the H\,{\sc ii} regions as it does not show any emission in MIR. We report a flux density of $S_{144~\mathrm{MHz}}$= 1.35$\pm$0.28~Jy and assign it a confidence code I. This candidate is within $\sim5'$ of the pulsar J1910+1026 (formerly known as J1910+1027). J1910+1026 was first discovered by \cite{Lazarus_2015}, and \cite{Parent_2022} later estimated its age to be $\sim 33$~kyr and its distance as $D\simeq$ 5.7~kpc. The angular separation of the pulsar and the candidate's approximate centre ($\sim5'$) correspond to $\sim$22.8 pc for the distance derived  by \cite{Parent_2022}. In order to cover this distance in 33~kyr, the pulsar must have had a mean velocity of $\sim$ 575~km s$^{-1}$. For reference, \citet{Hobbs2005} estimated  the 3D natal kicks of pulsars to be $\sim$ 400~km s$^{-1}$, with a 1D dispersion of $\sigma$ = 265~km s$^{-1}$. Further investigation of the two objects is needed to explore whether or not this candidate is associated with such a young pulsar.

G45.20+0.20 is a faint, filled-shell candidate with a poorly defined boundary, and is situated between H\,{\sc ii} regions that are particularly bright at 144~MHz. The candidate itself, however, lacks emission in MIR. We report a flux density of $S_{144~\mathrm{MHz}}$= 0.70$\pm$0.29~Jy. Given its faintness and its location in a noisy radio background we assign it a confidence code II. 

G46.60+0.20 is in a rich neighbourhood of the Galactic plane. It is a shell-type candidate whose central and southeastern parts are brighter. The emission in radio does not show a MIR counterpart (there is overlapping MIR emission but with different morphology). We report a flux density of $S_{144~\mathrm{MHz}}$= 2.80$\pm$0.57~Jy and assign it a confidence code I.

G47.78+2.02 is a well-defined shell-type candidate. It is one of the smallest SNRs in this study ($\sim3'$), and is found on a uniform and relatively quiet background. We report a flux density of $S_{144~\mathrm{MHz}}$= 0.59$\pm$0.13~Jy. It is coded I due to its clear morphology.

G52.90+0.10 is located south of the large  H\,{\sc ii} region G053.935+00.228 \citep{Anderson_2014}. The neighbourhood is also rich in SNRs. Such a rich region makes the identification of new SNRs tricky, as any source of radio emission can be easily misidentified. We detected the partial shell of G52.9+0.1, which stands out of its radio background but still remains quite faint. We report a flux density of $S_{144~\mathrm{MHz}}$= 0.30$\pm$0.06~Jy and assign it a confidence code II.

G53.80+0.80 is a bright, bow-like radio structure northwest of the G053.935+00.228 H\,{\sc ii} region \citep{Anderson_2014}. At its centre there is some MIR emission with vastly different orientation and morphology, suggesting that the spatial overlap could be coincidental. Nevertheless, due to the complex region that hosts this candidate, we assign it a code II. We report a flux density of $S_{144~\mathrm{MHz}}$= 0.55$\pm$0.12~Jy.

G55.54+1.96 is another filled-shell candidate without sharp boundaries. The MIR ring-like structure that is located in the northern part of the candidate (see left panel of figure \ref{fig:G55.54+1.96}) is the G055.590+02.029 H\,{\sc ii} region \citep{Anderson_2014} with a radius of $\sim6.5'$. The fact that the radio emission is more extended and does not show the shell structure of the H\,{\sc ii} region leads us to the conclusion that the two emitting regions could belong to different objects. We report a flux density of $S_{144~\mathrm{MHz}}$= 2.77$\pm$1.07~Jy and assign it a confidence code II. 

G65.38+2.06 is a well-defined, faint, and almost circular-shell SNR. We are highly confident in this candidate, mostly due to its classical morphology and the uniform radio background. We report a flux density of $S_{144~\mathrm{MHz}}$= 0.59$\pm$0.12~Jy and assign it a confidence code I.

G148.20+0.80 is a rather extended and extremely faint ring-like radio structure. It is interesting to note that within the radio ring, close to its approximate centre, there is a pulsar, B0355+54, which was associated with a PNW (G148.1+00.8) by \cite{Tepedelenlioglu_2007} and \cite{McGowan2006}. If the radio ring is indeed synchrotron emission, it could be the expanding shock of the supernova explosion that created B0355+54. We report a flux density of $S_{144~\mathrm{MHz}}$= 3.38$\pm$1.70~Jy and assign it a confidence code II.

G149.10+1.60 is yet another extended, ring-like structure. It falls next to candidate SNR G148.2+0.8. Its northern part is also the location of numerous H\,{\sc ii} regions (as evidenced in the false colour map, fig. \ref{fig:G149.1+1.6}); however, the radio emission continues with a faint bow to the southeast, giving the impression of a radio shell. We report a flux density of $S_{144~\mathrm{MHz}}$= 3.62$\pm$0.81~Jy and assign it a confidence code II.

\subsection{SNR candidates in the literature} \label{Ch:5.2}
The Galactic regions studied in this work host candidate SNRs that are proposed in the literature but have not been included in the G22. In this section, we probe those candidate SNRs in the 144~MHz LoTSS images.

G39.50+0.50 was first mentioned as a possible SNR by \cite{1990ApJ...364..187G}. In \cite{1996ApJ...458..257G}, the candidate was reported to have a flux density of 3~Jy at 1~GHz. This measurement combined with ours, of namely S$_{144~\mathrm{MHz}}$ = 4.48$\pm$0.93~Jy, yields a spectral index of $\alpha$ = $-0.2\pm$0.1. The negative index in addition to the lack of MIR emission (figure \ref{fig:G39.5+0.5}) grants this candidate a confidence code I, and we confirm its status as a candidate SNR.

G43.50+0.60 is one of the three candidate SNRs proposed by \cite{Kaplan_2002}. The other two, G41.5+0.4 and G42.0$-$0.1, are included in G22 in light of the follow-up study by \cite{Alves2012}, which showed that their emission is non-thermal. \cite{Kaplan_2002} report a flux density of  1.05$\pm$0.08~Jy at 332~MHz. This result, combined with our measurement of $S_{144~\mathrm{MHz}}$= 3.34$\pm$0.67~Jy, yields $\alpha=-1.4 \pm$0.3, a particularly steep spectral index value that evidences non-thermal emission. The small, radio-bright source south of the candidate is the H\,{\sc ii} region G043.432+00.521 \citep{Anderson_2014}, which was excluded from the flux density measurements. The steepness of the source's spectrum lead us to confirm G43.50+0.60 as a  candidate SNR.

\subsection{THOR SNR candidates in the LoTSS survey} \label{Ch:5.2.1}

\begin{table*}
\caption{Follow-up of THOR SNR candidates with LoTSS.} 
\centering          
\begin{tabular}{l l l l l l l }     
\hline\hline       
Name & J2000 RA (hms) & J2000 DEC (dms) & S$_{144MHz}$ [Jy] & S$_{1420MHz}$ [Jy]& $\alpha$ & SNR status\\ 
\hline                    
G36.90+0.49   & 18h56m56.27s & +03d45m16.89s &  0.76 $\pm$ 0.17 & 0.50    $\pm$0.08 & -0.2$\pm$ 0.1 & Inconclusive         \\
G37.62$-$0.22   & 19h00m47.34s & +04d04m14.66s & 0.68 $\pm$ 0.17 & 0.41  $\pm$ 0.12 & -0.2 $\pm$ 0.2   & Inconclusive        \\
G37.88+0.32   & 18h59m20.36s & +04d32m55.91s &  1.96 $\pm$ 0.50 & 3.05    $\pm$ 4.74  & 0.2 $\pm$ 0.7  &  Inconclusive   \\
G38.17+0.09   & 19h00m41.53s & +04d42m05.85s & 6.44 $\pm$ 1.37  & -        &     -  & Inconclusive        \\
G38.62$-$0.24   & 19h02m41.81s & +04d57m02.41s &  0.29 $\pm$ 0.08 & 0.10 $\pm$ 0.03  & -0.5 $\pm$ 0.2   & Confident       \\
G38.68$-$0.43   & 19h03m29.11s & +04d55m01.00s & 0.54 $\pm$ 0.13 & 0.44  $\pm$ 0.12 & -0.1 $\pm$ 0.2  &  Inconclusive       \\
G38.72$-$0.87   & 19h05m07.71s & +04d45m02.53s &  3.01 $\pm$ 0.76  & 0.70  $\pm$ 0.80 & -0.6 $\pm$ 0.5  & Inconclusive       \\
G38.83$-$0.01  & 19h02m15.72s & +05d14m33.68s &  0.01 $\pm$ 0.01  & 0.01 $\pm$ 0.00& 0.0 $\pm$ 0.2 & Inconclusive         \\
G39.19+0.52  & 19h01m01.82s & +05d48m19.21s & 0.36 $\pm$ 0.80 & 0.17 $\pm$0.21& -0.3 $\pm$ 1.1     & Inconclusive      \\
G39.56$-$0.32  & 19h04m42.83s & +05d44m57.77s &  1.29 $\pm$ 0.31 & 1.19  $\pm$ 1.64 & 0.0 $\pm$ 0.6   & Inconclusive       \\
G41.95$-$0.18  & 19h08m38.42s & +07d56m09.95s & 2.16 $\pm$ 0.44 & 0.44 $\pm$ 0.50  & -0.3 $\pm$ 0.2  & Inconclusive         \\
G42.62+0.14  & 19h08m44.43s & +08d40m41.44s & 0.18$\pm$ 0.04 & 0.5  $\pm$ 0.05 & 0.4 $\pm$ 0.1   & Reject        \\
G45.35$-$0.37  & 19h15m42.43s & +10d51m41.74s &  2.13 $\pm$ 0.51 & 0.91  $\pm$0.43 & -0.4 $\pm$ 0.2 & Confident    \\
G45.51$-$0.03  & 19h14m47.04s & +11d09m41.04s & 0.97 $\pm$ 0.21  & 1.63 $\pm$ 0.42  & 0.2 $\pm$ 0.1 & Reject \\
G46.18$-$0.02  & 19h16m01.23s & +11d45m32.72s &  1.50 $\pm$ 0.31   & 0.47       $\pm$ 0.44 & -0.5 $\pm$ 0.4 & Inconclusive      \\
G46.54$-$0.03  & 19h16m44.56s & +12d04m22.54s &  0.57 $\pm$ 0.13  & 0.85       $\pm$ 0.51 & 0.2 $\pm$ 0.3  & Reject     \\
G47.15+0.73  & 19h15m09.05s & +12d58m00.05s & 0.03 $\pm$ 0.01 & 0.01       $\pm$ 0.00& -0.5 $\pm$ 0.2  & Confident      \\
G53.07+0.49  & 19h27m35.93s & +18d04m37.60s & 0.06 $\pm$ 0.01  & 0.06       $\pm$ 0.00  & 0.00 $\pm$ 0.03   & Reject  \\
G53.84$-$0.75  & 19h33m43.97s & +18d09m23.44s &  4.20 $\pm$ 0.85 & 1.31        $\pm$ 3.43 &  -0.5 $\pm$ 1.2  & Inconclusive      \\
G54.11+0.25  & 19h30m35.21s & +18d52m31.27s & 1.31 $\pm$ 0.81  & 1.46        $\pm$ 0.28 & 0.0 $\pm$ 0.3 & Inconclusive   \\
G56.56$-$0.75 & 19h39m21.57s & +20d31m49.80s & 0.93 $\pm$ 0.27 & 0.94         $\pm$0.61& 0.0 $\pm$ 0.3 & Inconclussive    \\
G57.12+0.35  & 19h36m25.52s & +21d33m27.25s & 1.93 $\pm$ 0.65 & 0.60         $\pm$ 0.22 & -0.5 $\pm$ 0.2 & Confirmed     \\
\hline                  
\end{tabular} 
 \tablefoot{Reading from left to right, the columns contain: the name of the candidate, the Right Ascension and Declination of the candidate centre, its measured integrated flux density in the LoTSS map at 144~MHz, its 
 integrated flux density at 1420~MHz as reported by \cite{Anderson_2017}, its estimated spectral index, and the status of the candidate (based on the criterion discussed in Sect. \ref{Ch:5.2.1}).}
\label{tbl:followup} 
\end{table*}

\cite{Anderson_2017} reported the discovery of 76 SNR candidates using a combined dataset from the 1420~MHz H\,{\sc i}/OH/Recombination line survey of the VLA \citep[THOR,][]{beuther16} and the 1420~MHz VLA Galactic Plane Survey \citep[VGPS,][]{VGPS}. This combined dataset is known as THOR+VGPS and its angular resolution is $25''$ at 1420~MHz \citep{beuther16}.

Of these 76 candidates, 22 are located in the region within $39^\mathrm{o} < l < 66^\mathrm{o}$ and $|b|< 2.5^\mathrm{o}$ covered by our study. We measured the integrated flux densities of those 22 SNR candidates using the method described in Sect. \ref{Ch:4.4}. We drew a circular region with the centre and radius reported in \citet{Anderson_2017} to measure their integrated flux densities. Table \ref{tbl:followup} summarises our findings.

We measured the spectral indices, $\alpha$, of these candidate SNRs by  taking the ratio of the logarithms of their fluxes at the different frequencies ($S_{\nu} \propto \nu^{\alpha}$). The errors of the measured spectral indices were calculated by fully propagating the errors in the reported integrated flux densities. The derived spectral index uses only two datapoints, but spans an order of magnitude in frequency, making it possible, in some cases, to discriminate between thermal and non-thermal emission. 

We update the status of nine of the candidates based on the derived value of their spectral index (listed in Table \ref{tbl:followup}, sixth column). We note that the quoted errors can be large, and prevent us from discriminating between thermal and non-thermal emission in 14 out of the 22 cases. We take $\alpha = -0.1$ as the value differentiating between thermal and non-thermal emission \citep[H\,{\sc ii} regions radiate with $\alpha \gtrsim -0.1$, see][]{Condon_Ransom_2016}. For values of $\alpha$ that ---within the error bars--- include this boundary value, we neither confirm nor reject the SNR nature of the candidate. If the spectral index ---including errors--- is non-thermal ($\alpha < -0.1$), we are confident in the nature of the source as a candidate SNR. We reject the candidates whose indices ---including errors--- are $\alpha > -0.1,$ as they exhibit emission that is clearly thermal.

Based on the above criteria, we confirm the SNR candidate status of four sources
(G38.62$-$0.24, G45.35$-$0.37, G47.15+0.73 and G57.12+0.35, labelled as `Confident' in Table \ref{tbl:followup}). This study was inconclusive for 14 sources (G36.90+0.49, G37.62$-$0.22, G37.88+0.32, G38.17+0.09, G38.68$-$0.43, G38.72$-$0.87, G38.83$-$0.01, G39.19+0.52, G39.56$-$0.32, G41.95$-$0.18, G46.18$-$0.02, G53.84$-$0.75, G54.11+0.25, and G56.56$-$0.75, labelled `Inconclusive' in Table \ref{tbl:followup}). We find that four of the THOR candidates are not SNRs (G42.62+0.14, G45.51$-$0.03, G46.54$-$0.03, and G53.07+0.49, labelled ``Reject" in Table \ref{tbl:followup}) in light of their thermal spectral index. We now proceed to discuss noteworthy aspects of some of the candidates. The reader can refer to the figures in Appendix \ref{appendix_thor} for the LoTSS maps of the sources. 

G38.17+0.09 does not have a reported flux density given by \cite{Anderson_2017}. Since the sole discriminating factor for altering the status of a candidate in our study is the inferred spectral index, we do not update its status and consider it a candidate SNR.

G38.72$-$0.87 is a diffuse, filled-centre candidate whose inferred spectral index is quite steep ($\alpha$ = $-0.7\pm$0.5), indicating non-thermal emission. It is located close to the known SNR G38.7$-$1.3 (in the G22 catalogue). The morphology of the two SNRs in figure \ref{fig:G38.72-0.87} resembles that of a single, barrel-shaped shell \citep[see][]{Dubner_2015}. Based on the non-thermal nature of candidate G38.72$-$0.8, we propose that this candidate is the western part of G38.7$-$1.3.

G54.11+0.25 is a diffuse shell of radio emission around PWN G54.1+0.3. \citet{Green_2022} reports the spectral index of the PWN to be $\alpha$ = 0.1. \citet{Lang_2010} first suggested that a faint radio shell (radius of $\sim$8$'$) around G54.1+0.3 could be its SNR. \citet{Anderson_2017} later suggested that the outer parts of the ring belong to the G053.935+00.228 H\,{\sc ii} region. Howeer, these latter authors proposed that a fainter radio ring with smaller radius ($\sim7.2'$) could be an SNR associated with the PWN. Contrary to the LOFAR HBA data from \citet{Driessen_2018}, the radio shell is visible in the LoTSS images of the region. We measured the integrated flux density of this candidate without excluding the contribution of the central PWN. The calculated index is 0.0$\pm$0.3. We neither reject nor confirm this candidate as it is possible that both our measurements and those reported by \citet{Anderson_2017} are influenced by the flux density of the PWN.

The SNR candidates proposed by \cite{Anderson_2017} were identified based on their lack of MIR emission, which indicates that they are not likely to be H\,{\sc ii} regions. However, our follow-up observations of four of those candidates revealed a flat spectral index. We rejected the SNR scenario, based on this evidence. Some spatially compact sources, such as G53.07+0.49, could be PWNs. The synchrotron radiation in radio from  PWNs is found to be flatter than that from SNRs \citep[$-03 \lesssim \alpha \lesssim 0.0$,][]{Alsabti2017}. Nevertheless, additional and detailed observations are required to reveal the true nature of those candidates.


\section{Discussion}\label{Ch:6}

We have detected 14 new SNR candidates, and confirm the SNR candidate status of another 6 previously identified sources in two Galactic regions observed as part of the LoTSS survey. The region within $l=39^\mathrm{o}-60^\mathrm{o}$ is accessible to telescopes at a variety of latitudes, and so has been the focus of several SNR searches \citep{brogan06,Anderson_2017,Driessen_2018}, which show it to be heavily populated by known SNRs. Yet, even for such a thoroughly studied, busy, and low-elevation region (for LOFAR) of the Galactic plane, the sensitivity and resolution of LoTSS have resulted in a significant number of new SNR candidates. 

\subsection{Properties of the new candidate SNRs}

\begin{figure*}[h!]
   \centering
\includegraphics[width=19 cm]{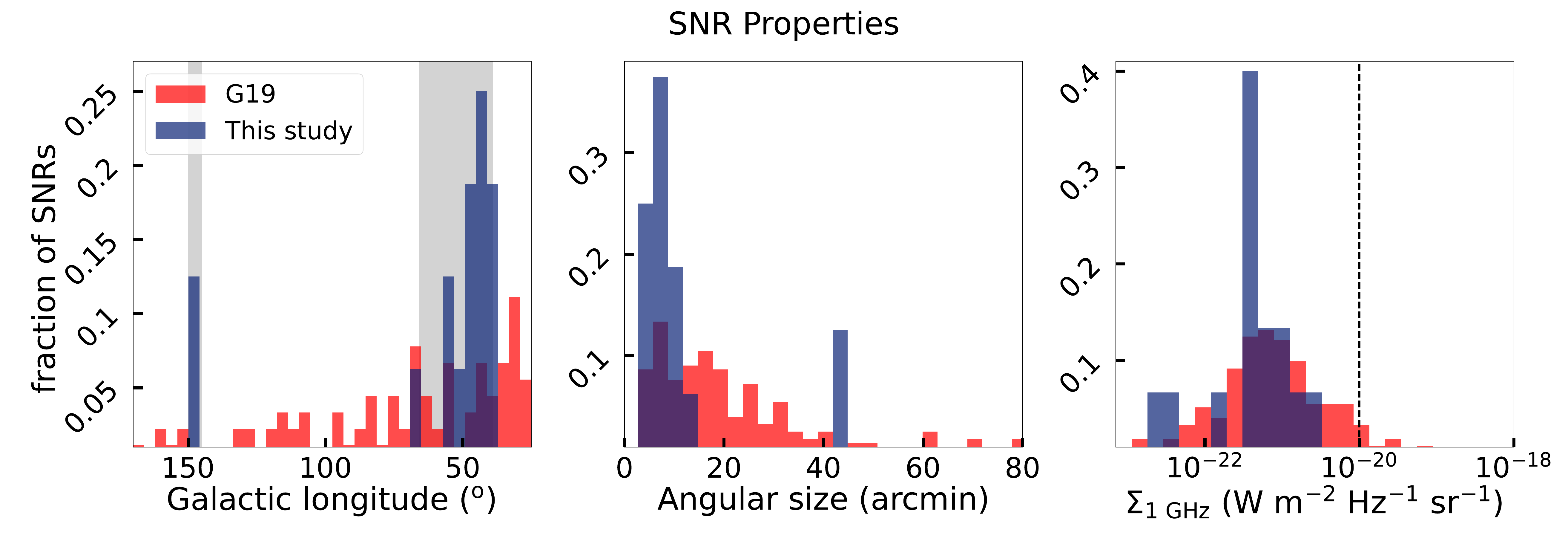}
    \caption{Properties of the candidate SNRs reported in this work, compared to the Galactic census of SNRs in the G19 catalogue. Depicted in blue is the sample of the candidate SNRs reported in this work. In red are the SNRs listed in the G19 catalogue. The height of the bars indicates the fraction of SNRs or SNR candidates within each bin, according to the colour scheme of their corresponding sample. The left panel shows the histogram of Galactic longitude of known SNRs and the new SNR candidates. The grey shaded background indicates the longitude range of the two Galactic regions used in this study. The middle panel shows the histogram of the angular size (semimajor axis) of known SNRs and the new SNR candidates. The right panel shows the histogram of the surface brightness of known SNRs and the new SNR candidates, at 1~GHz. The surface brightness of the candidate SNRs at 1~GHz was estimated using a typical SNR spectral index of $\alpha = -0.5$ and the measured integrated flux density at 144~MHz. The dashed black line indicates the completeness limit proposed by \cite{green04} for the catalogue presented therein.}
    \label{fig:SNRproperties}
   \end{figure*}
   
The left panel of fig. \ref{fig:SNRproperties} shows the distribution in Galactic longitude of the new SNR candidates, compared with the population of known SNRs in the region from the 2019 version of the Green catalogue \citep[][ hereafter G19]{Green_2019}\footnote{We used the 2019 version of Green's catalogue, G19, for the statistical study of the Galactic census of SNRs, because the most recent version, G22, is not yet available in tabular format on VizieR \citep{vizier}. This 2019 version lists 294 Galactic SNRs.}. 

The middle panel of Fig. \ref{fig:SNRproperties} shows the distribution of the candidates' angular sizes. The majority of the candidates reported in this study are comparatively small in size. The high resolution of the LoTSS survey (even the $20''$ data products used in this study) aids in the discovery of small, extended sources of radio emission that would be unresolved in other surveys of the region.

The right panel of Fig. \ref{fig:SNRproperties} shows the distribution of the candidates' surface brightness at 1~GHz ($\Sigma_\mathrm{1 GHz}$). A typical SNR spectral index of $\alpha = -0.5$ \citep[see][]{Dubner_2015} was used to estimate this surface brightness $\Sigma_\mathrm{1 GHz}$. \cite{green04} argued that the Galactic census of SNRs with $\Sigma_\mathrm{1 GHz} > 10^{-20}$~W~m$^{-2}$~Hz$^{-1}$~sr$^{-1}$ is complete. None of the candidates proposed in the present work exceed this limit. As most of these candidates have never been observed before ---even though the field has been the target of multiple surveys--- they might well have a steeper spectral index than $-0.5$, which would shift the blue histograms further to the left.

The high resolution and sensitivity of LoTSS resulted in the detection of 14 new SNR candidates of relatively small angular sizes ($<20'$ in most cases) and relatively low surface brightness ($\lesssim 10^{-21}$~W~m$^{-2}$~Hz$^{-1}$~sr$^{-1}$ at 1~GHz). These results showcase LOFAR's ability to address the main selection effects that hinder the detection of Galactic SNRs \citep[see][]{Green_1991}. Furthermore, 11 of the candidate SNRs reported in the present work are located in a busy Galactic region within $l=39^\mathrm{o}-60^\mathrm{o}$. If all are confirmed as SNRs, the number of known SNRs in that region will increase by 55\%.

\subsection{Recovery of extended emission in LoTSS}
Recovery of extended emission in LOFAR data can, in principle, be highly accurate, because of the large number of short baselines. In practice, however, the recovery of such structures can be imperfect if they have been partly modelled out during the calibration steps \citep{Shimwell_2022}. While the LoTSS-DR2 pipelines attempt to automatically detect such structures and accurately include them in the derived sky models, it is possible to miss some structure, especially in complex environments like the Galactic plane \citep[see Sect. 3.4 of ][]{Shimwell_2022}.

To further evaluate the accuracy of the deconvolution of the SNRs in this study, we inspected the deconvolution residual images of each field. We did not find substantial residual emission from any of the SNRs except from G46.60+0.20. The left-over emission of G46.60+0.20 could lead to misleading flux density. We therefore advise caution when interpreting the integrated flux density of this particular candidate SNR.

\subsection{Confirming the true nature of these SNR candidates}

The new SNR candidates proposed here, as well as our confirmed candidates from the \cite{Anderson_2017} work, are selected based on their radio and infrared properties. In the case of the \cite{Anderson_2017} `confirmed' candidates, we have spectral evidence for non-thermal emission. In the case of the new LOFAR SNR candidates, we have no spectral information, because their only existing detection is the one with LOFAR at 144~MHz.

However, a lack of MIR emission and a prominence at 144~MHz are not sufficient to confirm that an extended source is an SNR. Additional radio observations can be used to better constrain the nature of emission. However, the most convincing evidence that a source is an SNR would be an X-ray detection of a non-equilibrium ionisation (NEI) plasma associated with a recent strong shock \citep{vink12}. Follow-up with X-ray instruments, and in particular the examination of \textit{eROSITA} data \citep{sunyaev21}, would be the most promising avenue for confirming whether or not these candidates are indeed SNRs. However, if a SNR is relatively old, distant, or absorbed, it could still be undetected in the X-rays; therefore, it might prove difficult to confirm or reject the SNR nature of some of these candidates.

\section{Conclusions}\label{sec:conclusions}

We examined LoTSS images at 144~MHz of the Galactic regions within $39^\mathrm{o} < l < 66^\mathrm{o}$ and $|b|< 2.5^\mathrm{o}$, and within $145^\mathrm{o} < l < 150^\mathrm{o}$ and $|b| < 3^\mathrm{o}$ in order to look for new SNR candidates. We consider that an extended source is a SNR candidate if it is visible in the 144~MHz LOFAR maps, and lacks MIR emission in \textrm{WISE} bands W3 and W4. We find 14 new SNR candidates. We also followed up on 24 candidates previously reported in the literature, confirming that 6 of them show non-thermal emission and 4 of them are actually thermal sources; for the remaining 14, our constraints on the nature of their emission are inconclusive. Follow-up X-ray observations of these candidate sources are necessary in order to unambiguously confirm whether or not they are indeed SNRs.

\begin{acknowledgements}
LOFAR \citep{vanhaarlem13} is the LOw Frequency ARray designed and constructed by ASTRON. It has observing, data processing, and data storage facilities in several countries, which are owned by various parties (each with their own funding sources), and are collectively operated by the ILT foundation under a joint scientific policy. The ILT resources have benefitted from the following recent major funding sources: CNRS-INSU, Observatoire de Paris and Universit\'{e} d'Orl\'{e}ans, France; BMBF, MIWF-NRW, MPG, Germany; Science Foundation Ireland (SFI), Department of Business, Enterprise and Innovation (DBEI), Ireland; NWO, The Netherlands; The Science and Technology Facilities Council, UK; Ministry of Science and Higher Education, Poland; Istituto Nazionale di Astrofisica (INAF), Italy.
This research made use of the Dutch national e-infrastructure with support of the SURF Cooperative (e-infra 180169) and the LOFAR e-infra group. The J\"ulich LOFAR Long Term Archive and the German LOFAR network are both coordinated and operated by the J\"ulich Supercomputing Centre (JSC), and computing resources on the supercomputer JUWELS at JSC were provided by the Gauss Centre for Supercomputing e.V. (grant CHTB00) through the John von Neumann Institute for Computing (NIC). This research made use of the University of Hertfordshire high-performance computing facility and the LOFAR-UK computing facility located at the University of Hertfordshire and supported by STFC [ST/P000096/1], and of the Italian LOFAR IT computing infrastructure supported and operated by INAF, and by the Physics Department of Turin University (under an agreement with Consorzio Interuniversitario per la Fisica Spaziale) at the C3S Supercomputing Centre, Italy. 
MA acknowledges support from the VENI research programme with project number 202.143, which is financed by the Netherlands Organisation for Scientific Research (NWO).  MJH thanks the UK STFC for support through grant [ST/V000624/1].
\end{acknowledgements}
\bibliographystyle{aa}
\bibliography{bib.bib}
\begin{appendix}


\section{New SNR candidates}
\label{appendix}

Here we present the images of the new SNR candidates discovered in LoTSS galactic pointings. Green solid line shows the candidates while orange solid lines show known H\,{\sc ii} regions in the proximity of the candidates. Unless stated otherwise, the radio maps are taken from the LoTSS survey. 
\begin{figure}[h!]
   \centering
\includegraphics[width=\hsize]{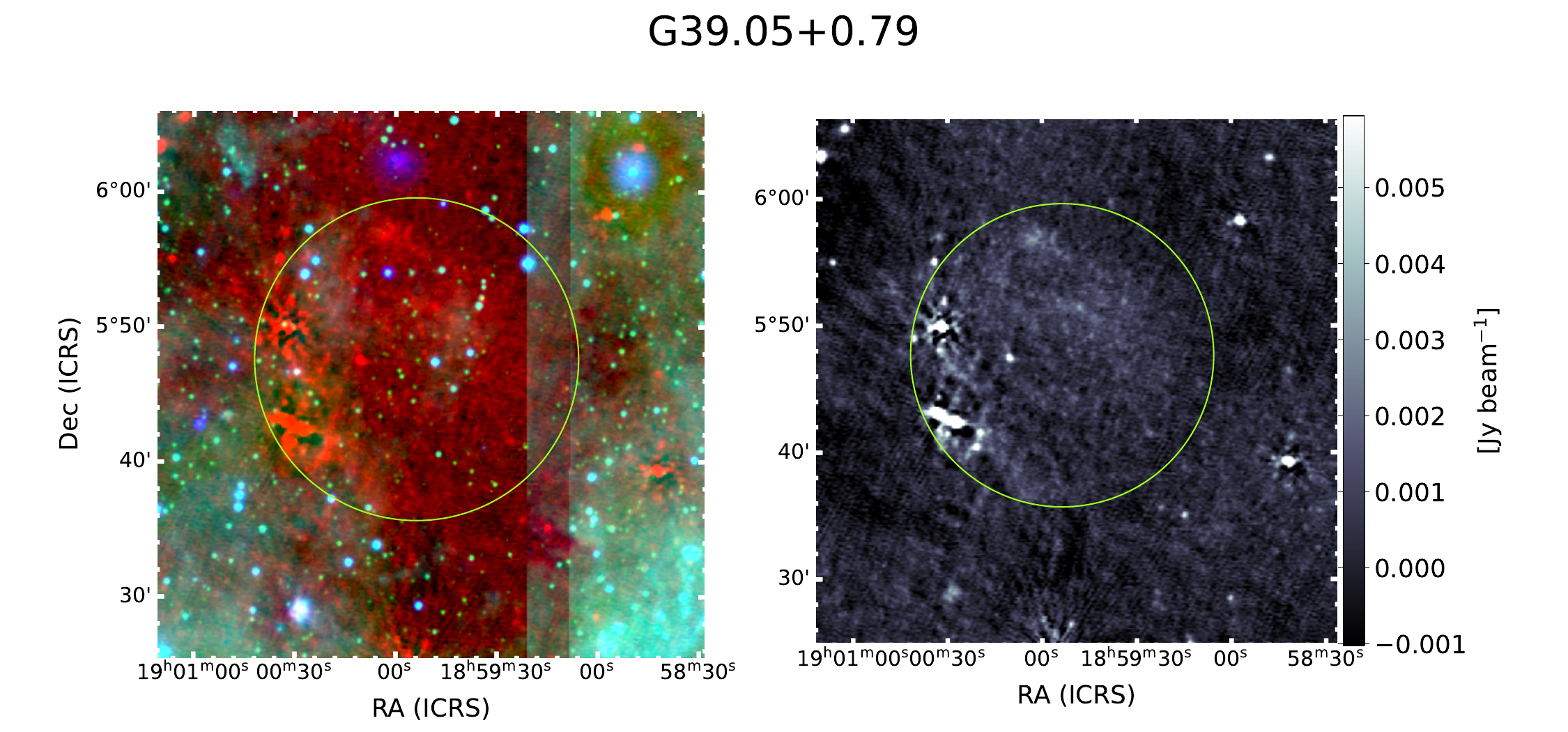}
    \caption{G39.05+0.79 in radio and in false colour map with MIR images from \textrm{WISE} W3 and W4 bands. Right panel shows the candidate at 144MHz in the low resolution image of the LoTSS survey. Left panel shows a false colour map of the region with LoTSS in red (2m), \textrm{WISE} W3 band in green (12 $\mu$m)  and \textrm{WISE} W4 band in blue (22 $\mu$m). North is upwards and West is leftwards. The green solid line indicates the candidate.}
    \label{fig:G39.05+0.79}
   \end{figure}
    \FloatBarrier

   \begin{figure}[h!]
   \centering
\includegraphics[width=\hsize]{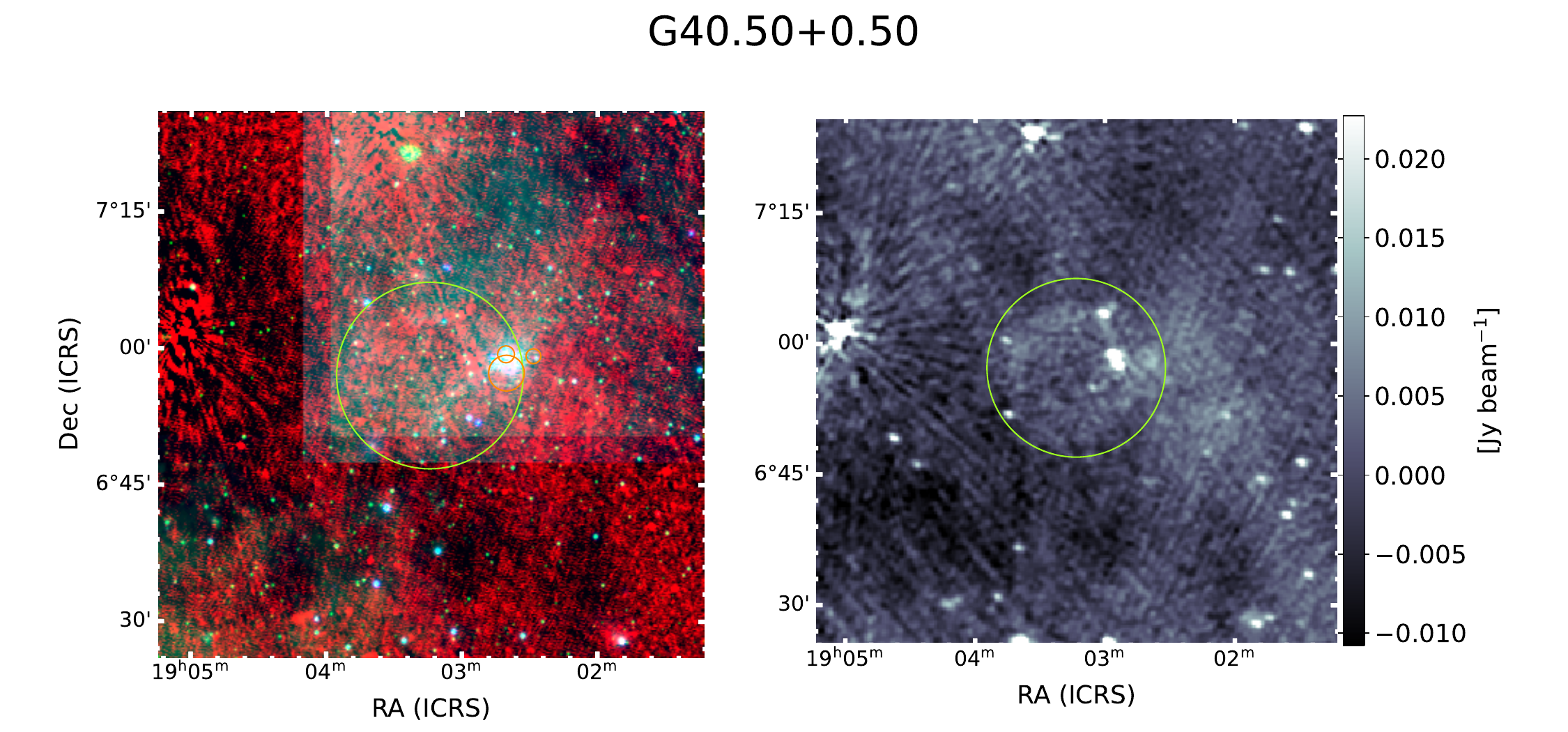}
    \caption{G40.50+0.50 in radio and in false colour map with MIR images from \textrm{WISE} W3 and W4 bands. The radio map in the right panel, is taken from the LOFAR project LC18\_027. The orange solid lines indicate the positions of known H\,{\sc ii} regions.}
    \label{fig:G40.5+0.5}
   \end{figure}
    \FloatBarrier
    
   \begin{figure}[h!]
   \centering
\includegraphics[width=\hsize]{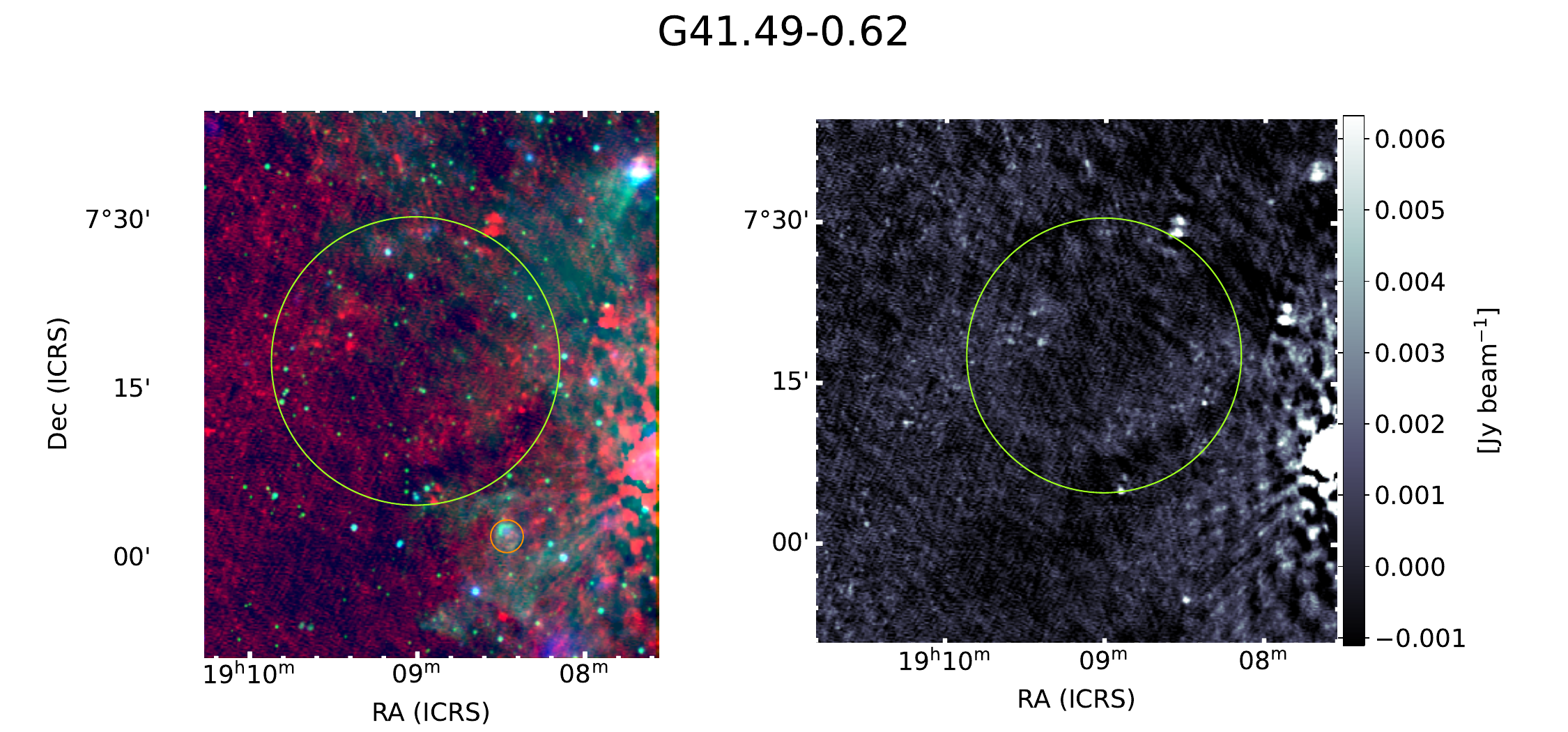}
    \caption{G41.49-0.62 in radio and in false colour map with MIR images from \textrm{WISE} W3 and W4 bands. The orange solid lines indicate the positions of known H\,{\sc ii} regions.}
    \label{fig:G41.49-0.62}
   \end{figure}
    \FloatBarrier
    
   \begin{figure}[h!]
   \centering
\includegraphics[width=\hsize]{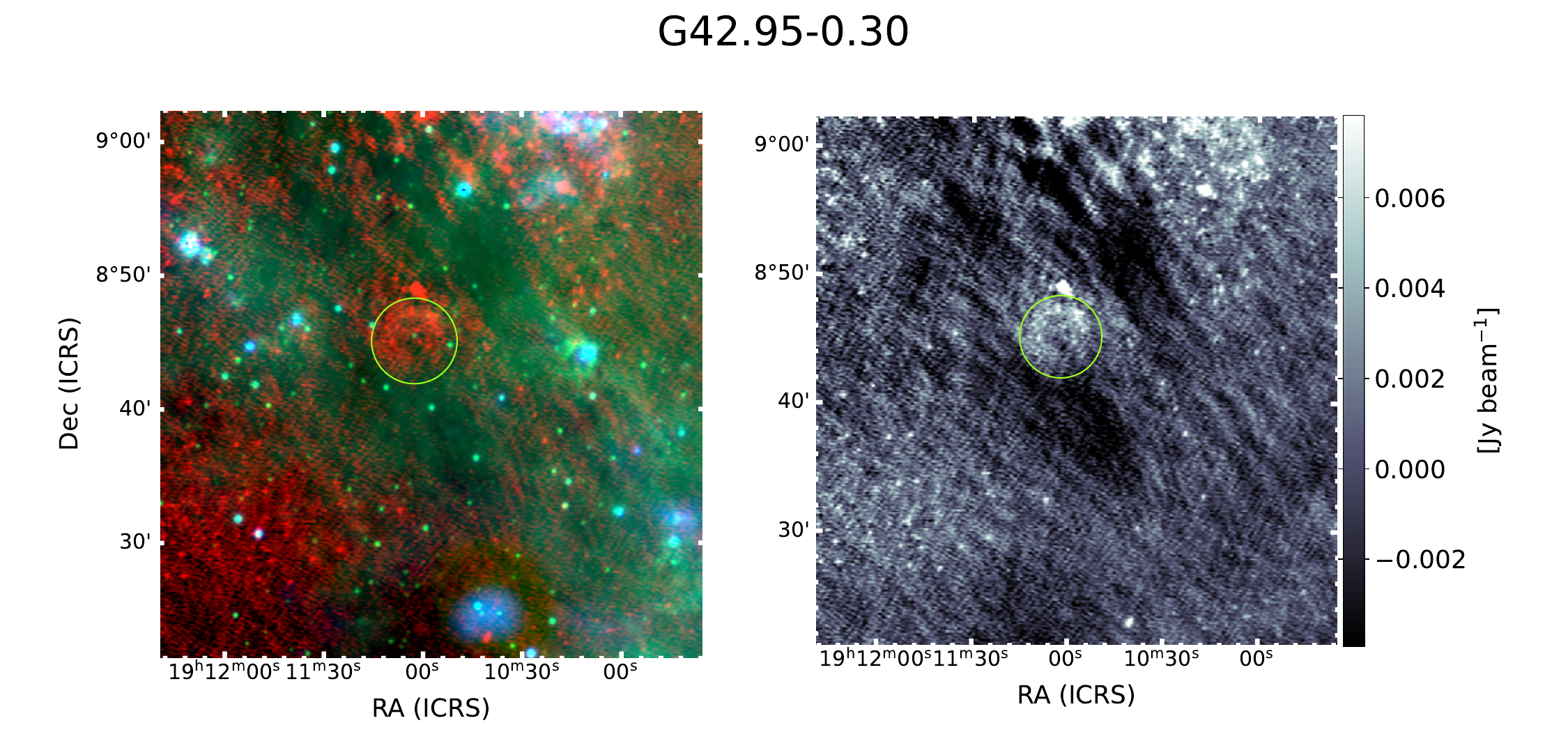}
    \caption{G42.95-0.30 in radio and in false colour map with MIR images from \textrm{WISE} W3 and W4 bands.}
    \label{fig:G42.95-0.30}
   \end{figure}
    \FloatBarrier   

   \begin{figure}[h!]
   \centering
\includegraphics[width=\hsize]{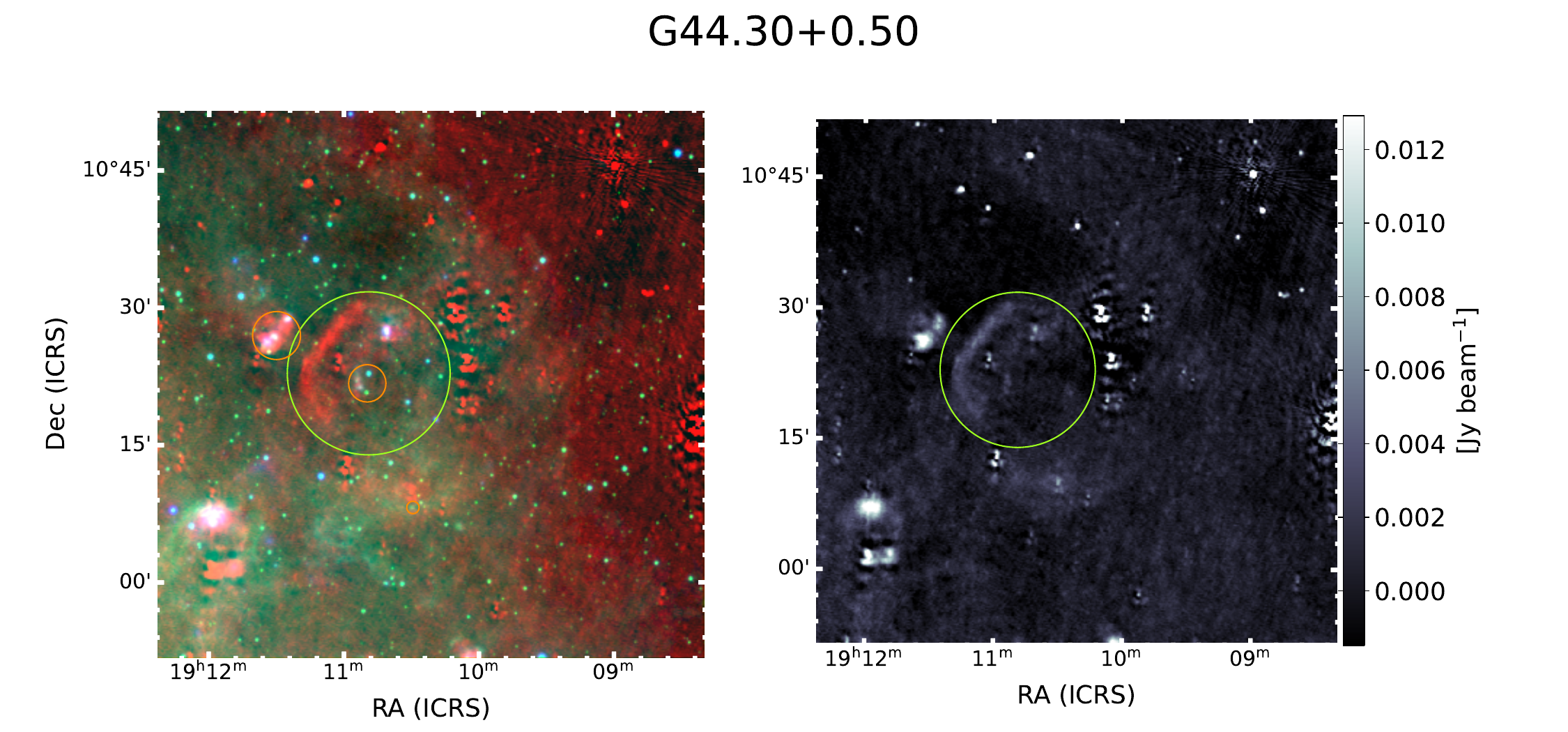}
    \caption{G44.30+0.50 in radio and in false colour map with MIR images from \textrm{WISE} W3 and W4 bands. The orange solid lines indicate the positions of known H\,{\sc ii} regions.}
    \label{fig:G44.3+0.5}
   \end{figure}
    \FloatBarrier
    
   \begin{figure}[h!]
   \centering
\includegraphics[width=\hsize]{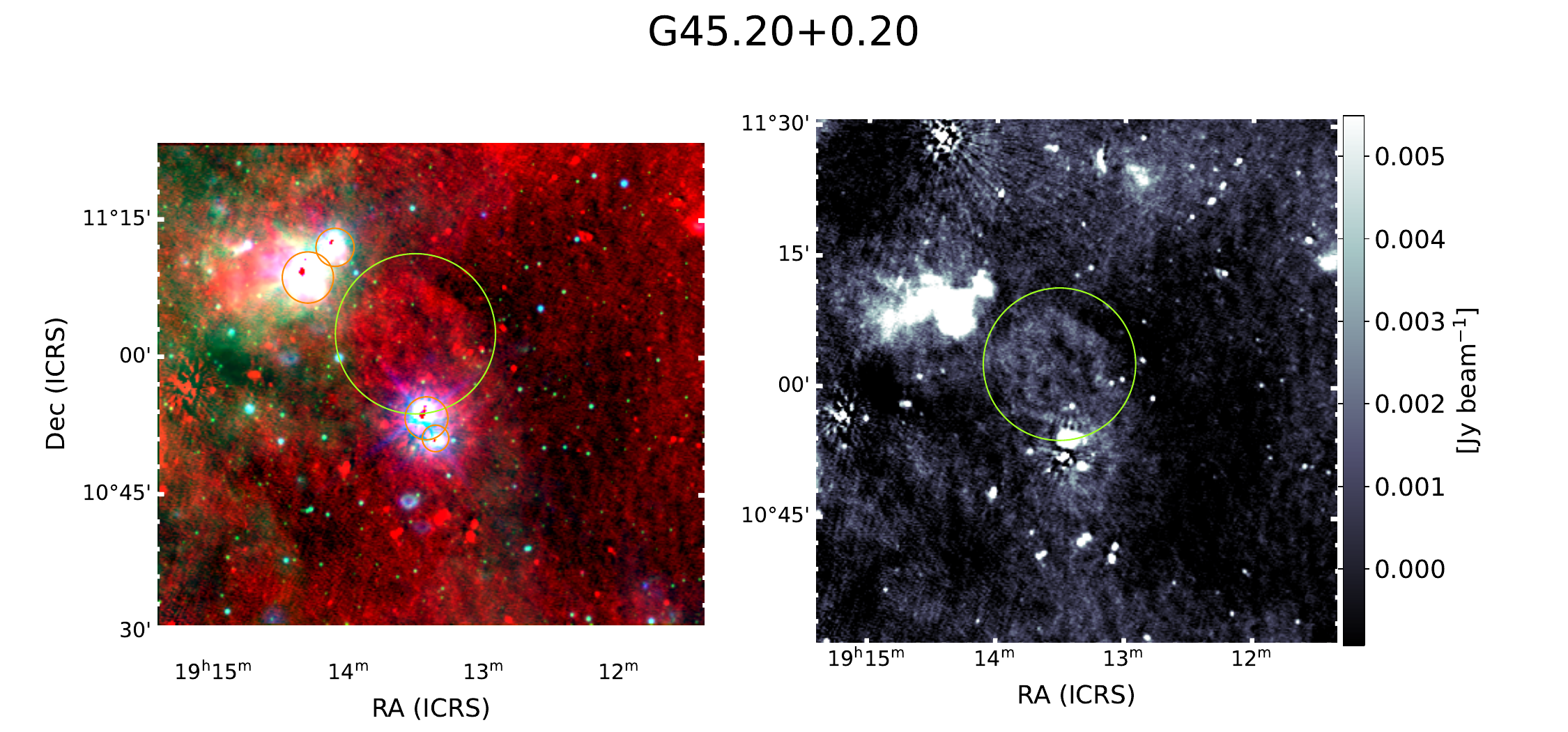}
    \caption{G45.20+0.20 in radio and in false colour map with MIR images from \textrm{WISE} W3 and W4 bands.The orange solid lines indicate the positions of known H\,{\sc ii} regions.}
    \label{fig:G45.2+0.2}
   \end{figure}
    \FloatBarrier
    
   \begin{figure}[h!]
   \centering
\includegraphics[width=\hsize]{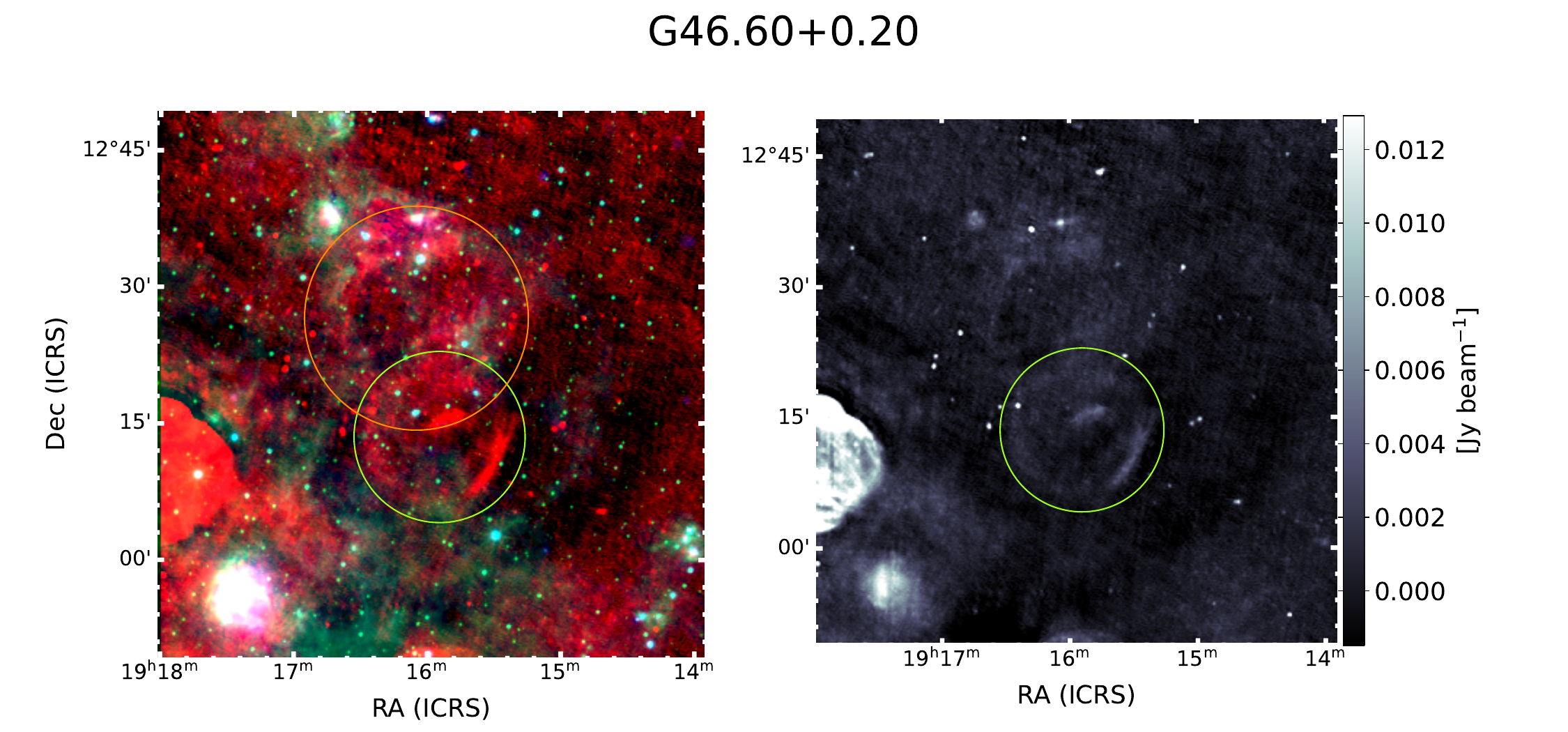}
    \caption{G46.60+0.20 in radio and in false colour map with MIR images from \textrm{WISE} W3 and W4 bands.The orange solid lines indicate the positions of known H\,{\sc ii} regions.}
    \label{fig:G46.6+0.2}
   \end{figure}
    \FloatBarrier
    
   \begin{figure}[h!]
   \centering
\includegraphics[width=\hsize]{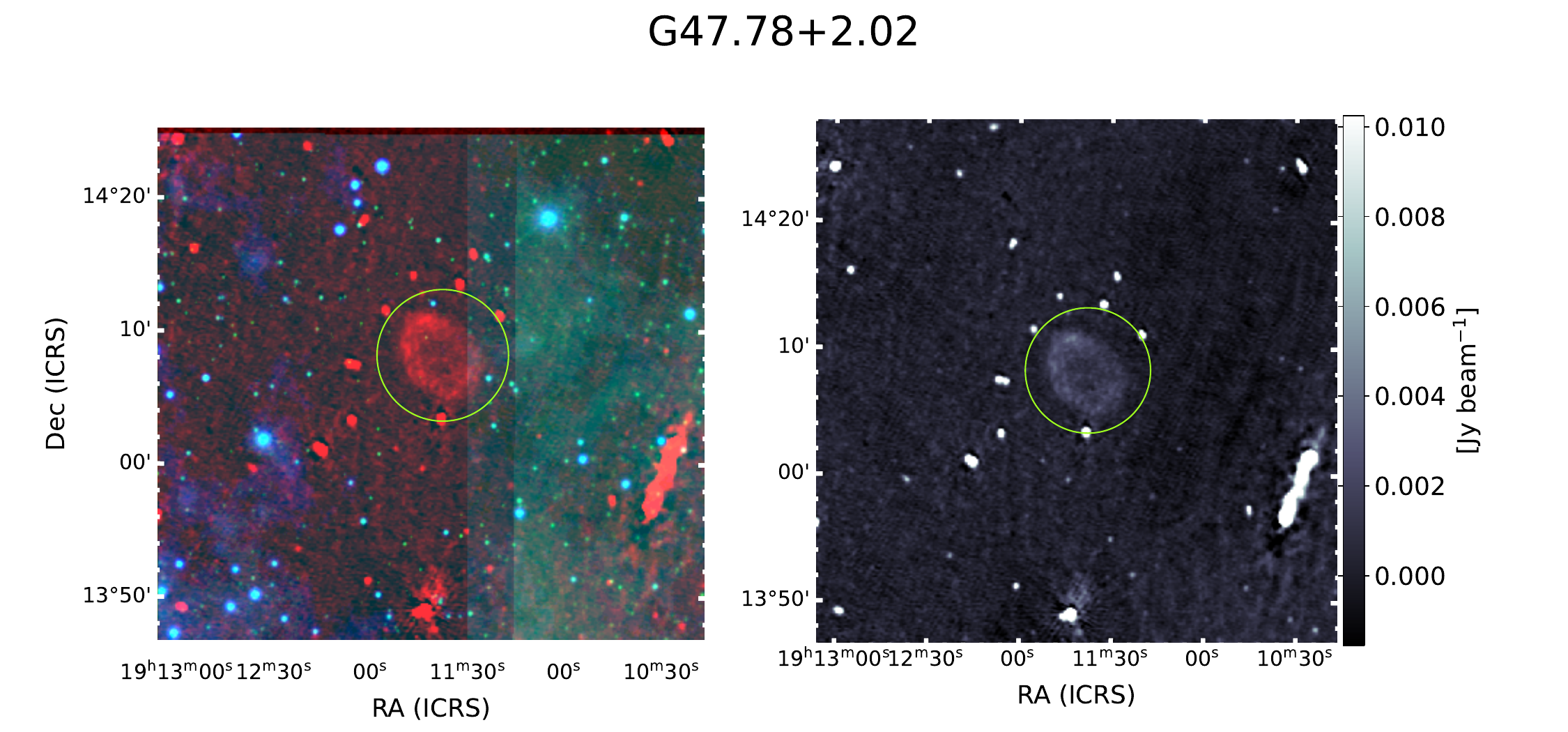}
    \caption{G47.78+2.02 in radio and in false colour map with MIR images from \textrm{WISE} W3 and W4 bands.}
    \label{fig:G47.78+2.02}
   \end{figure}
    \FloatBarrier
    
   \begin{figure}[h!]
   \centering
\includegraphics[width=\hsize]{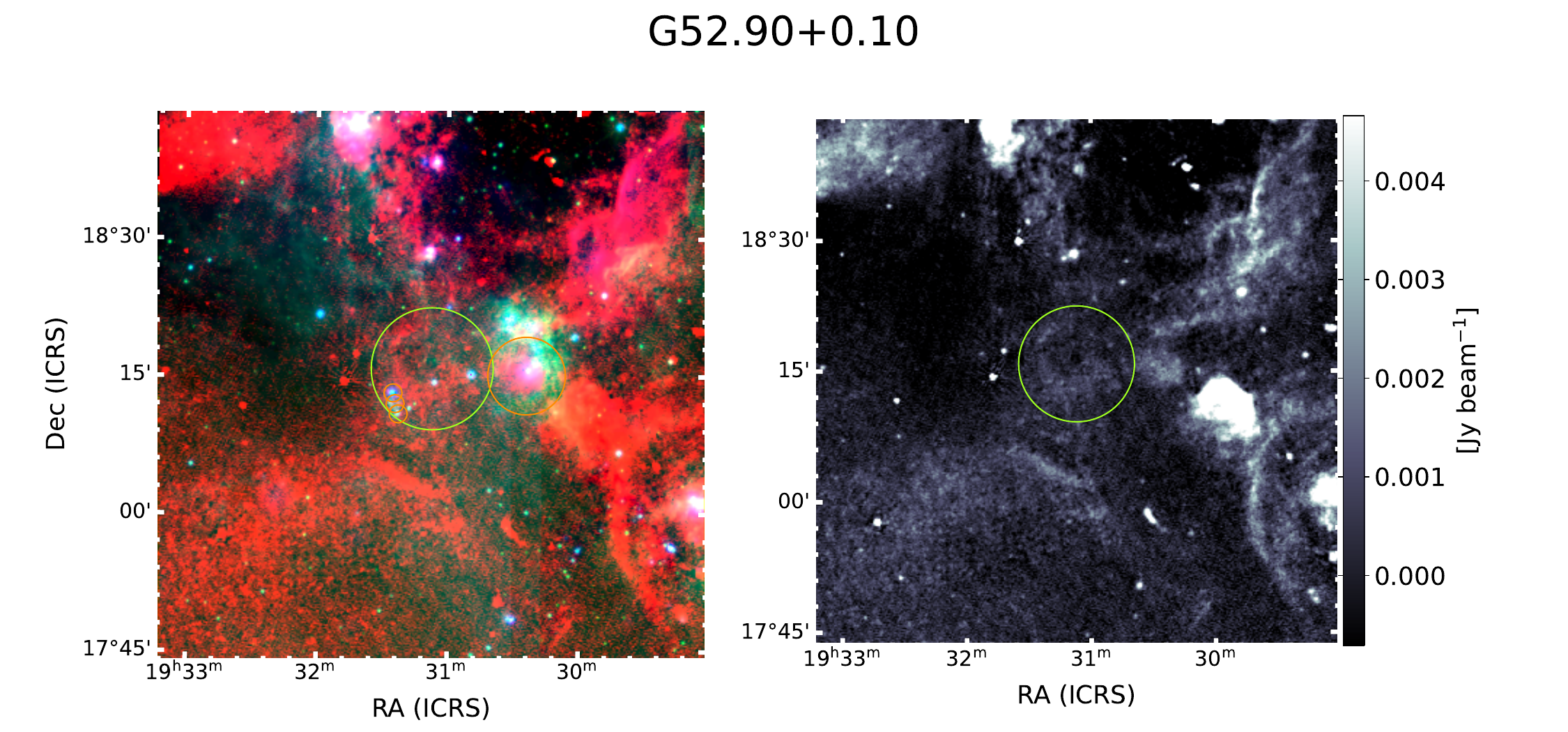}
    \caption{G52.90+0.10 in radio and in false colour map with MIR images from \textrm{WISE} W3 and W4 bands.The orange solid lines indicate the positions of known H\,{\sc ii} regions.}
    \label{fig:G52.9+0.1}
   \end{figure}
    \FloatBarrier
    
   \begin{figure}[h!]
   \centering
\includegraphics[width=\hsize]{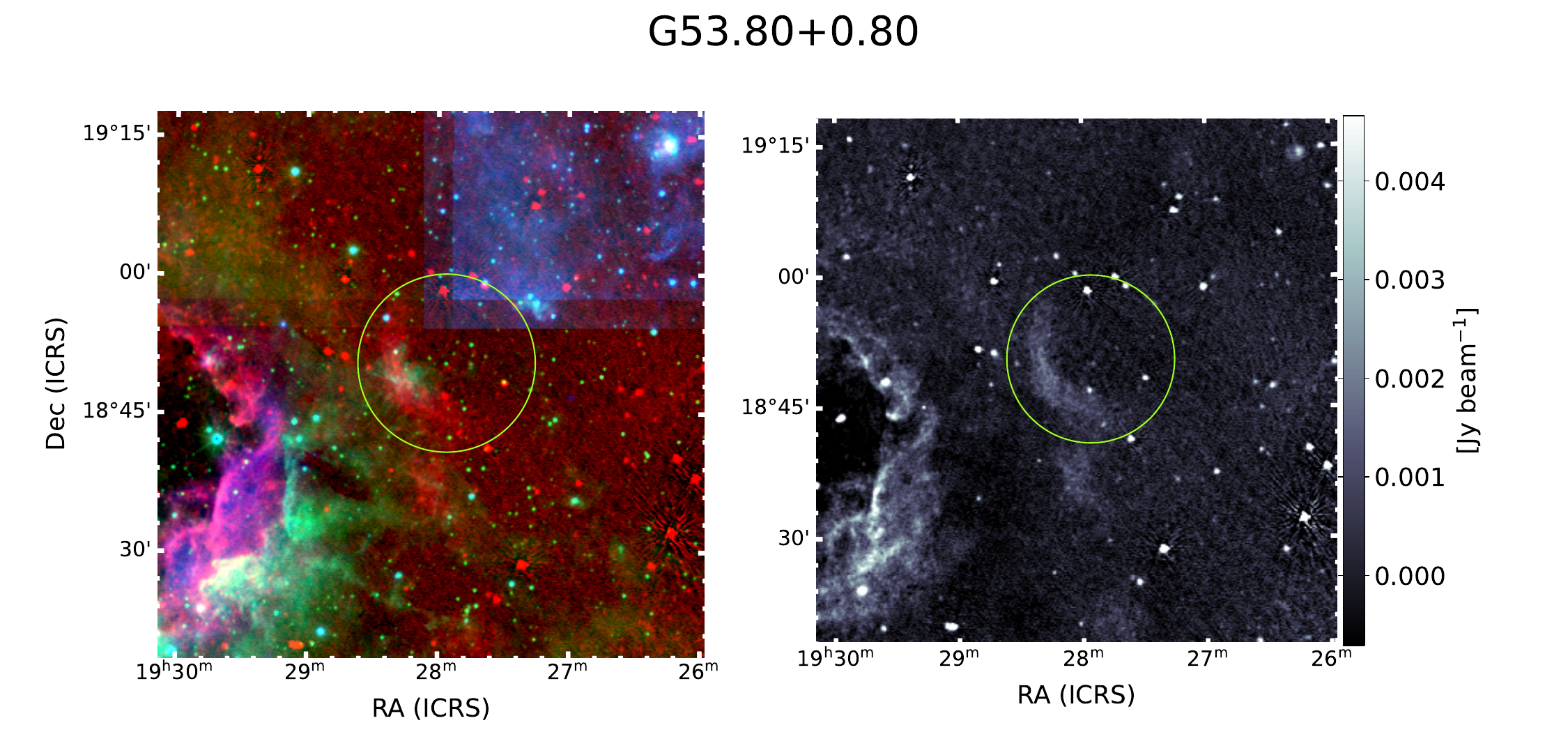}
    \caption{G53.80+0.80 in radio and in false colour map with MIR images from \textrm{WISE} W3 and W4 bands.}
    \label{fig:G53.8+0.8}
   \end{figure}
    \FloatBarrier
    
   \begin{figure}[h!]
   \centering
\includegraphics[width=\hsize]{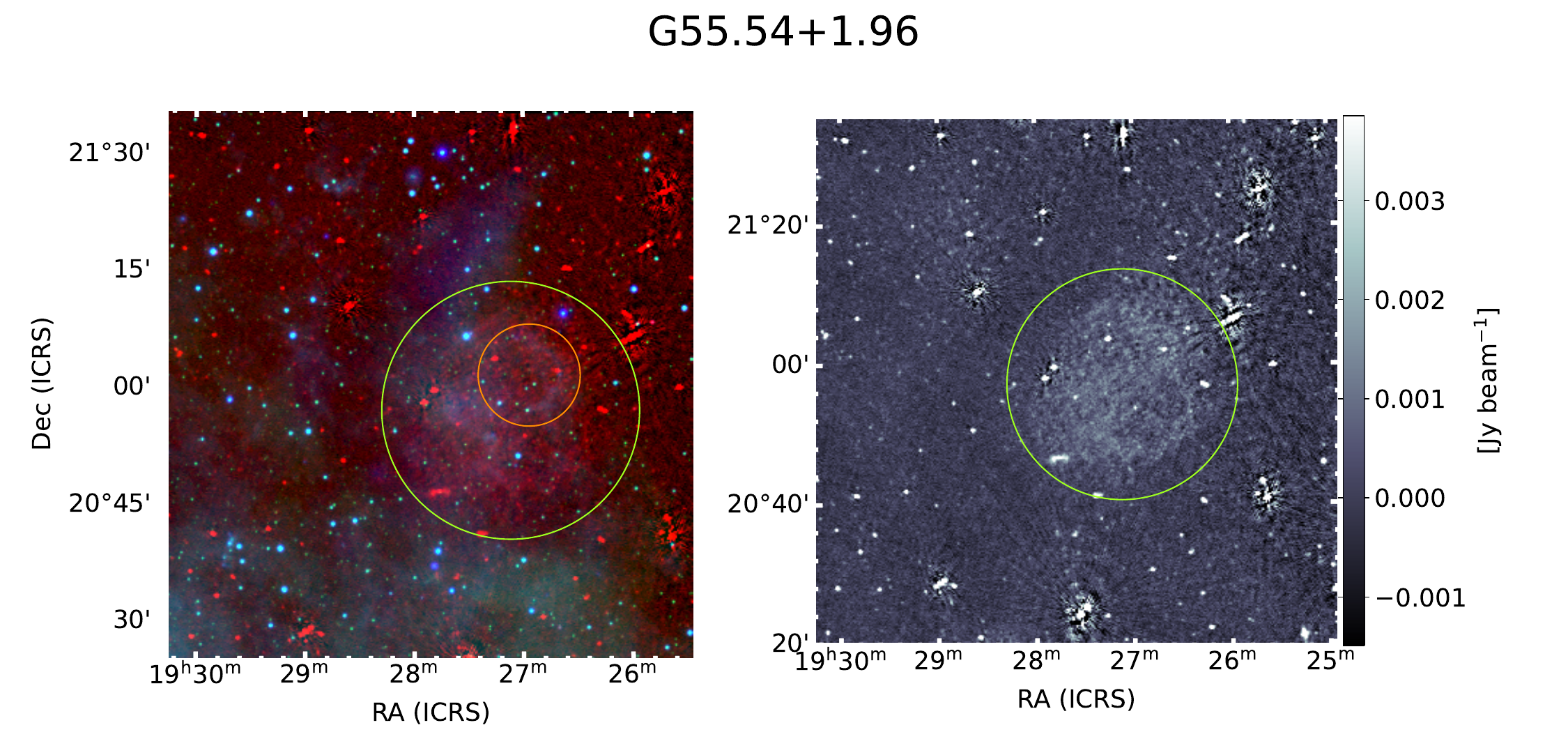}
    \caption{G55.54+1.96 in radio and in false colour map with MIR images from \textrm{WISE} W3 and W4 bands.The orange solid lines indicate the positions of known H\,{\sc ii} regions.}
    \label{fig:G55.54+1.96}
   \end{figure}   
    \FloatBarrier
    
   \begin{figure}[h!]
   \centering
\includegraphics[width=\hsize]{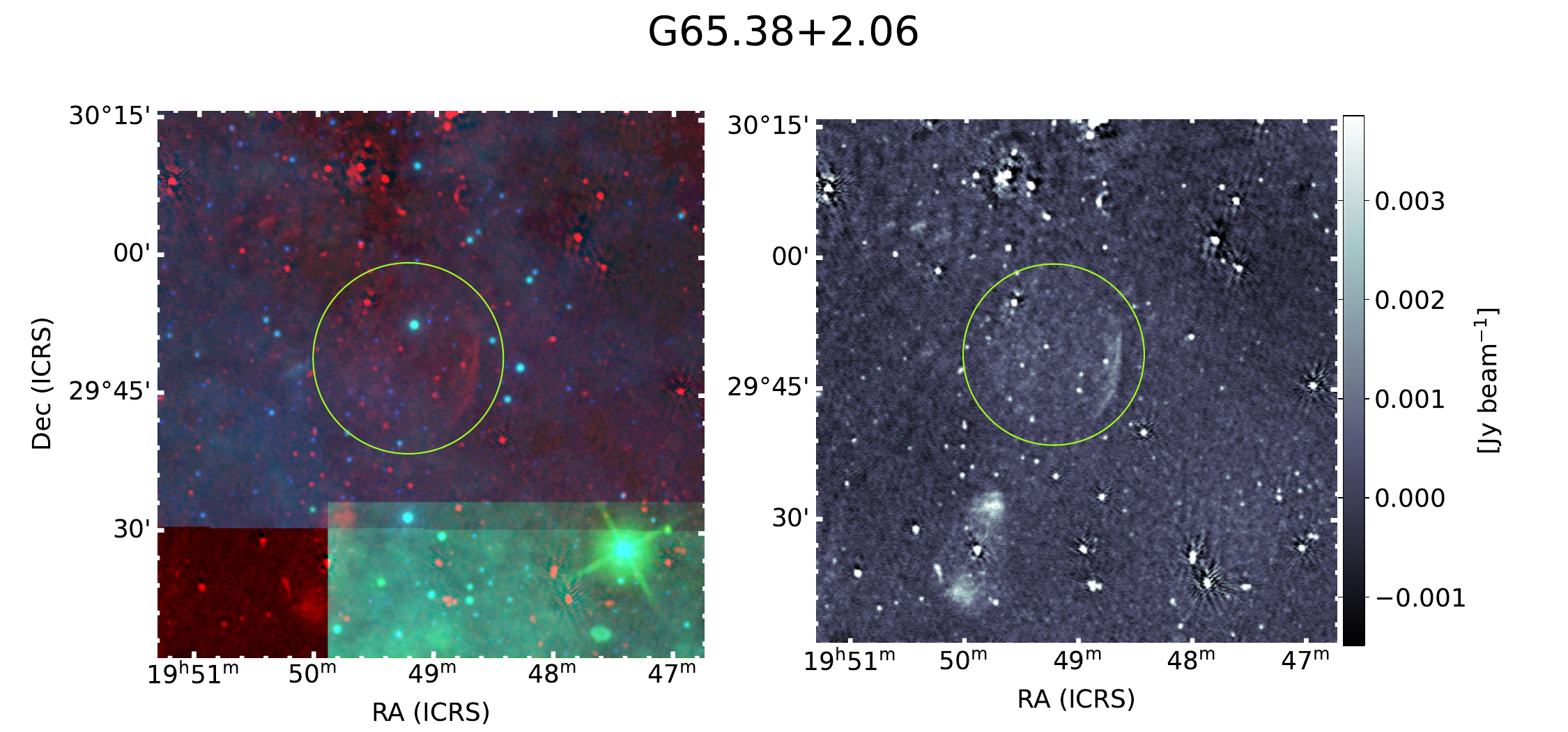}
    \caption{G65.38+2.06 in radio and in false colour map with MIR images from \textrm{WISE} W3 and W4 bands.}
    \label{fig:G65.38+2.06}
   \end{figure}   
    \FloatBarrier
    
   \begin{figure}[h!]
   \centering
\includegraphics[width=\hsize]{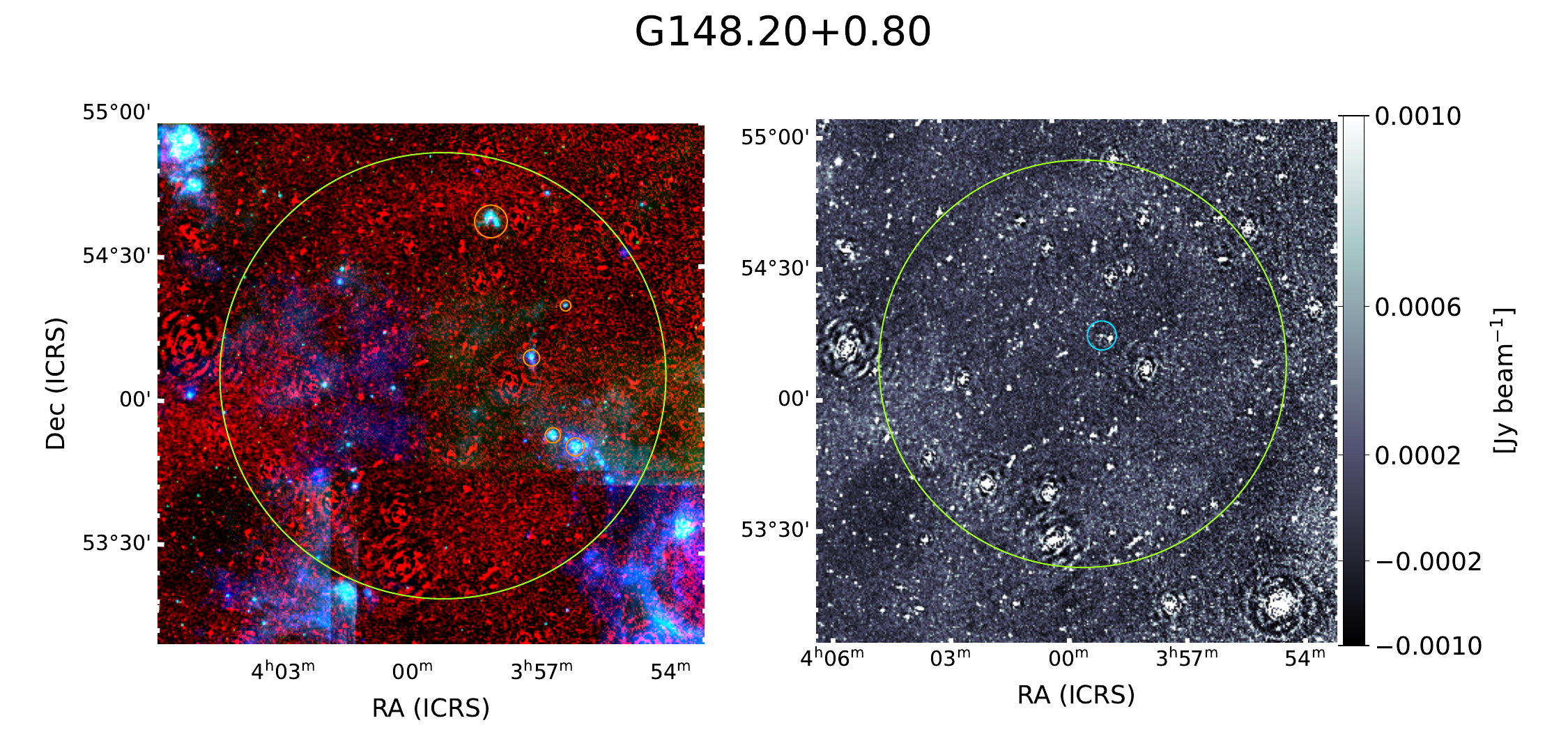}
    \caption{G148.2+0.8 in radio and in false colour map with MIR images from \textrm{WISE} W3 and W4 bands.The orange solid lines indicate the positions of known H\,{\sc ii} regions. The blue solid line indicates the position of the PWN G148.1+00.8.}
    \label{fig:G148.2+0.8}
   \end{figure}   
    \FloatBarrier
    
   \begin{figure}[h!]
   \centering
\includegraphics[width=\hsize]{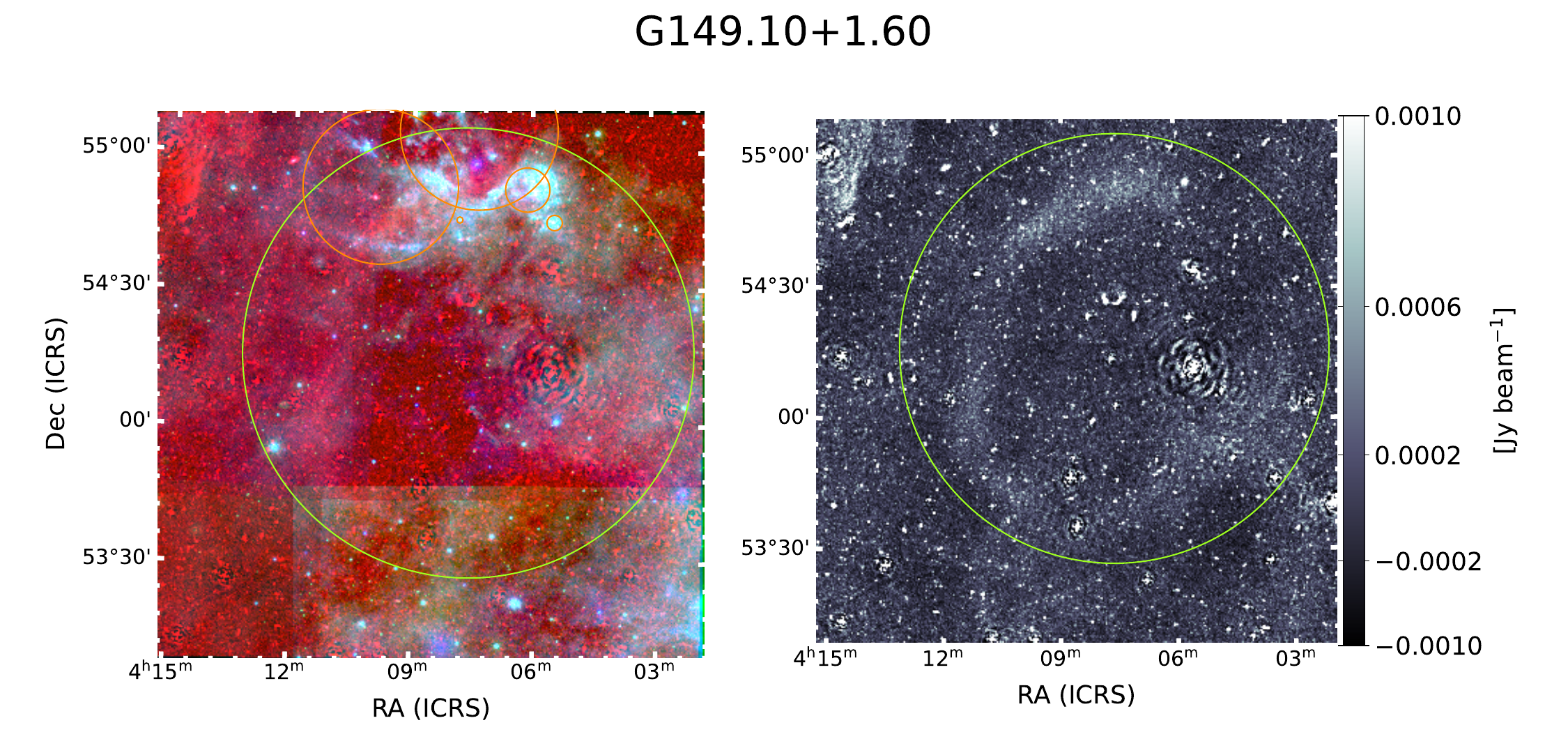}
    \caption{G149.1+1.6 in radio and in false colour map with MIR images from \textrm{WISE} W3 and W4 bands.The orange solid lines indicate the positions of known H\,{\sc ii} regions.}
    \label{fig:G149.1+1.6}
   \end{figure}   
    \FloatBarrier
\newpage
\newpage
\section{Known SNR candidates}
\label{appendix_thor}
In the following section we show the known SNR candidates investigated in this study. Green solid line shows the candidates while orange solid lines show known H\,{\sc ii} regions in the proximity of the candidates. We explicitly state which of the following radio maps are taken from the LC18\_027 project (see Sect. \ref{Ch:2.1.1}). The 22 SNR candidates from \cite{Anderson_2017} are only shown in radio, since their detection was based on the lack of MIR emission. We defined the green solid lines, indicating the candidates, using the reported centre and radius in \citet{Anderson_2017}.

\begin{figure}[h!]
   \centering
\includegraphics[width=\hsize]{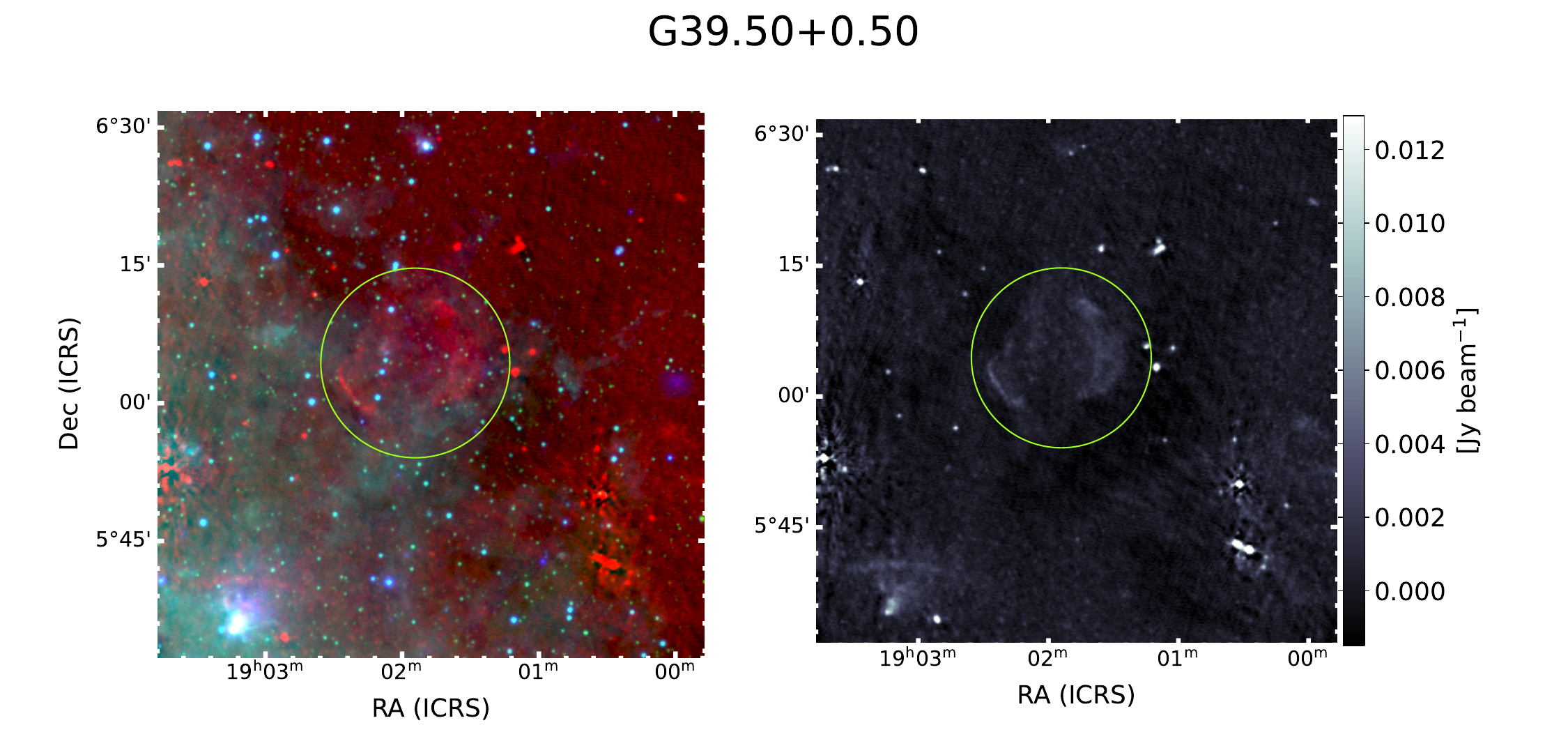}
    \caption{G39.50+0.50 in radio and in false colour map with MIR images from \textrm{WISE} W3 and W4 bands.}
    \label{fig:G39.5+0.5}
   \end{figure}
\FloatBarrier

\begin{figure}[h!]
   \centering
\includegraphics[width=\hsize]{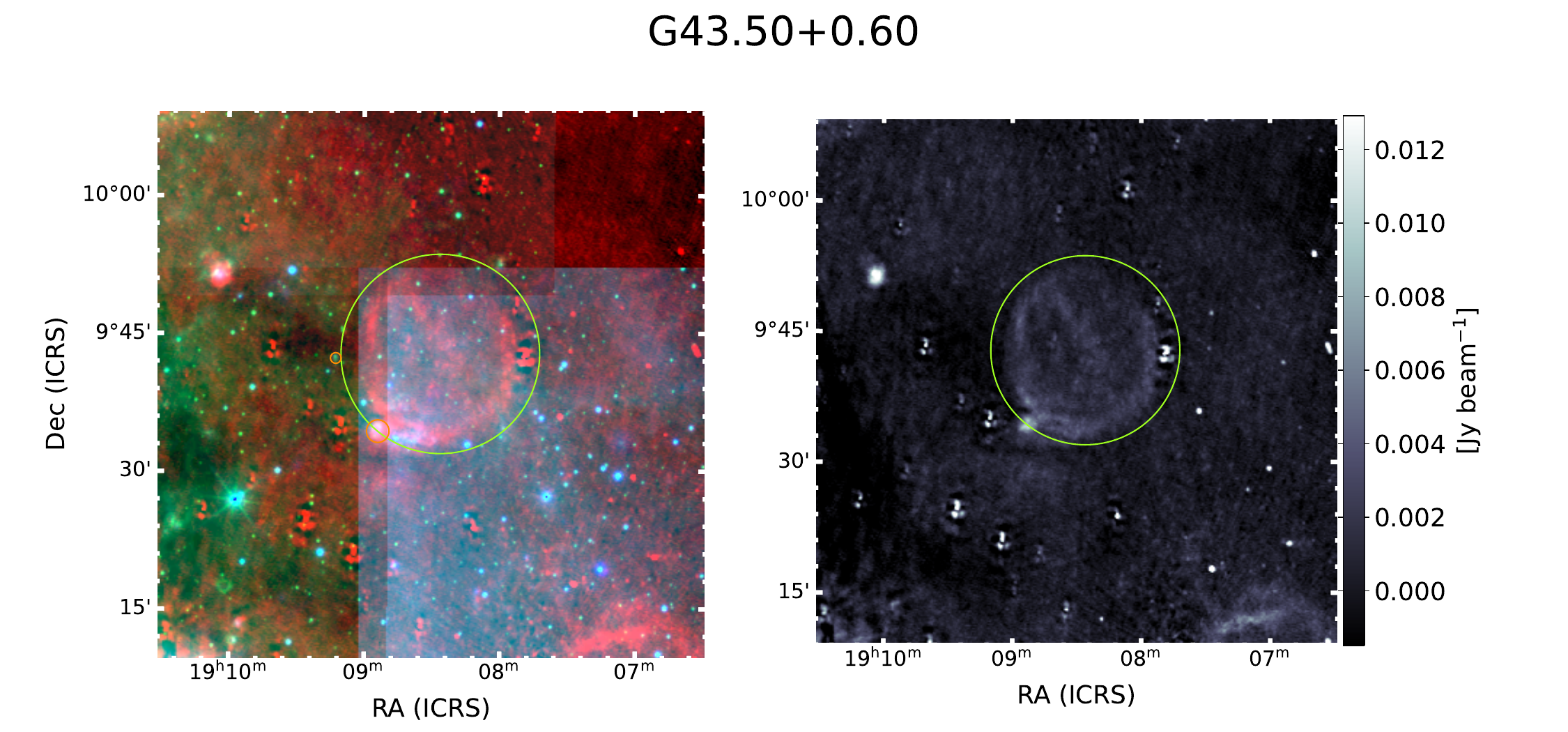}
    \caption{G43.50+0.60 in radio and in false colour map with MIR images from \textrm{WISE} W3 and W4 bands. The orange solid lines indicate the positions of known H\,{\sc ii} regions.}
    \label{fig:G43.5+0.6}
   \end{figure}
    \FloatBarrier

\begin{figure}[h!]
   \centering
\includegraphics[width=5.5 cm]{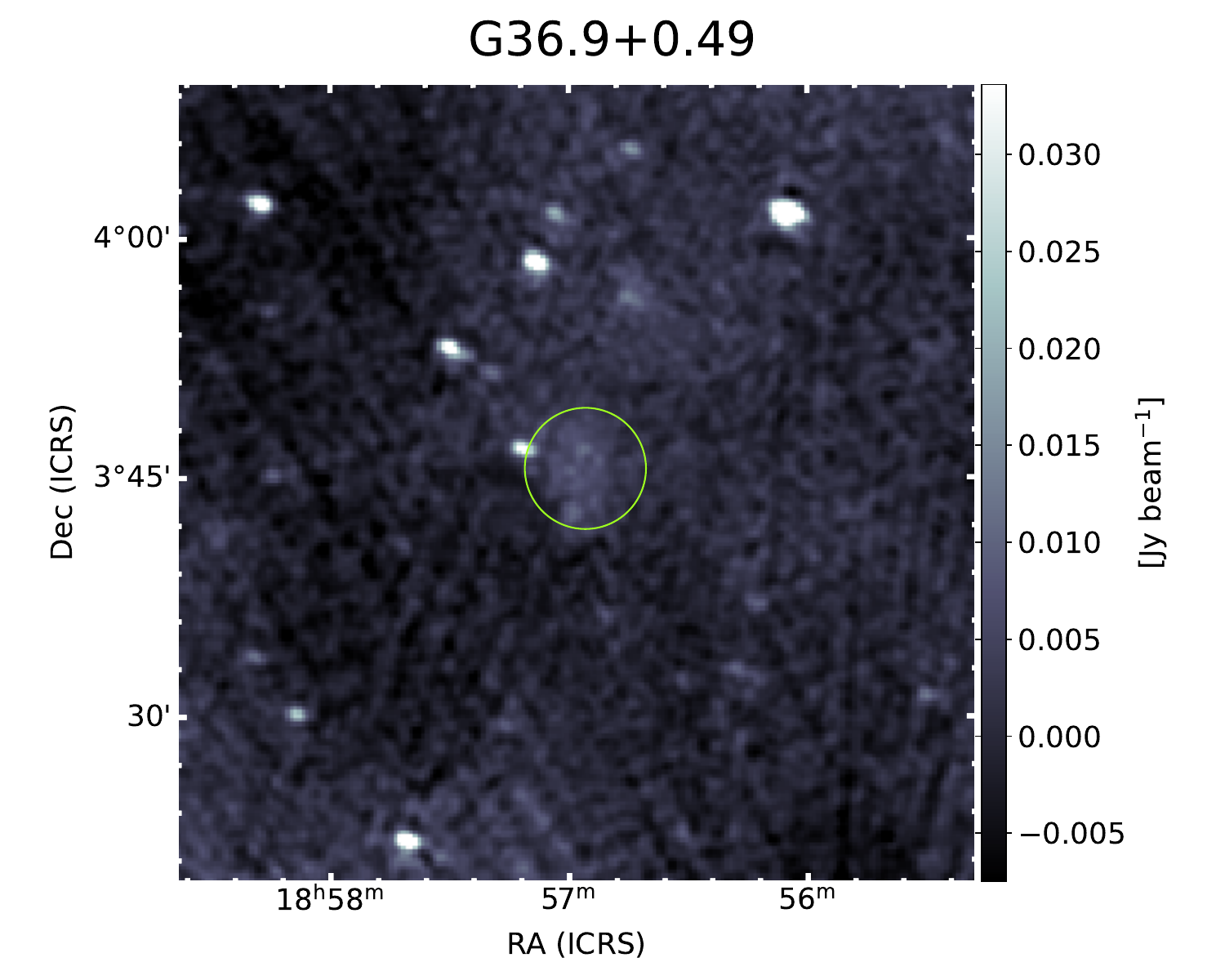}
    \caption{G36.9+0.49 in radio. The radio map is from the LOFAR project LC18\_027. North is
upwards and East is leftwards.The green solid line indicates the candidate.}
    \label{fig:G36.9+0.49}
   \end{figure}
    \FloatBarrier
    
\begin{figure}[h!]
   \centering
\includegraphics[width=5.5 cm]{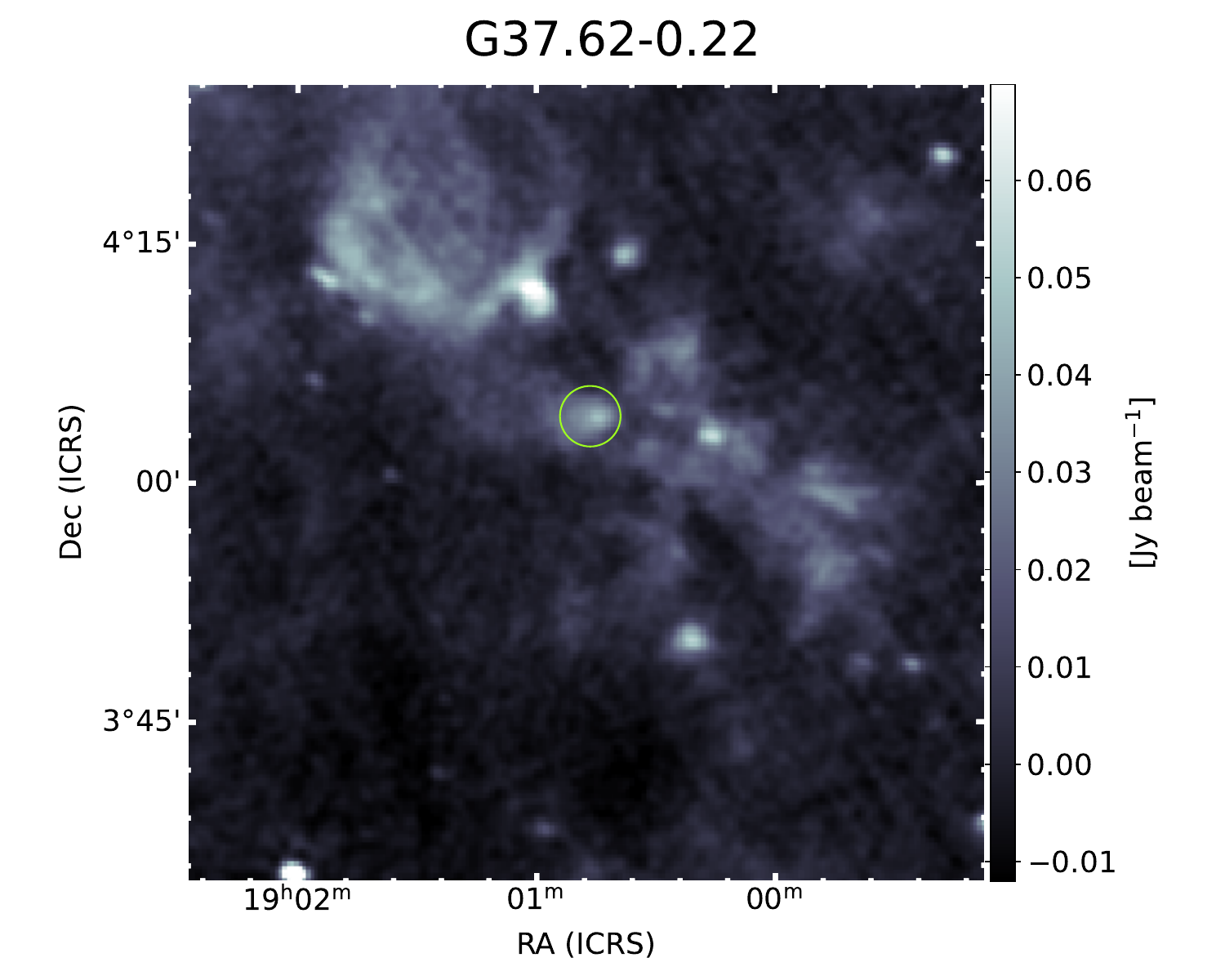}
    \caption{G37.62-0.22 in radio. The radio map is from the LOFAR project LC18\_027.}
    \label{fig:G37.62-0.22}
   \end{figure}
    \FloatBarrier
    
\begin{figure}[h!]
   \centering
\includegraphics[width=5.5 cm]{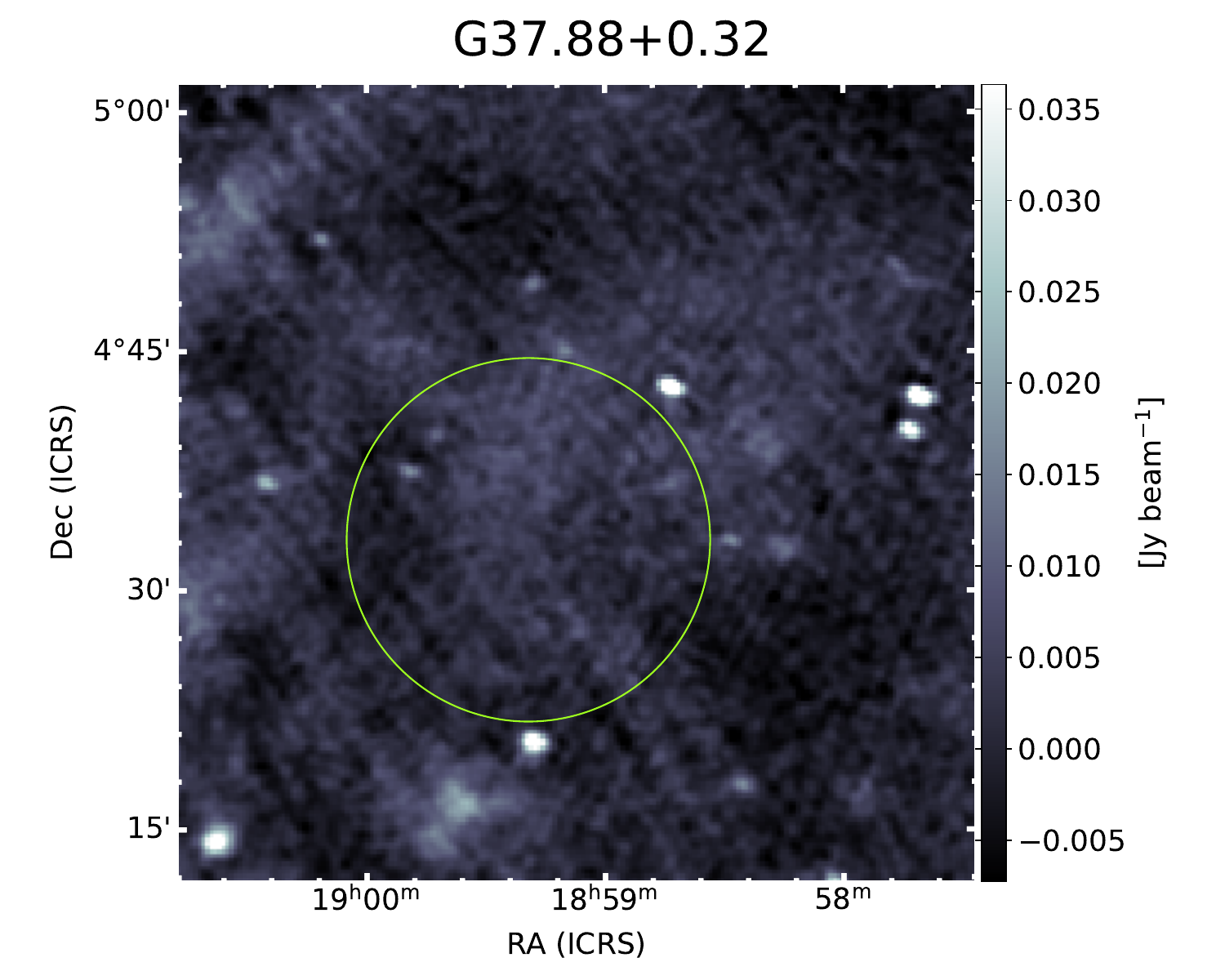}
    \caption{G37.88+0.32 in radio. The radio map is from the LOFAR project LC18\_027.}
    \label{fig:G37.88+0.32}
   \end{figure}
    \FloatBarrier
    
\begin{figure}[h!]
   \centering
\includegraphics[width=5.5 cm]{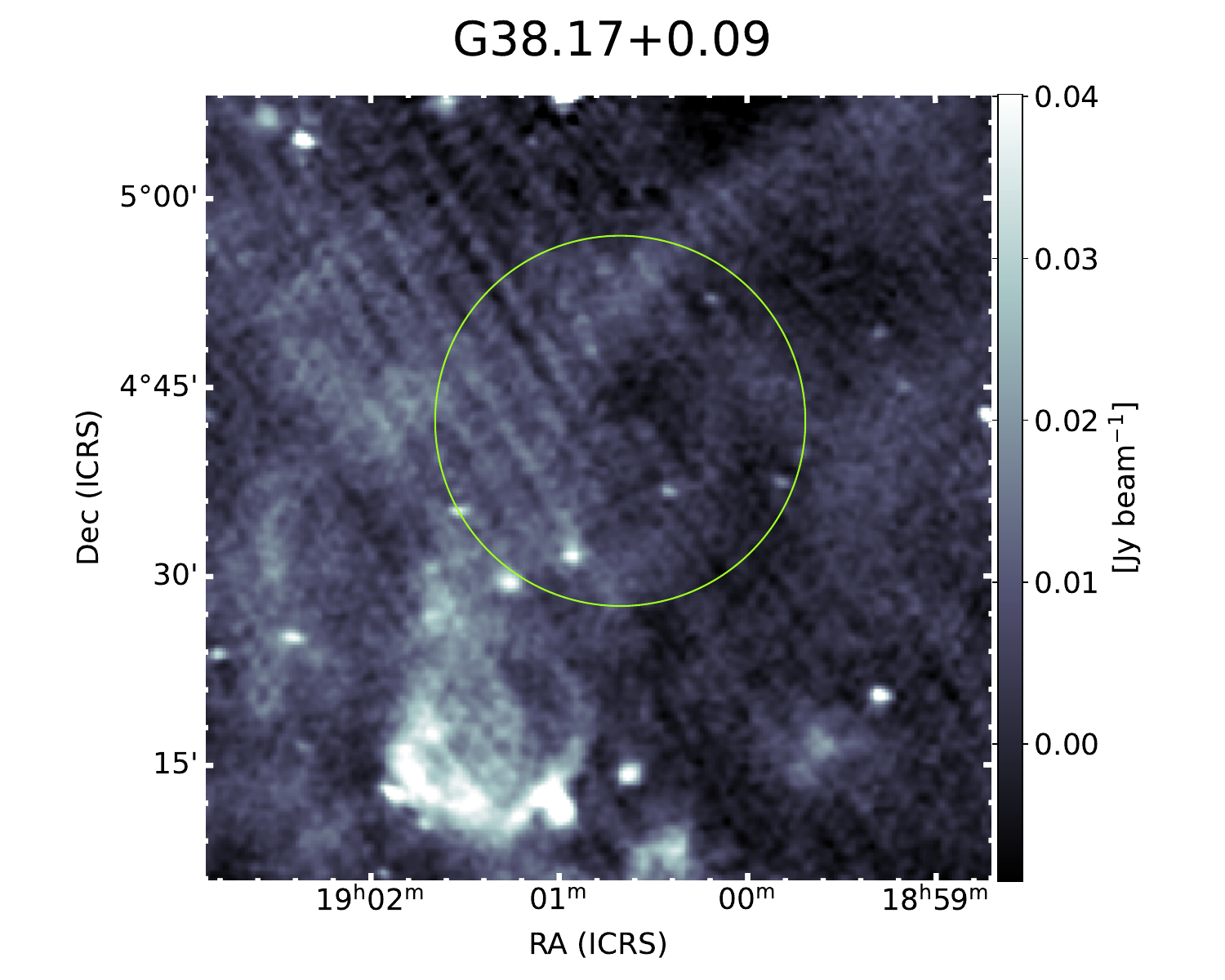}
    \caption{G38.17 in radio. The radio map is from the LOFAR project LC18\_027.}
    \label{fig:G38.17+0.09}
   \end{figure}
    \FloatBarrier
    
\begin{figure}[h!]
   \centering
\includegraphics[width=5.5 cm]{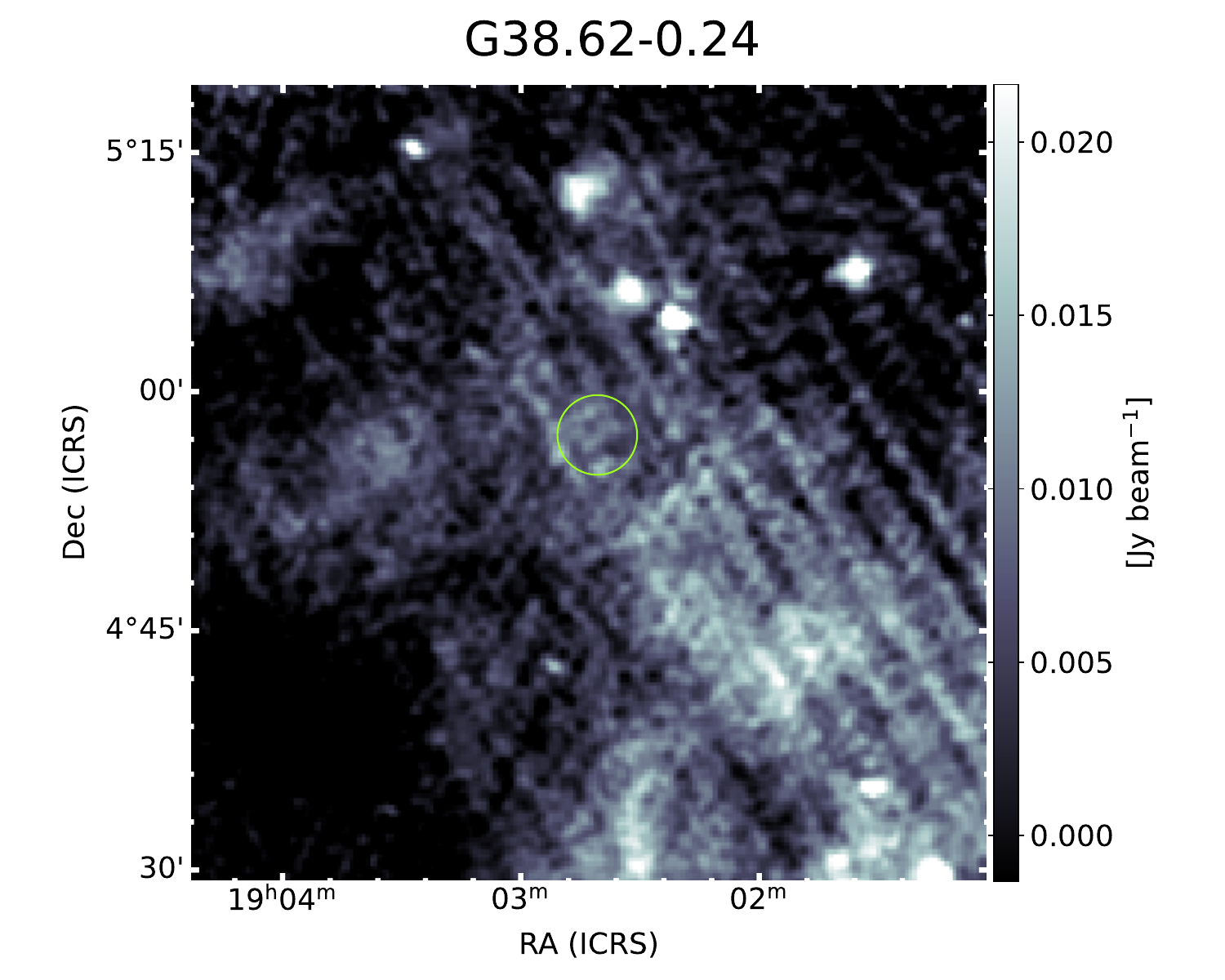}
    \caption{G38.62-0.24 in radio. The radio map is from the LOFAR project LC18\_027.}
    \label{fig:G38.62-0.24}
   \end{figure}
    \FloatBarrier
    
\begin{figure}[h!]
   \centering
\includegraphics[width=5.5 cm]{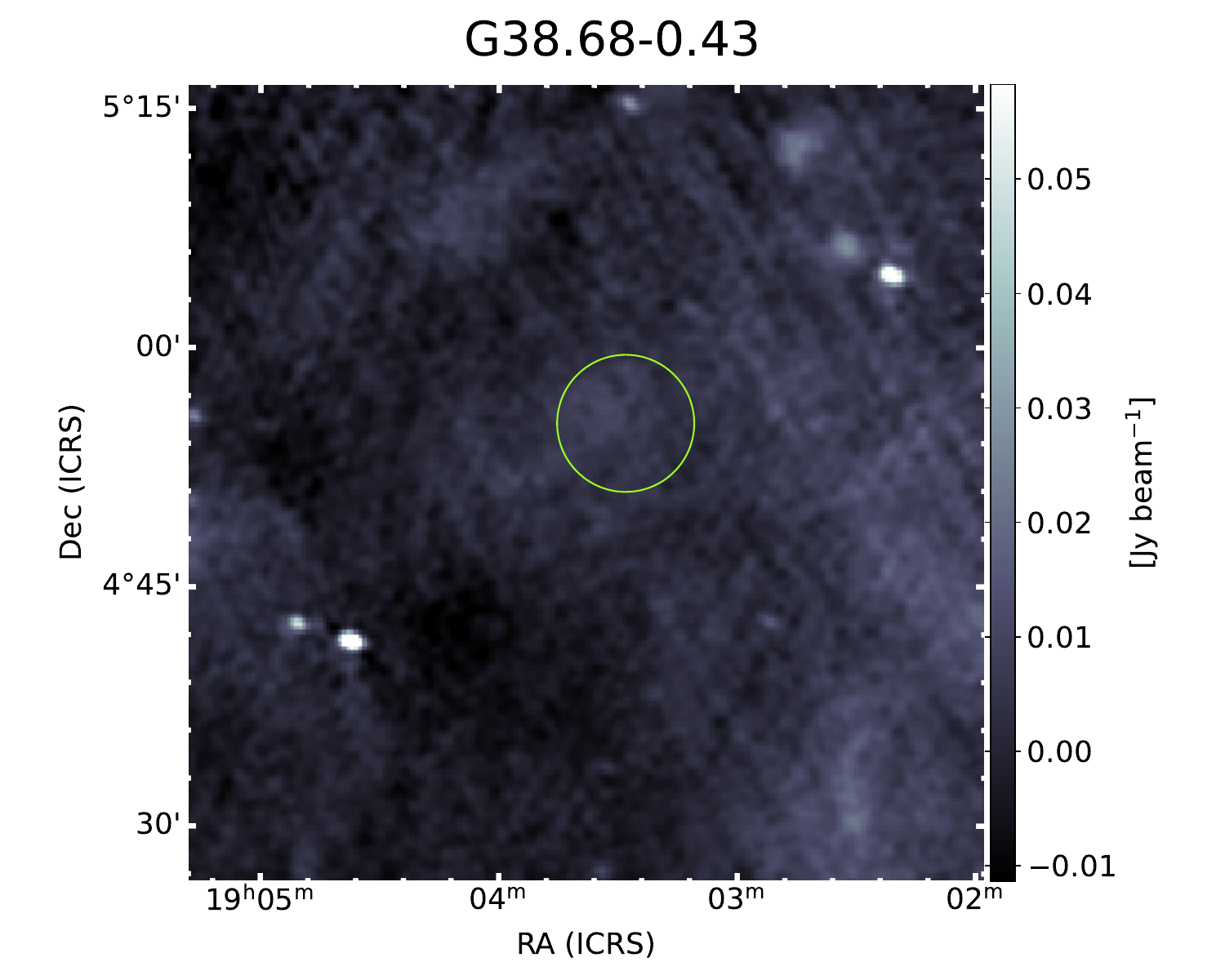}
    \caption{G38.68-0.43 in radio. The radio map is from the LOFAR project LC18\_027.}
    \label{fig:G38.68-0.43}
   \end{figure}
    \FloatBarrier
    
\begin{figure}[h!]
   \centering
\includegraphics[width=5.5 cm]{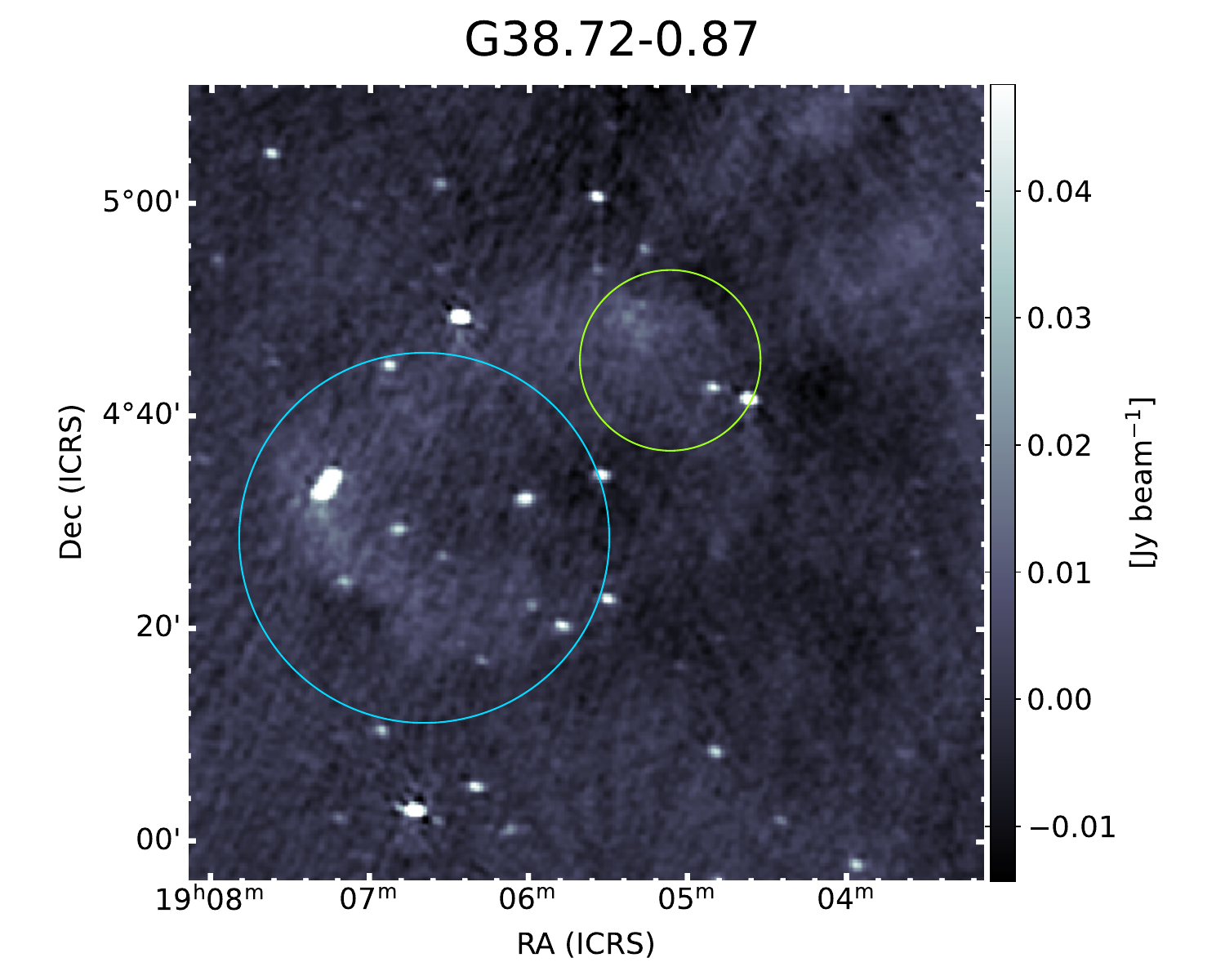}
    \caption{G38.72-0.87 in radio. The radio map is from the LOFAR project LC18\_027. The blue solid line indicates the position of the known SNR G38.7-1.3 (G22 catalogue).}
    \label{fig:G38.72-0.87}
   \end{figure}
    \FloatBarrier
    
\begin{figure}[h!]
   \centering
\includegraphics[width=5.5 cm]{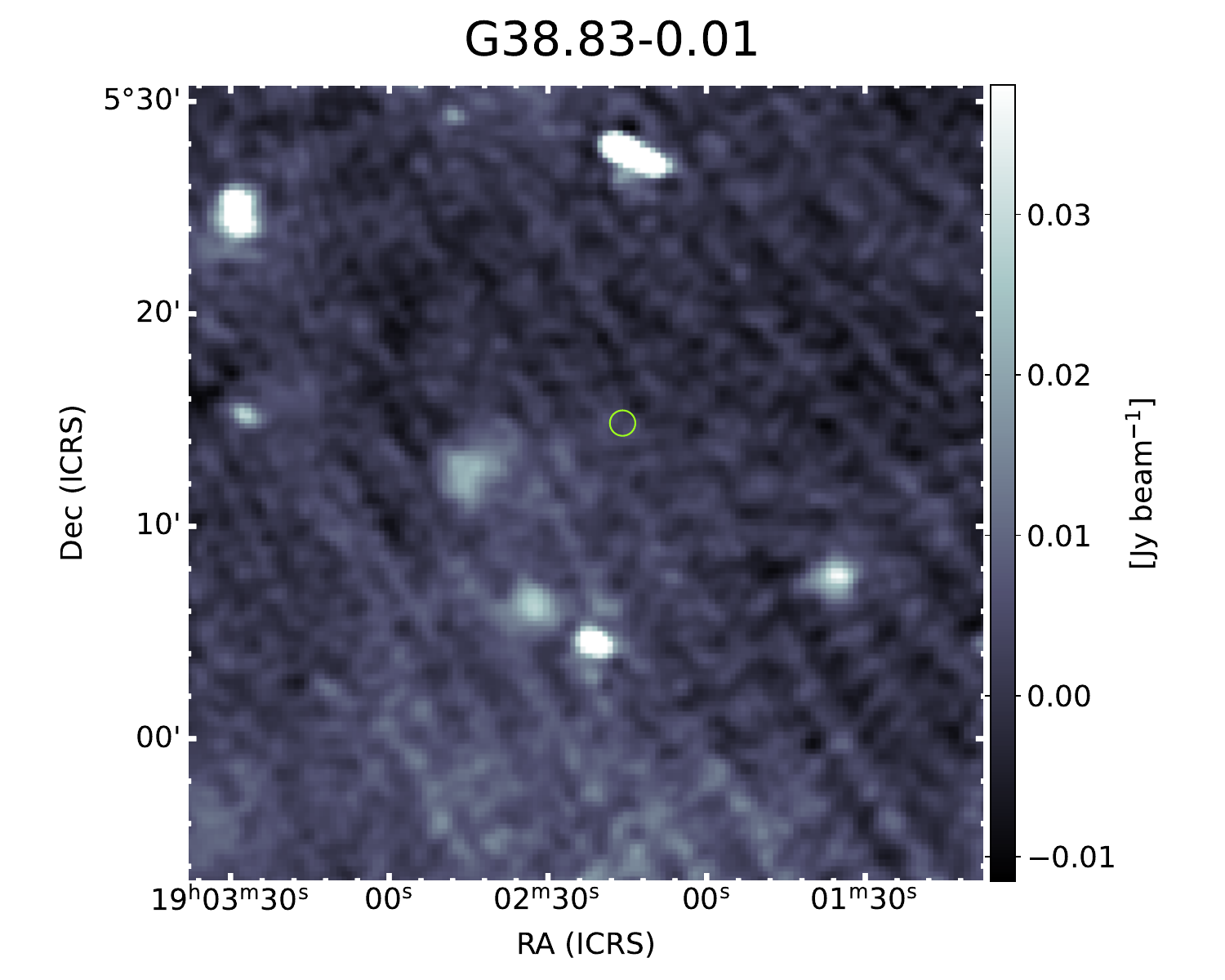}
    \caption{G38.83-0.01 in radio. The radio map is from the LOFAR project LC18\_027.}
    \label{fig:G38.83-0.01}
   \end{figure}
    \FloatBarrier
    
\begin{figure}[h!]
   \centering
\includegraphics[width=5.5 cm]{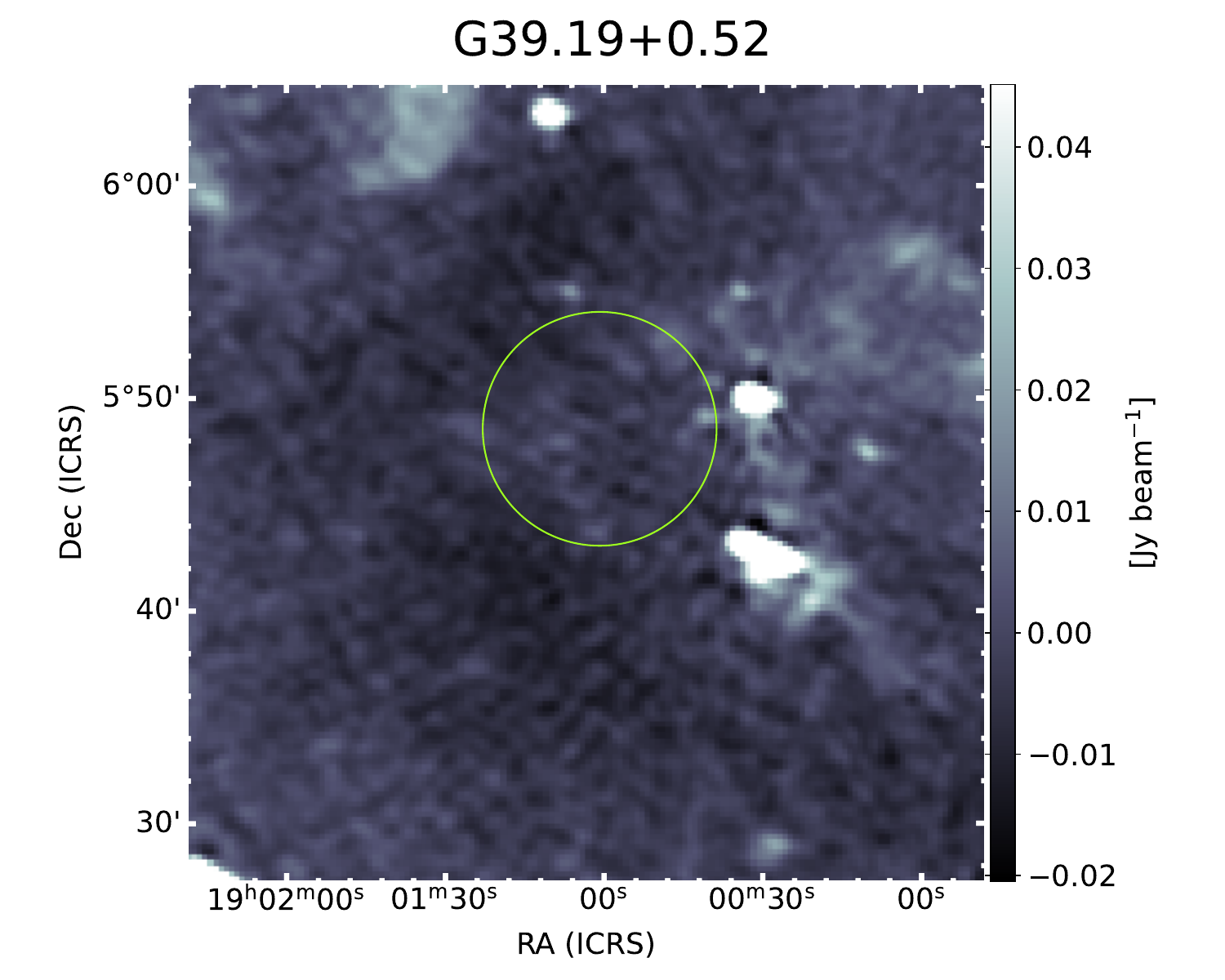}
    \caption{G39.19+0.52 in radio. The radio map is from the LOFAR project LC18\_027.}
    \label{fig:G39.19+0.52}
   \end{figure}
    \FloatBarrier
    
\begin{figure}[h!]
   \centering
\includegraphics[width=5.5 cm]{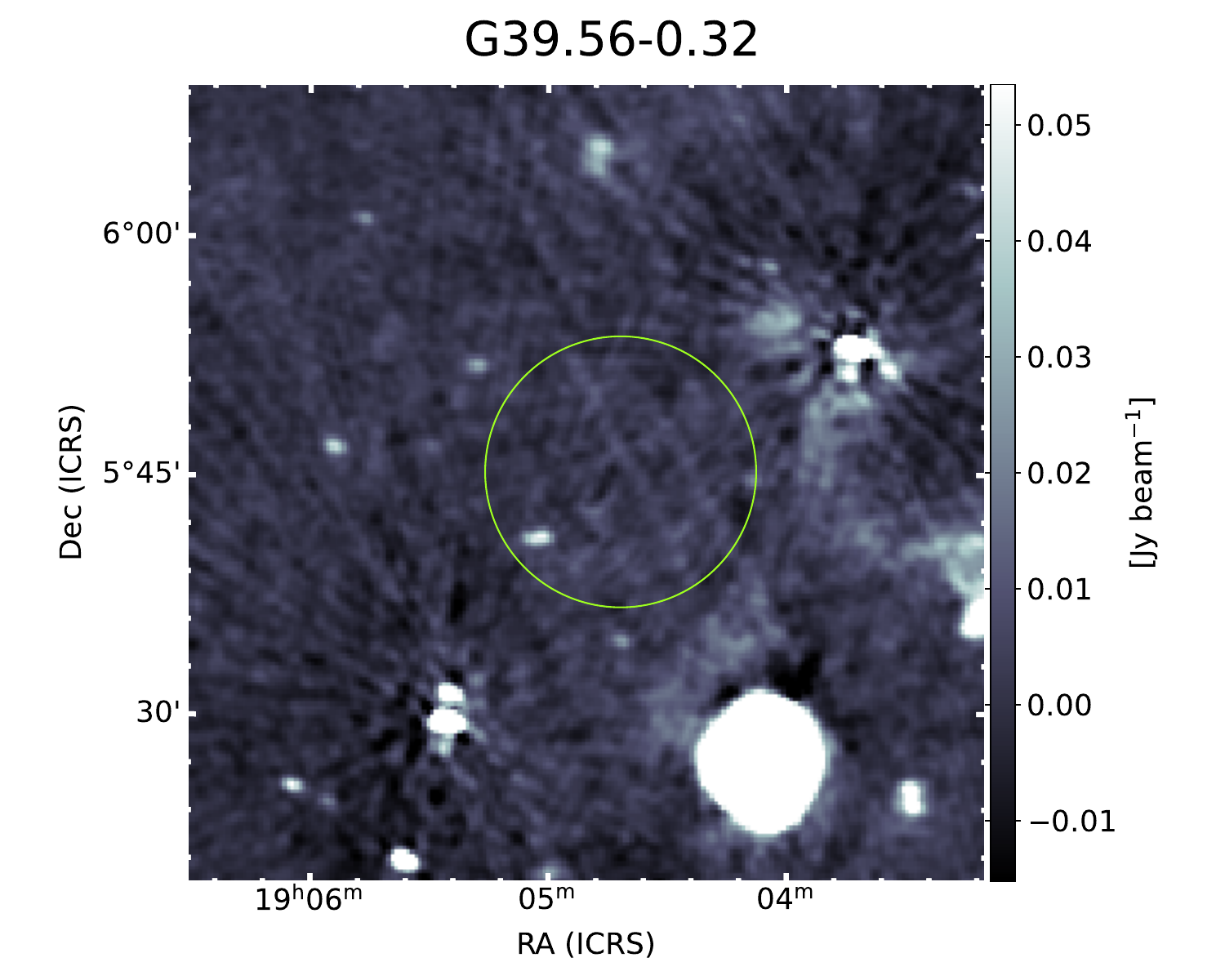}
    \caption{G39.56-0.32 in radio. The radio map is from the LOFAR project LC18\_027.}
    \label{fig:G39.56-0.32}
   \end{figure}
    \FloatBarrier

\begin{figure}[h!]
   \centering
\includegraphics[width=5.5 cm]{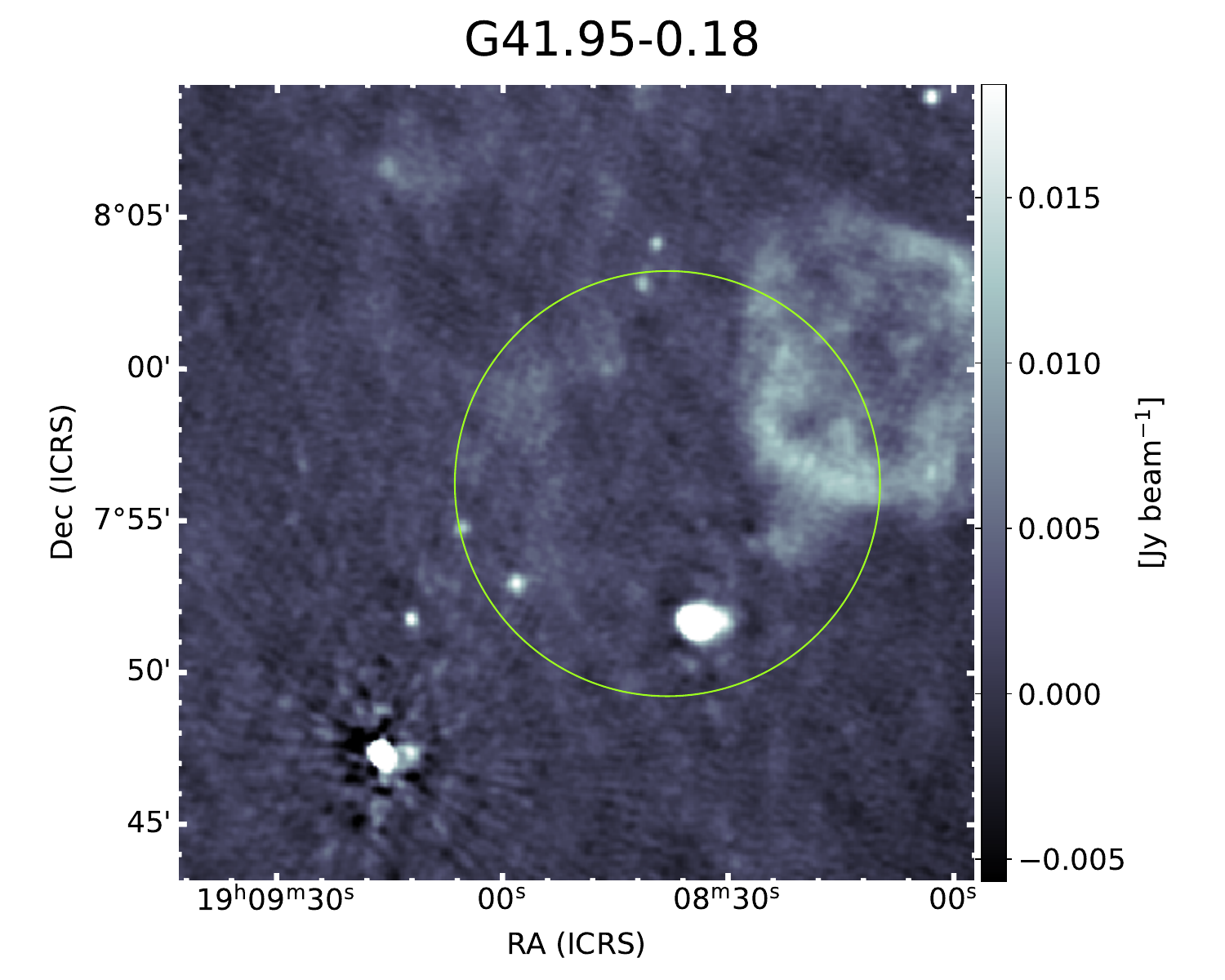}
    \caption{G41.95-0.18 in radio. The radio map is from the LoTSS survey.}
    \label{fig:G41.95-0.18}
   \end{figure}
    \FloatBarrier

\begin{figure}[h!]
   \centering
\includegraphics[width=5.5 cm]{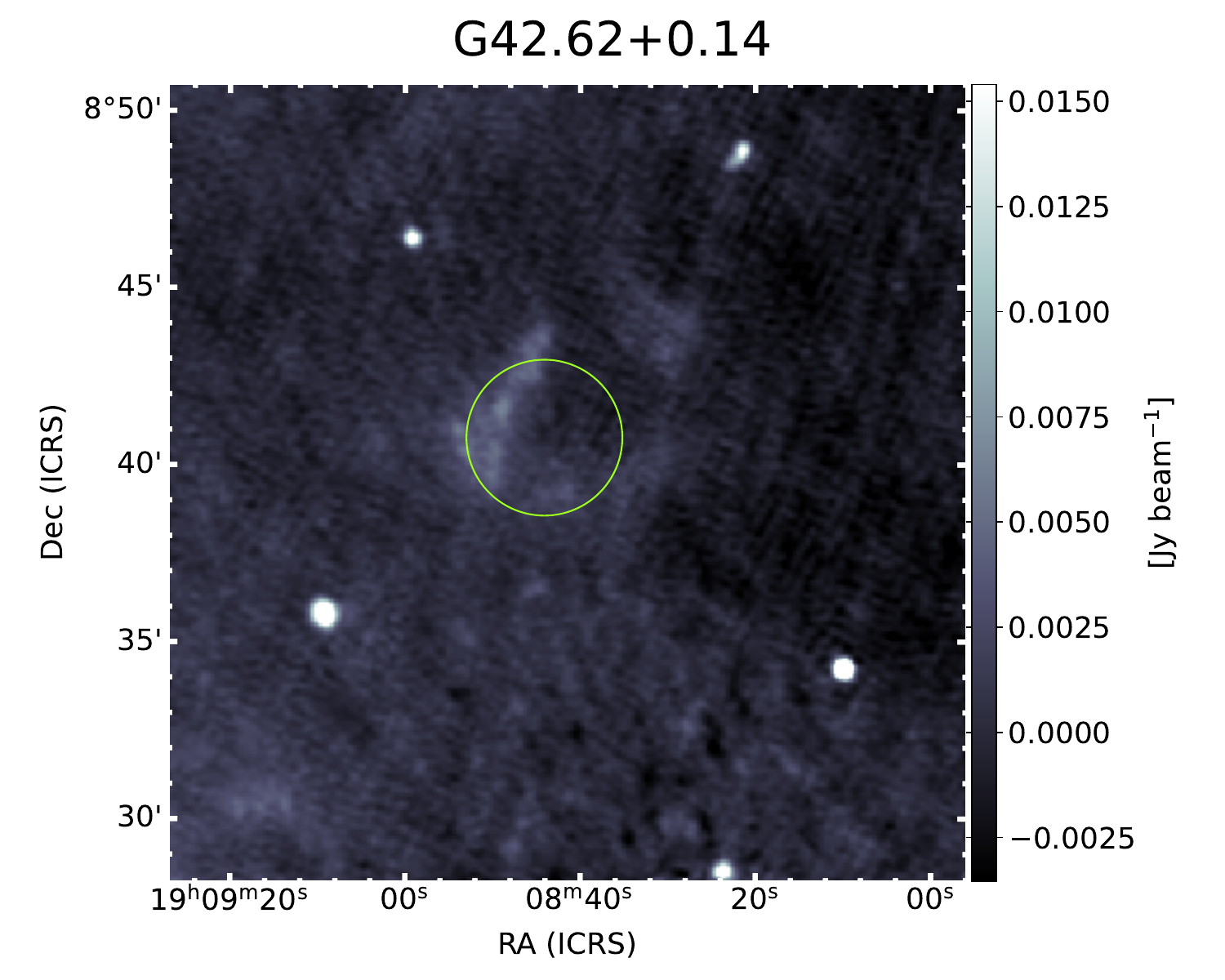}
    \caption{G42.62+0.14 in radio. The radio map is from the LoTSS survey.}
    \label{fig:G42.62+0.14}
   \end{figure}
    \FloatBarrier
    
\begin{figure}[h!]
   \centering
\includegraphics[width=5.5 cm]{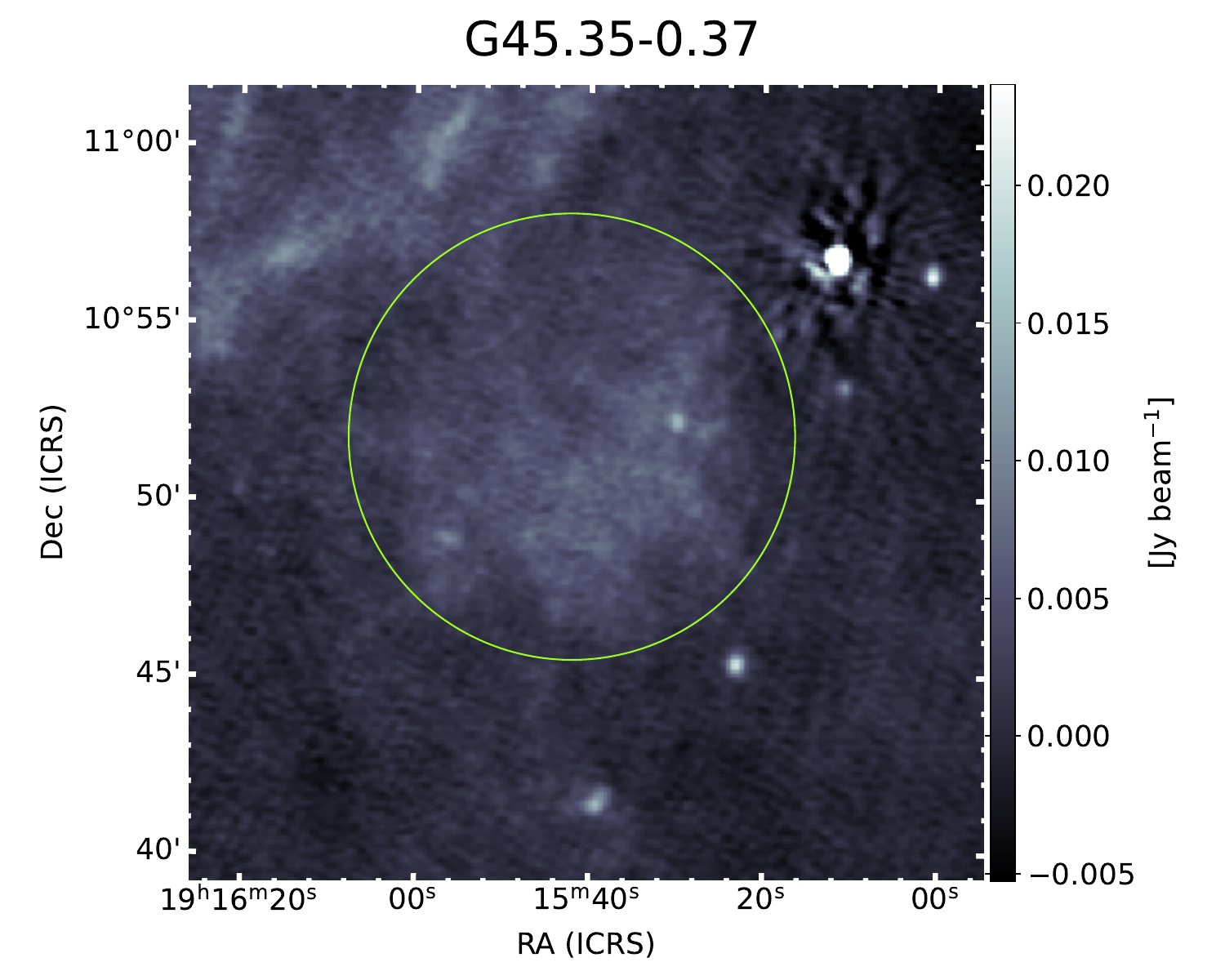}
    \caption{G45.35-0.37 in radio. The radio map is from the LoTSS survey.}
    \label{fig:G45.35-0.37}
   \end{figure}   
    \FloatBarrier
    
\begin{figure}[h!]
   \centering
\includegraphics[width=5.5 cm]{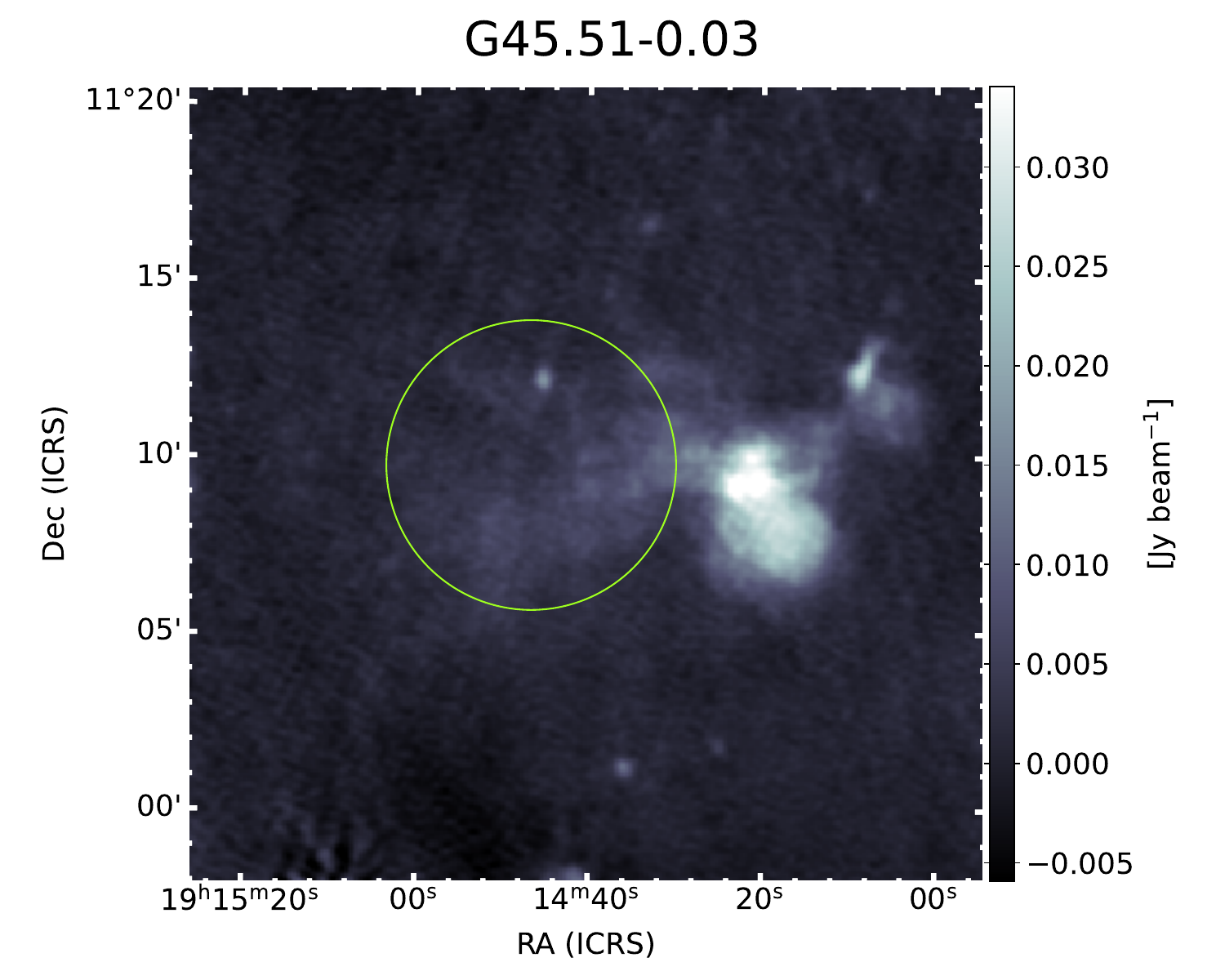}
    \caption{G45.51-0.03 in radio. The radio map is from the LoTSS survey.}
    \label{fig:G45.51-0.03}
   \end{figure} 
    \FloatBarrier
    
\begin{figure}[h!]
   \centering
\includegraphics[width=5.5 cm]{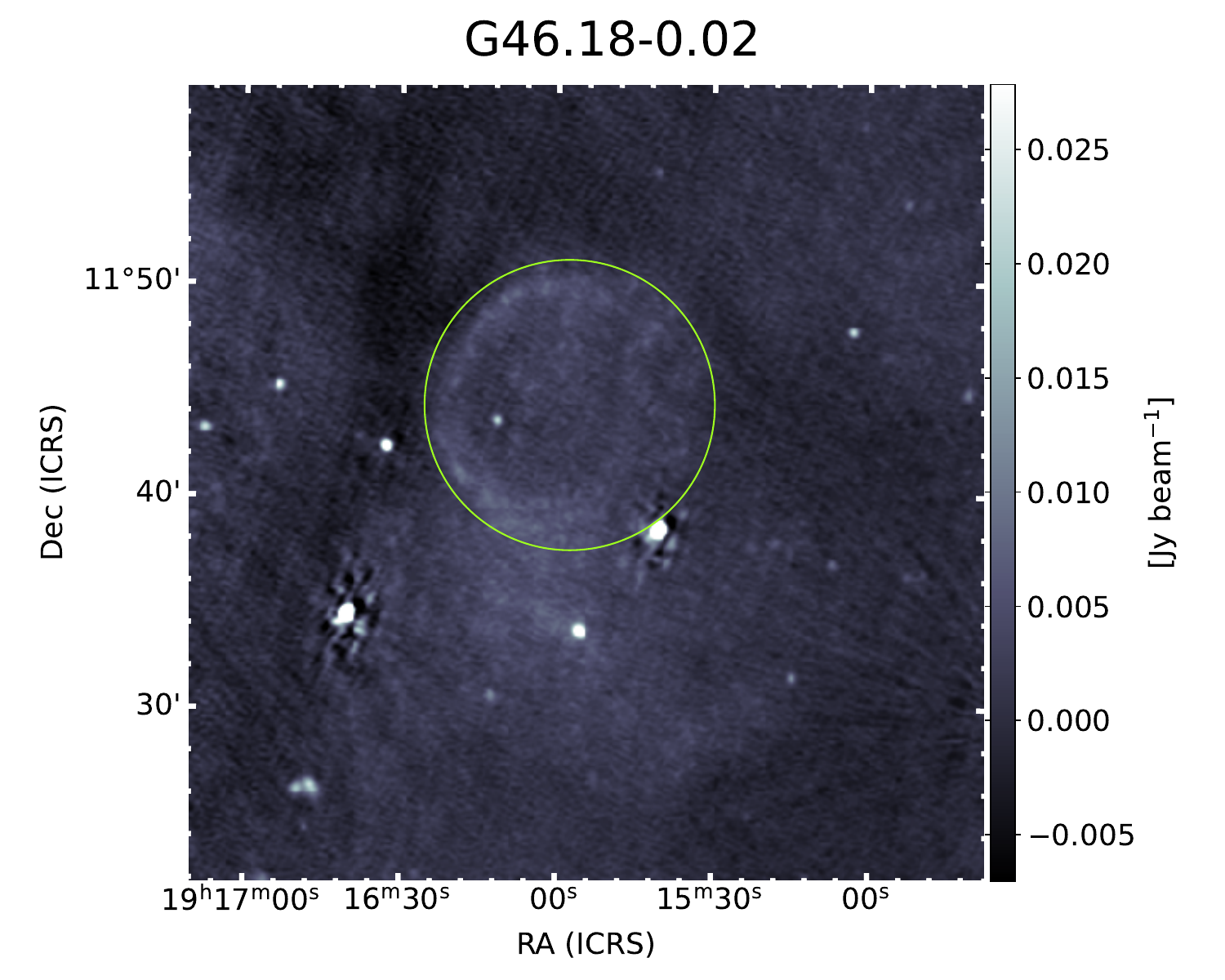}
    \caption{G46.18-0.02 in radio. The radio map is from the LoTSS survey.}
    \label{fig:G46.18-0.02}
   \end{figure} 
    \FloatBarrier
    
\begin{figure}[h!]
   \centering
\includegraphics[width=5.5 cm]{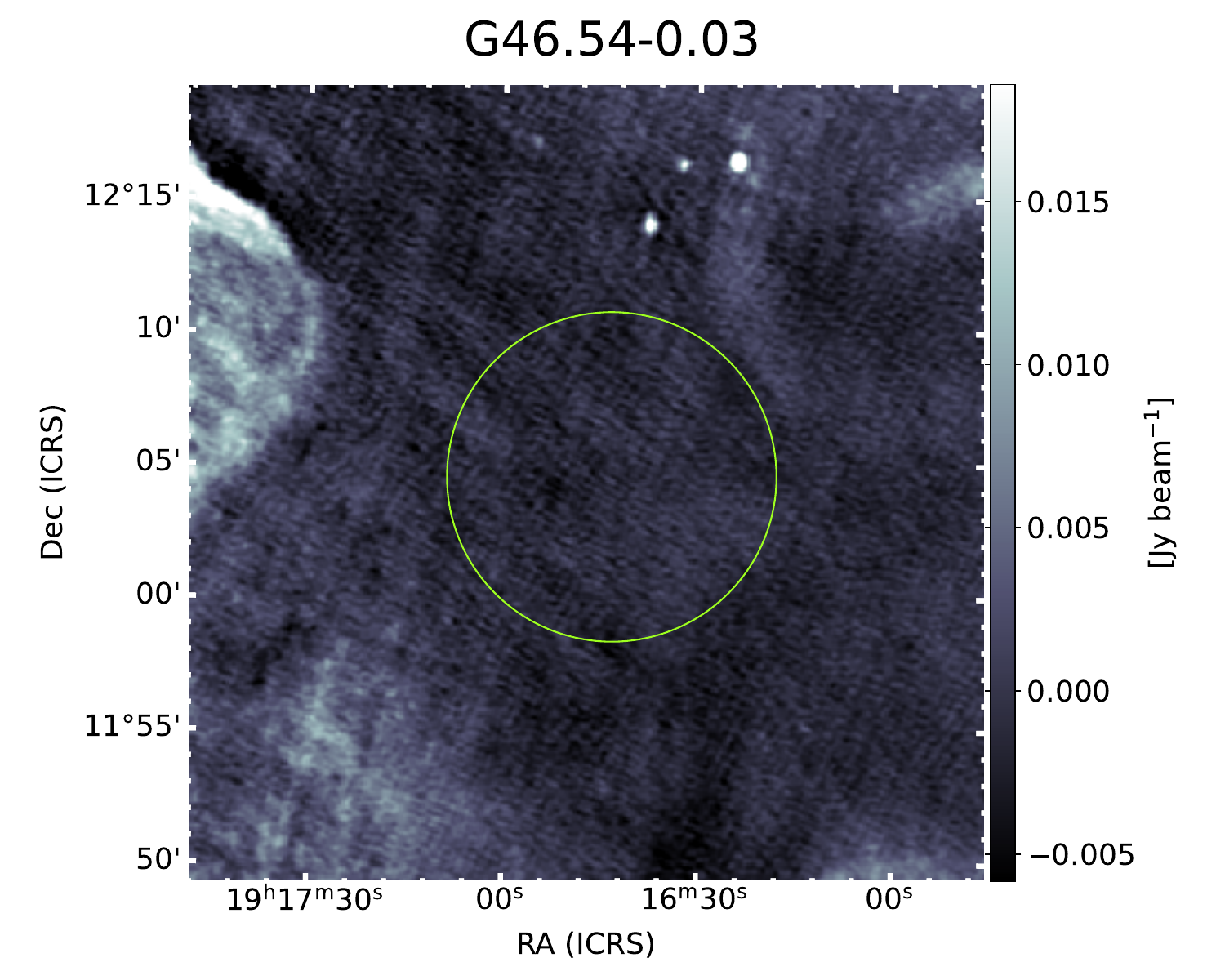}
    \caption{G46.54-0.03 in radio. The radio map is from the LoTSS survey.}
    \label{fig:G46.54-0.03}
   \end{figure} 
    \FloatBarrier
    
\begin{figure}[h!]
   \centering
\includegraphics[width=5.5 cm]{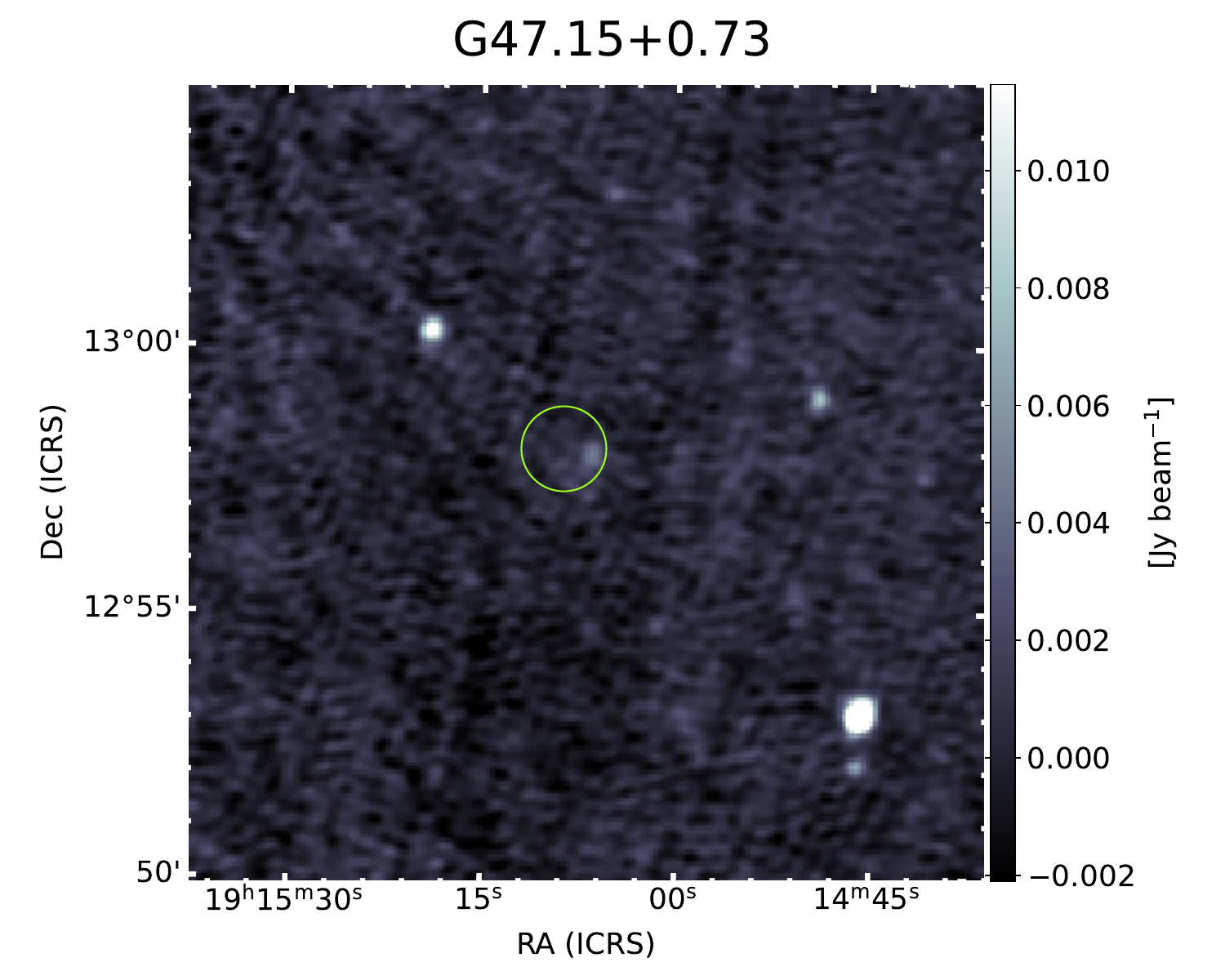}
    \caption{G47.15+0.73 in radio. The radio map is from the LoTSS survey.}
    \label{fig:G47.15+0.73}
   \end{figure} 
    \FloatBarrier
    
\begin{figure}[h!]
   \centering
\includegraphics[width=5.5 cm]{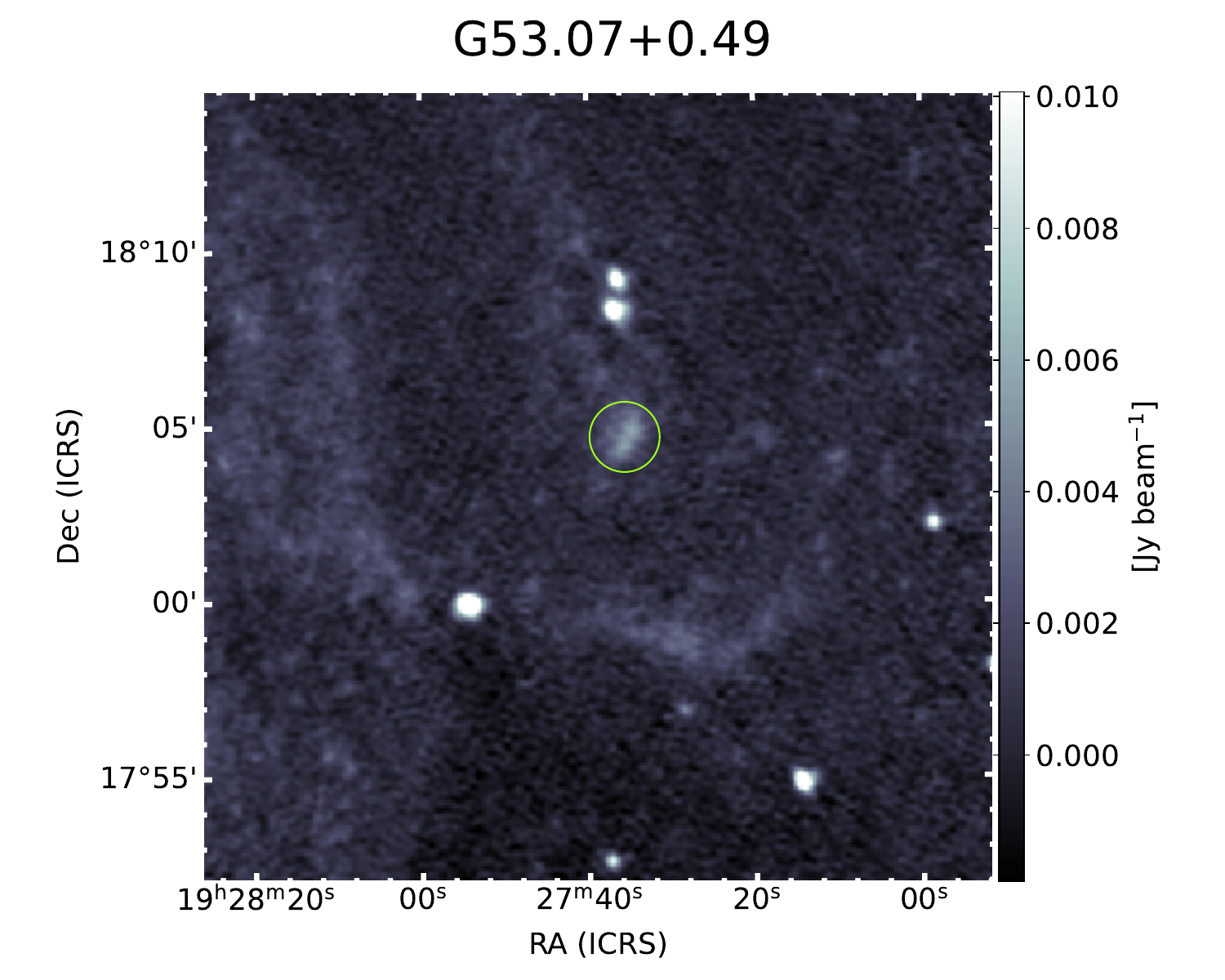}
    \caption{G53.07+0.49 in radio. The radio map is from the LoTSS survey.}
    \label{fig:G53.07+0.49}
   \end{figure} 
    \FloatBarrier
    
\begin{figure}[h!]
   \centering
\includegraphics[width=5.5 cm]{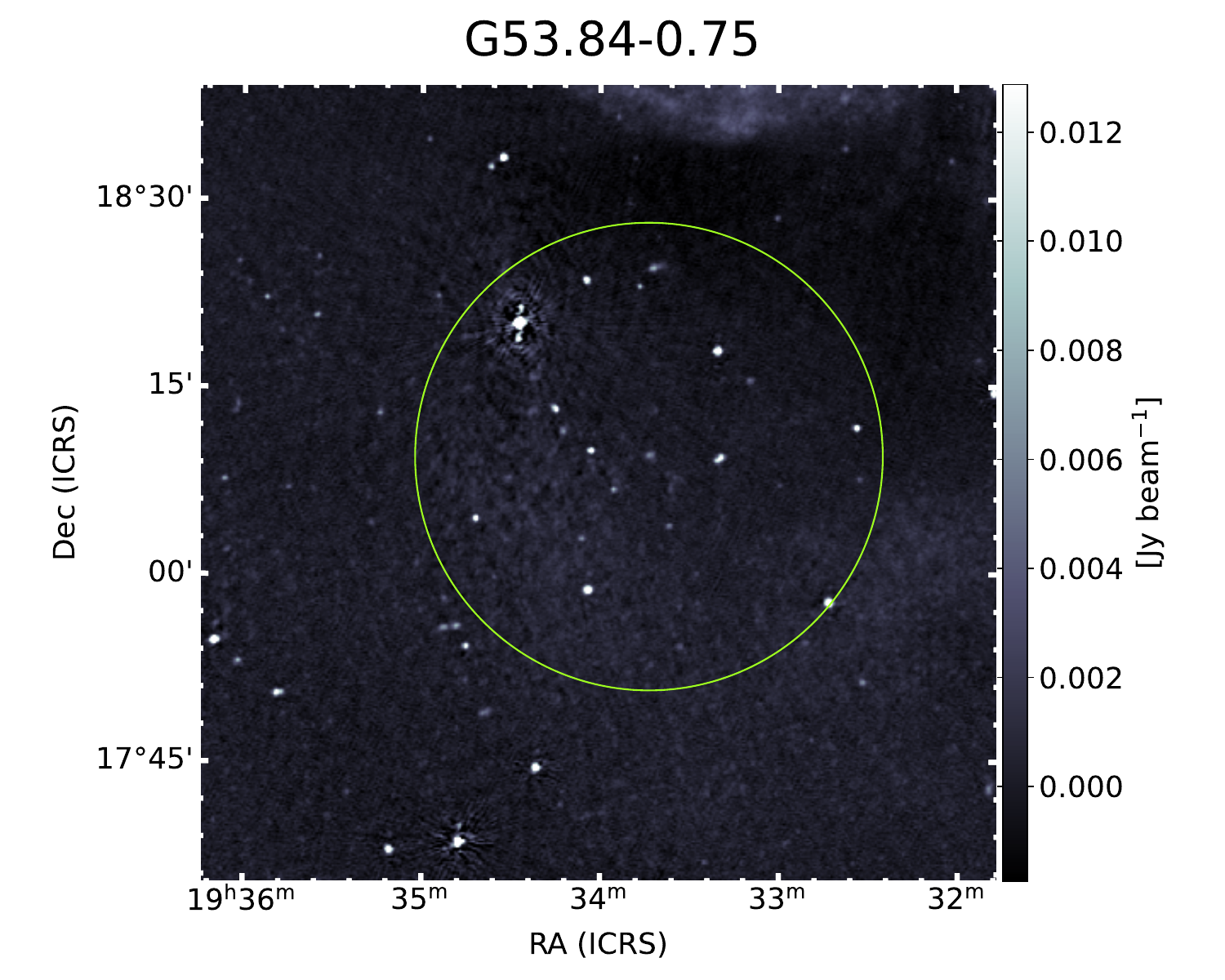}
    \caption{G53.84-0.75 in radio. The radio map is from the LoTSS survey.}
    \label{fig:G53.84-0.75}
   \end{figure} 
    \FloatBarrier
    
\begin{figure}[h!]
   \centering
\includegraphics[width=5.5 cm]{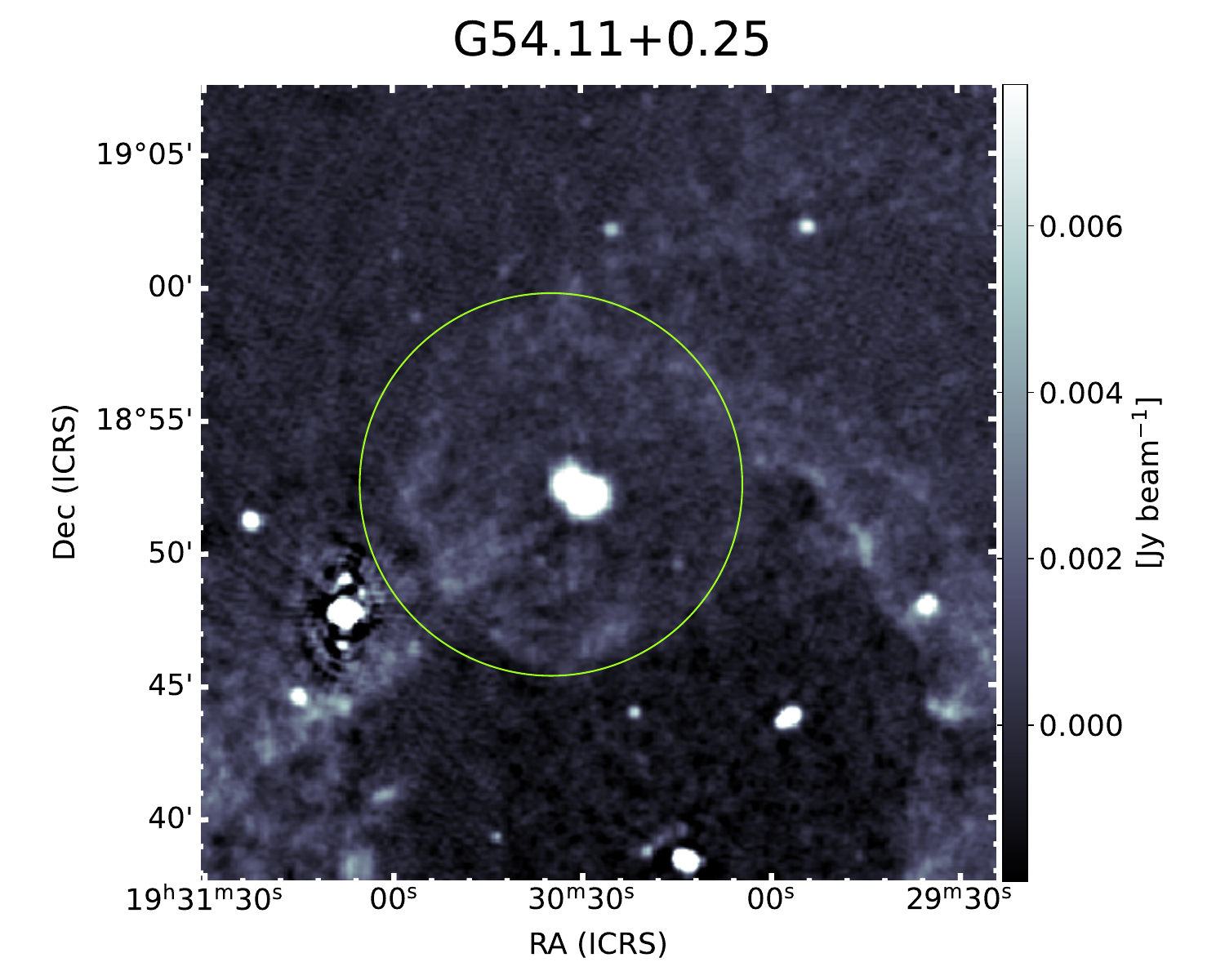}
    \caption{G54.11+0.25 in radio. The radio map is from the LoTSS survey.}
    \label{fig:G54.11+0.25}
   \end{figure}    
    \FloatBarrier
    
\begin{figure}[h!]
   \centering
\includegraphics[width=5.5 cm]{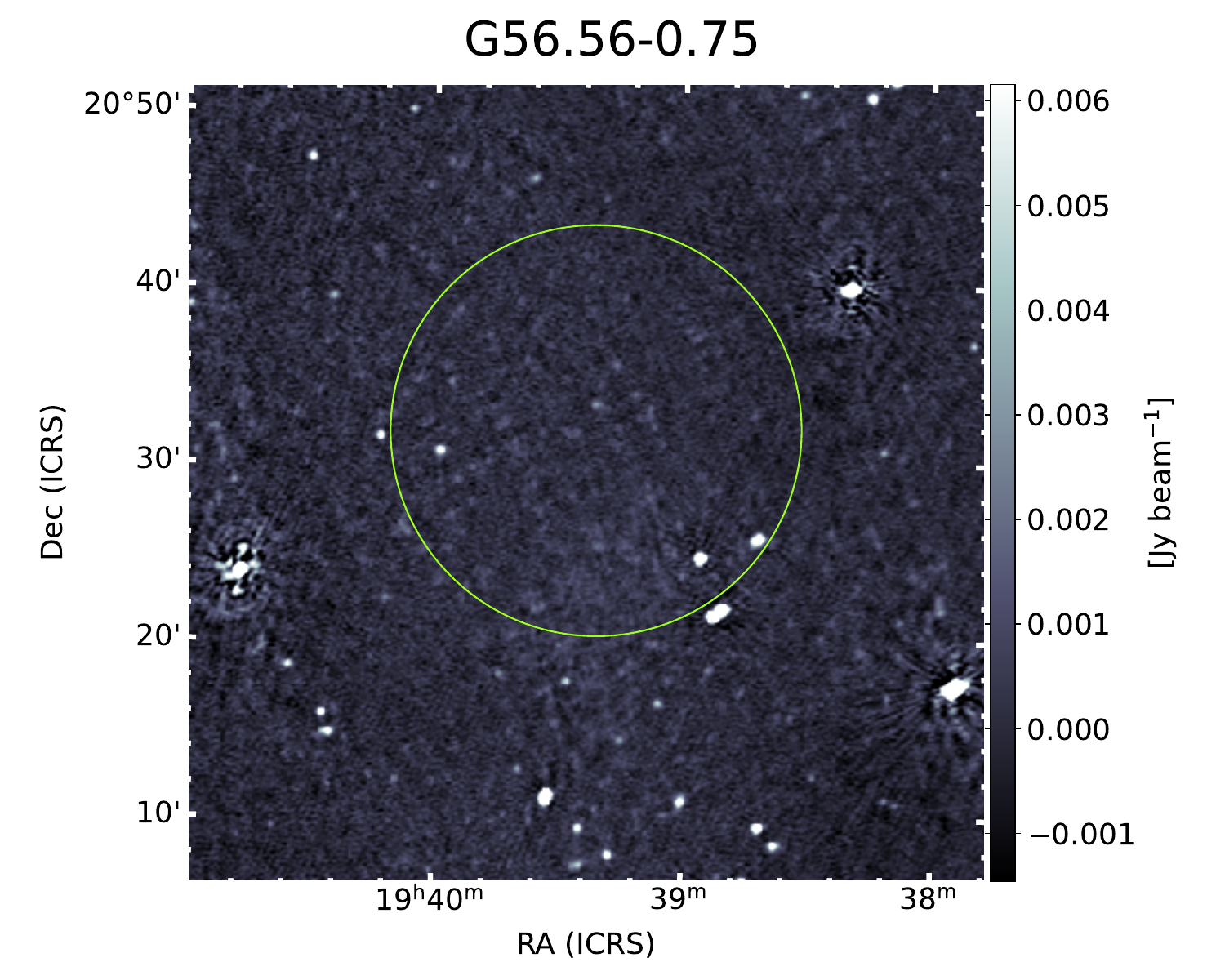}
    \caption{G56.56-0.75 in radio. The radio map is from the LoTSS survey.}
    \label{fig:G56.56-0.75}
   \end{figure} 
    \FloatBarrier
    
\begin{figure}[h!]
   \centering
\includegraphics[width=5.5 cm]{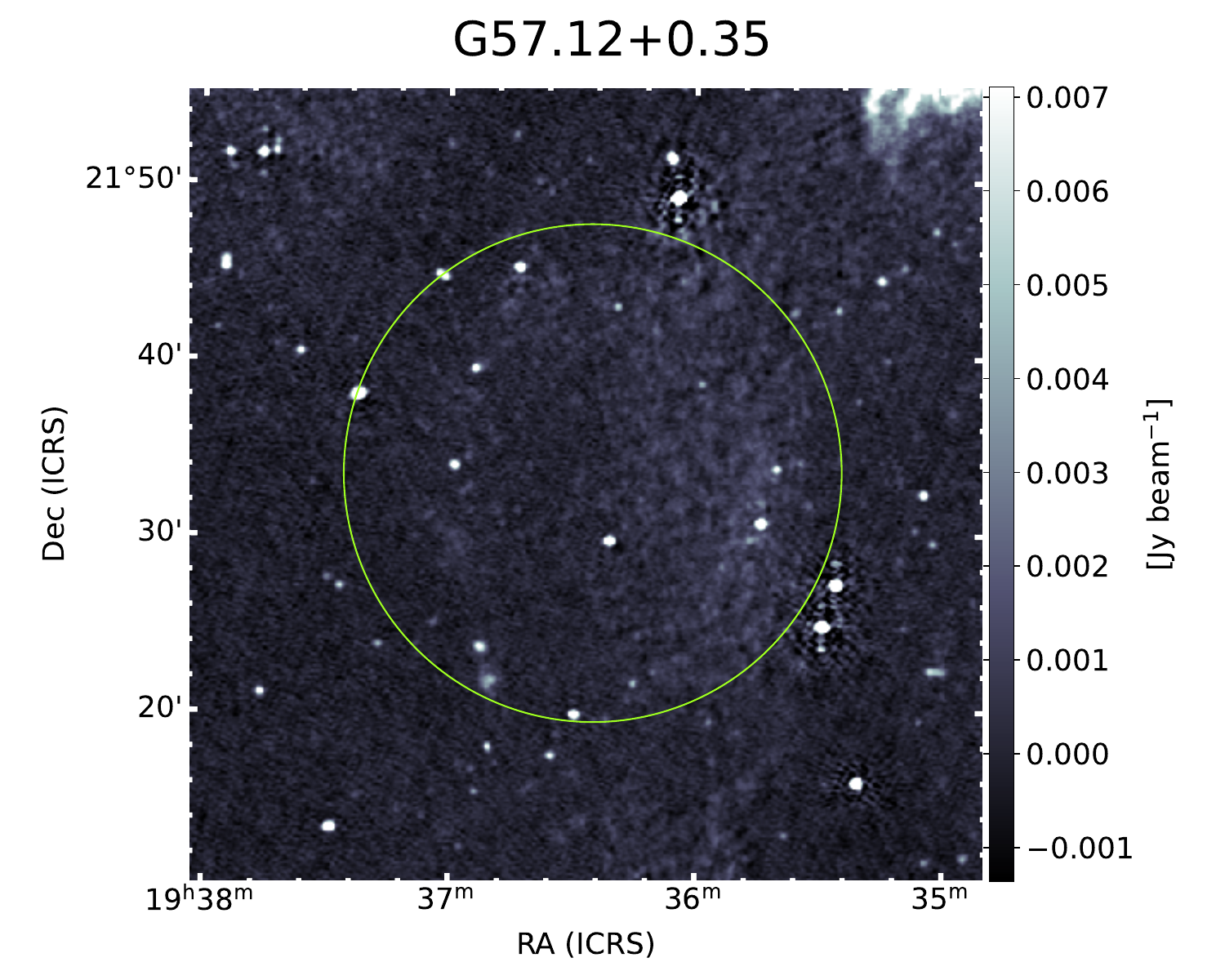}
    \caption{G57.12+0.35 in radio. The radio map is from the LoTSS survey.}
    \label{fig:G57.12+0.35}
   \end{figure}
    \FloatBarrier

\end{appendix}
\end{document}